\documentclass[pra,twocolumn]{revtex4-2}
\usepackage{graphicx}
\usepackage{amssymb}
\usepackage{amsmath}
\usepackage{bm}
\usepackage{color}
\usepackage{physics}

\usepackage[colorlinks,urlcolor=blue,citecolor=blue,linkcolor=blue]{hyperref}
\usepackage{comment}
\usepackage{cleveref}
\usepackage{mathtools}
\usepackage{mathrsfs}

\newcommand{\sch}{Schr{\"o}dinger }
\newcommand{\eb}{\varepsilon_B}
\newcommand{\nn}{\nonumber}
\renewcommand{\k}{\mathbf{k}}
\newcommand{\0}{\mathbf{0}}
\newcommand{\p}{\mathbf{p}}
\newcommand{\q}{\mathbf{q}}
\newcommand{\Q}{\mathbf{Q}}
\newcommand{\pone}{\mathbf{p}_1}

\newcommand{\ptwo}{\mathbf{p}_2}

\newcommand{\Qe}{\mathbf{Q}_e}
\newcommand{\Qh}{\mathbf{Q}_h}

\newcommand{\kP}{\mathbf{k}^\prime}

\newcommand{\ok}{\bar{\epsilon}_\k}

\begin{document}

\title{Theory of polariton-electron interactions in 
semiconductor microcavities}

\author{Guangyao Li}
\affiliation{School of Physics and Astronomy, Monash University, Victoria 3800, Australia}
\affiliation{ARC Centre of Excellence in Future Low-Energy Electronics Technologies, Monash University, Victoria 3800, Australia}

\author{Olivier Bleu}
\affiliation{School of Physics and Astronomy, Monash University, Victoria 3800, Australia}
\affiliation{ARC Centre of Excellence in Future Low-Energy Electronics Technologies, Monash University, Victoria 3800, Australia}

\author{Jesper Levinsen}
\affiliation{School of Physics and Astronomy, Monash University, Victoria 3800, Australia}
\affiliation{ARC Centre of Excellence in Future Low-Energy Electronics Technologies, Monash University, Victoria 3800, Australia}

\author{Meera M. Parish}
\affiliation{School of Physics and Astronomy, Monash University, Victoria 3800, Australia}
\affiliation{ARC Centre of Excellence in Future Low-Energy Electronics Technologies, Monash University, Victoria 3800, Australia}

\begin{abstract}
We develop a microscopic description of an electron-doped two-dimensional semiconductor embedded in a microcavity.
Specifically, we investigate the interactions between exciton-polaritons and electrons for the case where the interactions between charges are strongly screened and the system is spin polarized. As a starting point, we obtain an analytic expression for the exciton-polariton wave function,
and we relate the microscopic parameters of the light-matter system to experimentally measurable quantities, such as the Rabi coupling and the cavity photon frequency. We then derive the polariton-electron interaction within the standard Born approximation and compare it with the exact polariton-electron scattering $T$ matrix that we obtain from a diagrammatic approach that has proven highly successful in the context of nuclear physics and ultracold atomic gases. In particular, we show that the Born approximation provides an upper bound on the polariton-electron coupling strength at vanishing momentum. Using our exact microscopic calculation, we demonstrate that polariton-electron scattering can be strongly enhanced compared to the exciton-electron case, which is the opposite of that expected from the Born approximation. We furthermore expose a resonance-like peak at scattering momenta near the polariton inflection point, whose size is set by the strength of the light-matter coupling. Our results arise from the non-Galilean nature of the polariton system and should thus be applicable to a range of semiconductor microcavities such as GaAs quantum wells and atomically thin materials.
\end{abstract}

\maketitle

\section{Introduction}
By embedding a two-dimensional (2D) semiconductor in a microcavity, one can achieve a strong coupling between photons and bound electron-hole pairs (i.e., excitons). This in turn gives rise to exciton-polaritons, quasiparticles which are superpositions of both light and matter~\cite{DengRevMod10,CiutiRevMod13,ByrnesNatPhys14}. A key advantage of polaritons over ordinary photons is that they can pairwise interact with other particles via their excitonic component, which is important for a range of applications including polariton superfluidity \cite{Amo2009,Sanvitto2010}, ultra-fast polariton spin switching \cite{Amo2010}, and the generation of photon correlations~\cite{Munoz2019,Delteil2019}. Moreover, there is the prospect of enhancing the interactions in the polariton system by coupling to few-body bound states such as biexcitons~\cite{TakemuraNPhys2014} or trions~\cite{Tan2020}.

Interaction processes involving polaritons are conventionally described using the Born approximation~\cite{TassonePRB99,RamonPRB02,Malpuech2002,SnokePRB2010,Shahnazaryan2017}. This can essentially be viewed as a two step process. First, the matter component interactions are calculated within the Born approximation, and then the result is weighted by the matter fraction of the involved particles since only this component interacts. There have been various attempts \cite{TassonePRB99,RochatPRB00, Combescot_2007, COMBESCOTPhysRep2008, GlazovPRB2009} to obtain perturbative corrections beyond the standard Born approximation to include higher-order effects due to the light-matter coupling. However, until now there have been no exact microscopic calculations of few-body processes that involve the constituent electron-hole-photon components of the polaritons, which is mainly due to the complexity associated with treating both the electronic and the light-matter interactions. It is therefore an open and important question whether the assumptions in the Born approximation are valid even at a qualitative level. 

In this paper, which accompanies Ref.~\cite{ePshort}, we use a microscopic model that explicitly includes electrons, holes, and photons to investigate this question. The key simplification 
is to consider strongly screened electronic interactions. Such a simplified model has previously been used to theoretically investigate the crossover from a Bose-Einstein condensate (BEC) of polaritons to the Bardeen-Cooper-Schrieffer regime of electron-hole superfluidity~\cite{Yamaguchi2012,Hanai2017,Hu2019QuantumFI}. Here we show that this approximation allows us to formulate a diagrammatic approach where we can analytically obtain the polariton wave function and propagator, as well as the associated spectrum and photon and exciton fractions. Our resulting microscopic description of exciton-polaritons complements our previous results for Coulomb electronic interactions~\cite{JesperPRR19}, and it can act as a starting point for further few- and many-body calculations within the model of strongly screened electronic interactions.

We then apply our diagrammatic approach to perform the first exact calculation of spin-polarized polariton-electron scattering within a microscopic model. Importantly, we find that the polariton-electron scattering is strongly enhanced up to the polariton dispersion inflection point compared with exciton-electron scattering, in stark contrast to the assumptions of the Born approximation. We argue that this unexpected feature is because the light-matter coupling shifts the collision energy, 
which has a stronger effect than the reduction in the interaction strength due the reduced exciton fraction of the polariton.
Given that this argument relies only on standard 2D 
scattering theory~\cite{AdhikariAJP86} which is valid also for unscreened electronic interactions, we expect the enhanced polariton-electron interactions to be a generic feature of 2D 
semiconductors embedded in microcavities, and we expect our results to be quite accurate even at a quantitative level in the case of more realistic electronic interactions.

Of particular interest is the scattering at low momentum, due to the potential implications for polariton BECs~\cite{BEC06} in charge-doped semiconductors. It is well known~\cite{landau2013quantum} that the scattering of 2D quantum particles with short range interactions approaches zero logarithmically with momentum. In the case of exciton-electron scattering, the relevant momentum scale is the inverse exciton Bohr radius. Remarkably, in the case of polariton-electron scattering, we find that the resulting momentum scale is suppressed exponentially by the large electron-photon mass ratio, and consequently the polariton-electron scattering only approaches zero in systems that greatly exceed the size of the universe. This is a striking consequence of the polariton being formed from a superposition of particles with extremely different masses, and it allows us to define a finite coupling constant at vanishing momentum.

We furthermore find that the Born approximation severely overestimates the polariton-electron interaction constant for typical experimental parameters. We argue that this is because 
the Born approximation represents an \textit{upper bound} on the interaction constant. While this implies that this simple approximation cannot in general be trusted, we show that it can be replaced by a similarly simple expression that calculates the interaction constant instead from the low-energy exciton-electron scattering amplitude, which represents an excellent approximation in the case of a transition metal dichalcogenide (TMD) monolayer or a single GaAs quantum well embedded in a microcavity.

The paper is organized as follows. In Sec.~\ref{sec:Hamiltonian}, we present the microscopic description of a single polariton, taking care to appropriately renormalize the model. We obtain analytic expressions for the polariton wave function, spectrum, and excitonic and photonic fractions, and demonstrate that our results can be obtained within both an operator and a diagrammatic approach. In Sec.~\ref{sec:e-P scattering zero k} we outline the Born approximation of polariton-electron scattering, 
 which provides an upper bound 
 for our exact calculation 
 presented in Sec.~\ref{sec:e-P_scattering}. We discuss how our results depend on system parameters relevant to both semiconductor quantum wells and atomically thin semiconductors. In Sec.~\ref{sec:conclusion} we conclude. Technical details are given in the appendices.

\section{Theoretical description of exciton-polaritons}\label{sec:Hamiltonian}

We consider a system consisting of spin-polarized electrons, holes, and photons in a 2D semiconductor, 
such as a quantum well or an atomically thin transition metal dichalcogenide material. The Hamiltonian reads
\begin{align}\label{eq:Hamiltonian}
H=&\sum_\k \left(\epsilon^e_{\k}\,e^\dagger_\k e_\k+\epsilon^h_{\k}\,h^\dagger_\k h_\k \right)+\sum_\k(\omega+\epsilon^c_{\k})c^\dagger_{\k}c_{\k}\nn\\
&-V_0 \sum_{\k\, \k^\prime\, \q} e^\dagger_{\k}h^\dagger_{\q-\k}h_{\q-\kP}e_{\kP}
\nn \\ & +g\sum_{\k\,\q}\left( e^\dagger_{\k}h^\dagger_{\q-\k} c_{\q} +  c^\dag_{\q}h_{\q-\k} e_{\k} \right).
\end{align}
The first line describes the kinetic energies of the particles, where $e_\k$, $h_\k$, and $c_\k$ are electron, hole, and photon annihilation operators, respectively, with momentum $\k$ and kinetic energy $\epsilon^{e,h,c}_{\k}=|\k|^2/(2 m_{e,h,c})\equiv k^2/(2 m_{e,h,c})$, and corresponding masses $m_e$, $m_h$, and $m_c$. 
$\omega$ is the cavity photon frequency measured from the bandgap. In the second line, we have the electron-hole interactions, which we take 
to be strongly screened contact interactions of strength $V_0>0$. 
Note that there are no electron-electron or hole-hole interactions in this screened case since the interactions between identical fermions formally vanish due to Pauli exclusion. 
The last term corresponds to the creation (annihilation) of an electron-hole pair through the absorption (emission) of a photon
within the rotating wave approximation, 
and $g$ denotes the strength of the (unrenormalized) light-matter coupling. This term explicitly breaks Galilean invariance by coupling the light photon with much heavier matter particles, which has important consequences for the few-body properties of the system.
Note that here and in the following we adopt units where $\hbar=1$ and the system area $A=1$.

The Hamiltonian contains the bare parameters $V_0$ and $g$ that describe the strength of matter-matter and light-matter contact interactions, respectively. We take these to be constant up to an ultraviolet momentum cutoff which is set by the detailed band structure of the 2D semiconductor. For our discussion of polariton physics, the precise value of the momentum cutoff will be irrelevant, since we aim to develop a low-energy theory that is independent of the short-distance physics~\cite{JesperPRR19}. Therefore, both $V_0$ and $g$ need to be renormalized such that all observable quantities are independent of the momentum cutoff. For simplicity, we take the cutoffs related to $V_0$ and $g$ to be the same, denoted $\Lambda$, since they both drop out after the renormalization.

\subsection{Exciton problem}

Let us first discuss how the formation of an exciton is described within the contact interaction model~\eqref{eq:Hamiltonian}. To this end, we consider the most general state of an electron-hole pair with zero  center-of-mass momentum:
\begin{equation}\label{eq:exciton}
\ket{\Phi} =\sum_\k \phi_\k e^\dag_\k h^\dag_{-\k}\ket{0},
\end{equation}
with the normalization condition $\bra{\Phi}\ket{\Phi}=\sum_\k\abs{\phi_\k}^2=1.$ Here, $\ket{0}$ is the electron-hole vacuum. In the absence of coupling to light, i.e., at $g=0$, the wave function $\phi_\k$ satisfies the \sch equation
\begin{align} 
(E-\ok)\phi_\k &=-V_0\sum_{\k'}\phi_{\k'}, \label{eq:sch_contactX}
\end{align} 
where we define the electron-hole kinetic energy $\bar{\epsilon}_\k \equiv \epsilon^{e}_{\k} + \epsilon^{h}_{\k} = k^2/2m_r$ with $m_r=(1/m_e+1/m_h)^{-1}$ the electron-hole reduced mass.

Equation~\eqref{eq:sch_contactX} admits a single bound state with binding energy $\eb$, corresponding to the $1s$ exciton state~(see, e.g., Ref.~\cite{LevinsenBook15}), which has the associated effective Bohr radius $a_X\equiv 1/\sqrt{2m_r\eb}$. The corresponding wave function $\phi_{X\k}$ satisfies the equation
\begin{align} 
(\eb+\ok)\phi_{X\k} &=V_0\sum_{\k'}\phi_{X\k'}. \label{eq:sch_contactX2}
\end{align} 
Importantly, the right hand side of Eq.~\eqref{eq:sch_contactX2} does not depend on momentum, and therefore we immediately find that the $1s$ exciton has the wave function
\begin{align}
    \phi_{X\k}=\frac{\sqrt{Z_X}}{\eb+\ok},
    \label{eq:phiXk}
\end{align}
where $Z_X$ comes from the normalization condition
\begin{align}
    Z_X=\left[\sum_\k\frac{1}{(\eb +\bar{\epsilon}_{\k})^2} \right]^{-1}=\frac{2\pi\eb}{m_r}. 
    \label{eq:Zx}
\end{align}
Equation~\eqref{eq:sch_contactX2} also allows us to relate the bare coupling constant, $V_0$, to $\eb$: Acting with the operator $\sum_\k \frac{1}{\eb+\ok}(\cdot)$ on Eq.~\eqref{eq:sch_contactX2}, we find
\begin{align}
    \frac1{V_0}=\sum_\k^\Lambda \frac1{\eb+\ok}.
    \label{eq:1overV0}
\end{align}
Here, the sum on $\k$ is logarithmically divergent, and we have therefore explicitly introduced the ultraviolet momentum cutoff $\Lambda$. Once the coupling constant and $\Lambda$ are related to the exciton binding energy via Eq.~\eqref{eq:1overV0}, all dependence on these bare parameters is eliminated from the problem~\cite{LevinsenBook15}.

It is instructive to compare the operator formalism described above for the exciton problem with the electron-hole $T$ matrix~\cite{LevinsenBook15}:
\begin{equation}\label{eq:2d T matrix_copy_2}
\mathcal{T}_0(E)=\frac{2\pi/m_r}{-\ln\left[E/\eb+i0\right]+i\pi}.
\end{equation}
This is discussed further below in Sec.~\ref{sec:Diagrammatic formulation}.
In this formalism, the bound state emerges as a pole at the negative energy $E=-\eb$, where the presence of the infinitesimal positive imaginary part $+i0$ shifts the pole slightly into the lower half of the complex plane. In the vicinity of the pole, we expand the $T$ matrix to find  $\mathcal{T}_0(E)\simeq Z_X/(E+\eb+i0)$. Thus, the normalization $Z_X$ naturally emerges in both approaches.

\subsection{Operator approach to exciton-polaritons}
\label{sec:operator}

Let us now discuss how to obtain the polariton spectrum within the model~\eqref{eq:Hamiltonian}, and how to relate this to experimentally measurable quantities. As we will show, the Hamiltonian \eqref{eq:Hamiltonian} is analytically solvable in the case of the single-polariton problem. In the following, we provide a derivation of the exact wave function for a polariton at normal incidence (i.e., zero momentum), but our results can be straightforwardly extended to describe polaritons at finite momentum --- see Appendix~\ref{app:polaritonQ}. 

To proceed, we write the most general electron-hole-photon wave function as: 
\begin{equation}\label{eq:Q=0_general_wavefun}
\ket{\Psi} =\sum_\k \psi_\k e^\dag_\k h^\dag_{-\k}\ket{0}+\gamma c^\dag_{\0} \ket{0},
\end{equation}
with the normalization condition:
\begin{equation}\label{eq:normalization_condition}
\bra{\Psi}\ket{\Psi}=\sum_\k\abs{\psi_\k}^2+\abs{\gamma}^2=1. \end{equation}
The \sch equation can be obtained by projecting $(E-\hat{H})\ket{\Psi}=0$ onto the electron-hole and photon parts of Eq.~\eqref{eq:Q=0_general_wavefun}, which gives:
\begin{subequations}
\label{eq:sch_contact}
\begin{align} 
(E-\ok)\psi_\k &=-V_0\sum_{\k'}\psi_{\k'} +g\gamma, \label{eq:sch_contact1}\\ 
(E-\omega)\gamma & = g\sum_\k \psi_\k \label{eq:sch_contact2}.
\end{align} 
\end{subequations}
In the following, we assume that the exciton binding energy is larger than other relevant energy scales, such as the photon-exciton detuning and the light-matter Rabi coupling. This is a good approximation in the TMDs and in a single GaAs quantum well. As we show, this condition need not be strictly satisfied, and therefore our results also apply to structures containing multiple GaAs quantum wells.

\begin{figure*}
	\centering
	\includegraphics[width=\linewidth]{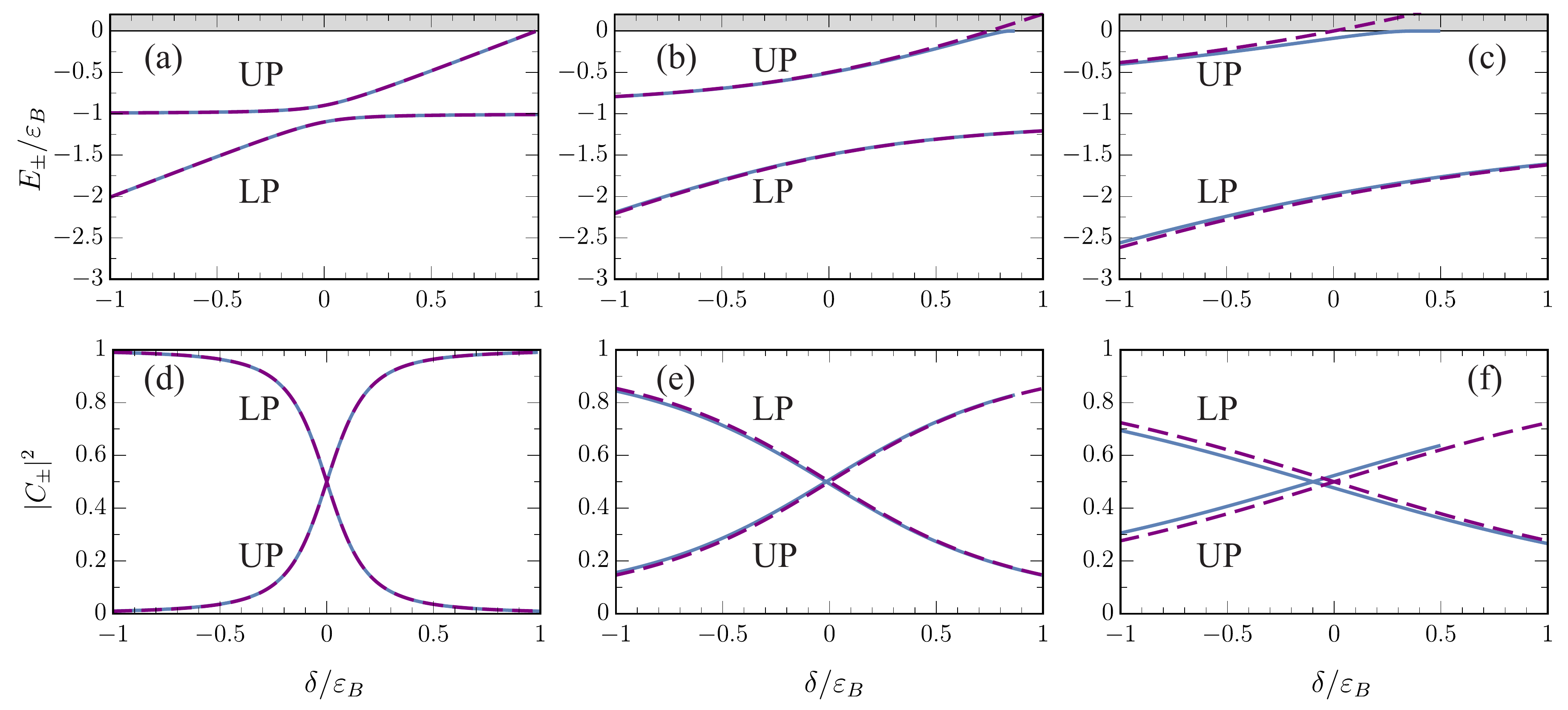}
	\caption{Energies (top row) and photon fractions (bottom row) of the lower and upper polaritons obtained within our electron-hole-photon model in Eq.~\eqref{eq:Ep_dispersion_Q=0}
(blue solid lines) and within the two coupled oscillator model Eq.~\eqref{eq:dispersion_two_level} (purple dashed lines). The shaded regions correspond to the electron-hole continuum. Parameters: (a) and (d) $\Omega/\eb=0.1$; (b) and (e) $\Omega/\eb=0.5$; (c) and (f) $\Omega/\eb=1$.}
	\label{fig:polaritoncheck}
\end{figure*}

Before delving into 
the technical details, 
we will briefly summarize our main conclusions. We emphasize that the results 
presented here act as a starting point for both few-body physics, as discussed in this paper, and many-body physics within the model \eqref{eq:Hamiltonian}. As we show in the following, the exact polariton energies $E_\pm$ satisfy the transcendental equation
\begin{equation}\label{eq:Ep_dispersion_Q=0}
(\omega-E_\pm)\ln\left(\frac{-E_\pm}{\eb }\right)=\frac{\Omega^2}{\eb },
\end{equation}
where
\begin{align}
    \Omega\equiv g\sum_\k^\Lambda \phi_{X\k} = \frac{g}{V_0} \sqrt{Z_X}
    \label{eq:g_renormalization0}
\end{align} 
is the effective exciton-photon Rabi coupling.
The subscripts $-$ and $+$ indicate the lower and upper polaritons, respectively, and we assume that $E_\pm < 0$~\footnote{If $\Omega\gtrsim\eb$ and/or $\delta\gtrsim\eb$, the upper polariton enters the continuum of unbound electron-hole states. In that case, one must analytically continue the energy slightly into the complex plane, $E\to E+i0$}. Like in the exciton problem, the right hand side of Eq.~\eqref{eq:sch_contact1} is independent of momentum, and hence the electron-hole wave function 
has the same functional form as the exciton wave function $\phi_{X\k}$ in Eq.~\eqref{eq:phiXk}~\footnote{The variational approach developed by Khurgin~\cite{Khurgin2001} approximated the functional form of the electron-hole wave function as unchanged in the presence of light-matter coupling. While this is an approximation in the Coulomb case studied in that work~\cite{JesperPRR19}, we see that it is exact for strongly screened electron-hole interactions.}:
\begin{align}
    \psi_{\pm,\k}=\frac{\sqrt{Z_\pm}\sqrt{1-|\gamma_\pm|^2}}{-E_\pm+\ok}.
    \label{eq:ehwave}
\end{align}
Here the numerator follows from the normalization condition in Eq.~\eqref{eq:normalization_condition}, and
\begin{align}
    Z_\pm=\left[\sum_\k\frac{1}{(E_\pm -\bar{\epsilon}_{\k})^2} \right]^{-1}=\frac{2\pi|E_\pm|}{m_r}
    \label{eq:Zpm}
\end{align}
is the generalization of the exciton normalization $Z_X$ in Eq.~\eqref{eq:Zx}. The corresponding photon and exciton Hopfield coefficients $C_\pm$ and $X_\pm$ take the form
\begin{subequations}\label{eq:Hopfields}
\begin{align}
    |C_{\pm}|^2 & \equiv |\gamma_\pm|^2=\frac1{1+
    \frac{\eb}{|E_\pm|}\frac{(E_\pm-\omega)^2}{\Omega^2}}, \\
    |X_{\pm}|^2 & \equiv 1-|\gamma_\pm|^2=
    \frac1{1+\frac{|E_\pm|}{\eb}
    \frac{\Omega^2}{(E_\pm-\omega)^2}}.
\end{align}
\end{subequations}
Finally, we note that the photon frequency relative to the exciton energy is shifted by a finite amount in the presence of the active medium, such that the effective photon-exciton detuning $\delta$ is
\begin{align}
    \delta=\omega-(-\eb)-\frac{\Omega^2}{2\eb}.
    \label{eq:finite_detuning}
\end{align}

Figure~\ref{fig:polaritoncheck} shows the polariton energies and the corresponding photon and exciton fractions according to Eqs.~\eqref{eq:Ep_dispersion_Q=0} and \eqref{eq:Hopfields}. We find that they agree extremely well
with the results of treating the excitons and photons as two coupled oscillators~\cite{HopfieldPR1958}, even when the Rabi coupling is comparable to the exciton binding energy as in structures containing multiple GaAs quantum wells. In particular, our scheme provides an improved agreement between the two models than that obtained in Ref.~\cite{Hu2019QuantumFI}, which considered a similar electron-hole-photon model. Moreover, our scheme is technically simpler to implement 
since Eqs.~\eqref{eq:Ep_dispersion_Q=0}-\eqref{eq:finite_detuning} are fully analytic and do not require the introduction of an additional infrared cutoff. We now proceed to derive these results.

\subsubsection{Renormalization procedure}
The central idea of the renormalization procedure is to relate the parameters in Eq.~\eqref{eq:sch_contact} to the experimental observables in the coupled-oscillator model of excitons and photons. At a technical level, we note that our renormalization of the electron-hole-photon model is conceptually similar to that of atoms interacting via both an open and a closed channel in three dimensions~\cite{Werner2009}. First, let us consider the case of small exciton-photon Rabi coupling such that the polariton energy is close to the exciton energy, i.e., $E=-\eb +\Delta E$ with $\abs{\Delta E}\ll \eb $ the energy correction. In the limit $E\to-\eb$, the electron-hole part of the polariton wave function in Eq.~\eqref{eq:ehwave} can then be approximated as proportional to the exciton wave function in Eq.~\eqref{eq:phiXk}, $\psi_\k\simeq\beta \phi_{X\k}$, where $\beta$ is a complex number. Within this approximation, we use the exciton \sch equation in Eq.~\eqref{eq:sch_contactX2} to find that Eq.~\eqref{eq:sch_contact} takes the form:
\begin{subequations}
\begin{align}
    (E+\eb )\beta&=\gamma\,g\sum_\k \phi_{X\k},   \label{eq:two-level_sch_1}\\
    (E-\omega)\gamma&=\beta\,g\sum_\k \phi_{X\k}. \label{eq:two-level_sch_2}
\end{align}
\end{subequations}
Written in matrix form we have
\begin{align}
    \begin{bmatrix}
-\eb  & \Omega \\[0.2em]
\Omega & \omega 
\end{bmatrix}\begin{bmatrix}
\beta \\[0.2em]
\gamma 
\end{bmatrix}=E \begin{bmatrix}
\beta \\[0.2em]
\gamma 
\end{bmatrix} \label{eq:matrix_eq_two_level},
\end{align}
where we have identified the off-diagonal term $g\sum_\k \phi_{X\k}$ as the experimentally measurable Rabi coupling $\Omega$ introduced in Eq.~\eqref{eq:g_renormalization0}. Evaluating the sum using Eqs.~\eqref{eq:phiXk} and \eqref{eq:1overV0} allows us to relate the bare coupling $g$ to $\Omega$ and the exciton parameters,
\begin{equation}\label{eq:g_renormalization}
g = \Omega \frac{V_0}{\sqrt{Z_X}}. 
\end{equation} 
Since $1/V_0$ diverges logarithmically with the cutoff $\Lambda$ in Eq.~\eqref{eq:1overV0}, 
we thus require $g\sim 1/\ln\Lambda$ to ensure that $\Omega$ is finite in our renormalization scheme.

Equation~\eqref{eq:matrix_eq_two_level} yields the spectrum of two coupled oscillators:
\begin{align}
    E_\pm^{\rm osc}
    &=-\eb +\frac{1}{2}\left(\delta\pm\sqrt{\delta^2+4\Omega^2} \right),\label{eq:dispersion_two_level}
\end{align}
where $\delta=\omega-(-\eb )$ is the (bare) photon-exciton detuning. Thus we see that in the limit of small Rabi coupling we recover the usual spectrum~\cite{DengRevMod10,CiutiRevMod13} of the lower $(-)$ and upper $(+)$ polaritons. We also note that the photon and exciton fractions can be written as
\begin{subequations}
\begin{align}
    |X_{\pm}^{\rm osc}|^2
    &=|\beta|^2=\frac1{1+\frac{\Omega^2}{(E_\pm^{\rm osc}-\omega)^2}},\\
    |C_{\pm}^{\rm osc}|^2
    &=1-|\beta|^2=\frac1{1+\frac{(E_\pm^{\rm osc}-\omega)^2}{\Omega^2}}.
\end{align}
\end{subequations}
Note the similarity to our exact equation~\eqref{eq:Hopfields}. In particular, the coefficients exactly match in the limit where the polariton energies approach the exciton energy. 

In general, the exciton wave function will be modified by the coupling to light, unlike what we have assumed 
in the above analysis. 
We therefore now proceed to find the exact spectrum of Eq.~\eqref{eq:sch_contact}, which also allows us to arrive at the modified Hopfield coefficients. To this end, we define $f\equiv V_0\sum_\k^\Lambda \psi_\k$ which is finite as $\Lambda\to\infty$  since $V_0\sim 1/\ln\Lambda$ and $\sum_\k\psi_\k\sim\ln\Lambda$. Equation~\eqref{eq:sch_contact} then becomes
\begin{subequations}
\label{eq:f_and_gamma}
\begin{align} \label{eq:feq}
    f&=-V_0\sum_\k\frac1{E-\ok}(f-g\gamma),\\ \label{eq:gameq}
    \gamma&=\frac1{E-\omega}\frac{gf}{V_0}.
\end{align}
\end{subequations}
Rearranging Eq.~\eqref{eq:feq} and then inserting \eqref{eq:gameq}, we find
\begin{align}
    \frac1{V_0}+\sum_\k\frac1{E-\ok}
    &=\frac{g\gamma}{f}\sum_\k\frac1{E-\ok} \nn \\
    & = \frac{1}{E-\omega}\frac{g^2}{V_0}\sum_\k\frac1{E-\ok} \nn \\
    & = \frac{1}{\omega-E} \frac{\Omega^2}{Z_X} 
    \label{eq:complicated}
\end{align}
where, in the last line, we have used  Eq.~\eqref{eq:g_renormalization} and the fact that $V_0\sum_\k\frac1{E-\ok} \to -1$ as $\Lambda\to\infty$.
Replacing $1/V_0$ on the left hand side using Eq.~\eqref{eq:1overV0},  
we finally obtain Eq.~\eqref{eq:Ep_dispersion_Q=0} for the polariton spectrum:
\begin{equation}\label{eq:Ep_dipserion_Q=0}
(\omega-E)\ln\left(\frac{-E}{\eb }\right)=\frac{\Omega^2}{\eb }.
\end{equation}
The solutions to Eq.~\eqref{eq:Ep_dipserion_Q=0} correspond to the exact polariton energies for the Hamiltonian~\eqref{eq:Hamiltonian}, under the assumption that $E<0$ (which is always true for the lower polariton). 
Expanding the logarithm to leading order around $E=-\eb$ 
allows us to recover the spectrum of two coupled oscillators in Eq.~\eqref{eq:dispersion_two_level}.

To extract the photon fraction, we use Eq.~\eqref{eq:gameq} and Eq.~\eqref{eq:g_renormalization} to obtain
\begin{align}
    \gamma = \frac{f}{\omega-E} \frac{\Omega}{\sqrt{Z_X}} .
\end{align}
In the limit $\Lambda \to \infty$, we have $f= \sqrt{Z} \sqrt{1-|\gamma|^2}$ according to its definition and the form of the electron-hole wave function in Eq.~\eqref{eq:ehwave} (dropping the $\pm$ subscripts). 
We can thus solve for $\gamma$ to obtain Eq.~\eqref{eq:Hopfields}.

\subsubsection{Effective photon-exciton detuning}

As noted above, the coupling of the photon to the active semiconductor medium can also 
shift the cavity photon frequency. In experiment, the Rabi coupling and the exciton-photon detuning are parameters that are fitted from the observed polariton spectrum. Therefore, we now define an effective detuning that would result from such a fitting procedure. This allows us to directly relate the bare photon frequency $\omega$ to a detuning $\delta$.
In contrast to the case of Coulomb interactions~\cite{JesperPRR19}, we find that
the shift is independent of the cutoff $\Lambda$ for the contact-interaction model.

To capture the leading order correction to the cavity frequency when $\Omega \ll \eb$,
we take $\omega=\delta+s-\eb$ with $s$ small compared to $\eb$, and then insert this into 
Eq.~\eqref{eq:Ep_dispersion_Q=0}:
\begin{equation}\label{eq:detuning_shift_equation}
    (\delta+s-\eb -E_\pm)\ln(\frac{-E_\pm}{\eb })=\frac{\Omega^2}{\eb}.
\end{equation}
Using the lowest-order expressions for the polariton energies, Eq.~\eqref{eq:dispersion_two_level}, and keeping terms only up to order $\Omega/\eb$ and $\delta/\eb$, we find
\begin{align}
    s=\frac{\Omega^2}{2\eb}.
\end{align}
This leads to the expression in Eq.~\eqref{eq:finite_detuning}. As seen in Fig.~\ref{fig:polaritoncheck}, with this definition we have an excellent agreement with the coupled oscillators model up to Rabi couplings $\Omega\sim\eb$, well beyond our initial assumption that $\Omega\ll\eb$. In particular, had we instead used the bare detuning $\omega-\eb$, our results would be shifted by $\eb/2$ in panels (c) and (f), and hence the definition \eqref{eq:finite_detuning} substantially improves the agreement with the coupled-oscillator model.

\subsection{Diagrammatic formulation}\label{sec:Diagrammatic formulation}

\begin{figure}
	\centering
	\includegraphics[width=\linewidth]{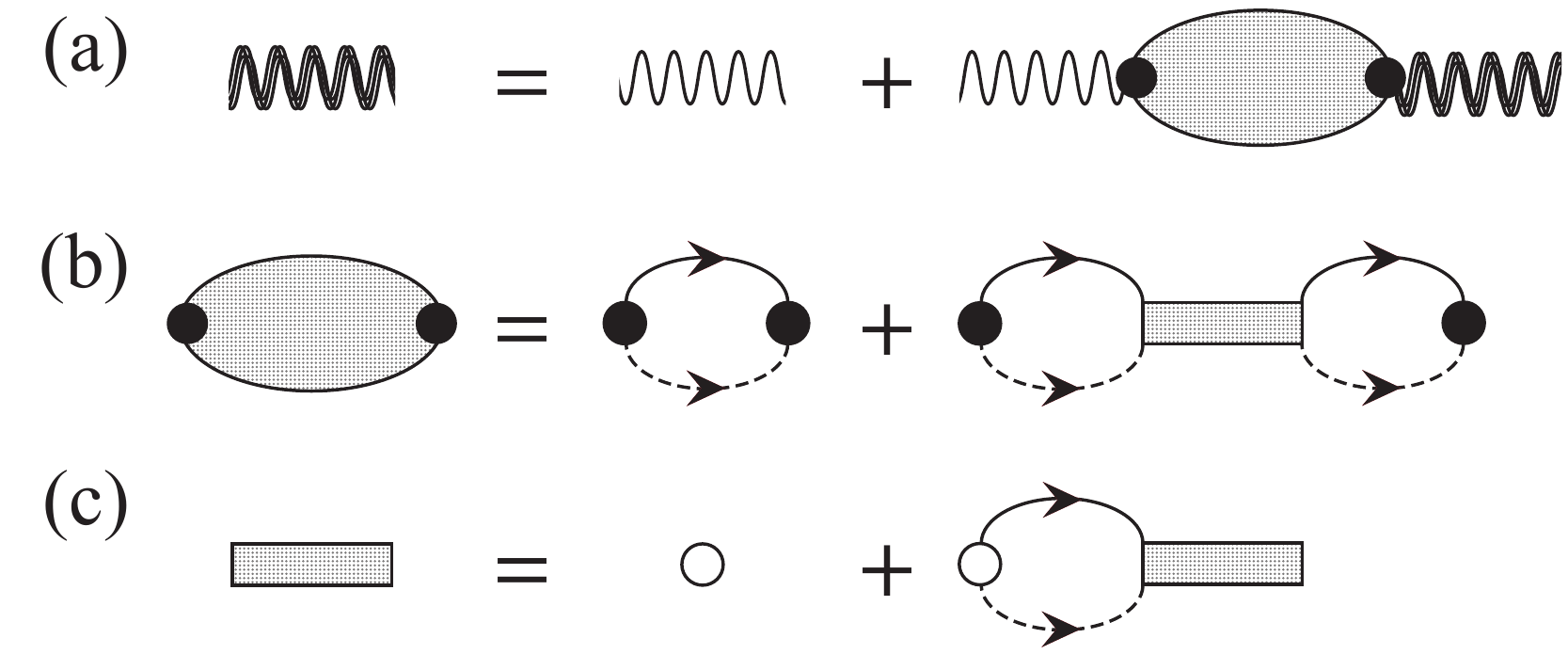}
	\caption{Feynman diagrams for the two-body scattering process. (a) Dressed photon propagator (double wavy line) in terms of the bare photon propagator (single wavy line) and the photon self-energy (shaded ellipse). The black dots represent $g$. (b) Photon self-energy consisting of all possible electron (solid line) and hole (dashed line) interaction terms. The loop consisting of an electron and a hole will be referred to as a polarization bubble, and is denoted $\Pi(E)$.
	(c) Electron-hole $T$ matrix (shaded rectangle), where the white dots represent $-V_0$.}
	\label{fig:diagram_dressphoton}
\end{figure}

In this section, we provide an alternative description of exciton-polaritons using a diagrammatic formulation. While this is equivalent to the operator approach described above, the advantage of the diagrammatic formulation is that it allows a straightforward characterization of repeated scattering processes.  
As we shall see, this enables us to obtain the polariton propagator in closed form, and it provides a convenient starting point for further few-body studies such as the polariton-electron scattering discussed in Sections~\ref{sec:e-P scattering zero k} and \ref{sec:e-P_scattering}.

We start by considering the properties of a photon interacting with an undoped 2D semiconductor in a microcavity. The microcavity photon will be modified by repeated interactions with electron-hole pairs, which includes both bound excitons and unbound electron-hole pairs in the continuum. 
Referring to Fig.~\ref{fig:diagram_dressphoton}, the resulting ``dressed'' photon is characterized by the self-energy $\Sigma$ \cite{FetterBook}. 
Following Ref.~\cite{JesperPRR19}, we obtain the dressed photon propagator $D(\Q,E)$ 
from the Dyson equation in Fig.~\ref{fig:diagram_dressphoton}(a):
\begin{align}\label{eq:dressed photon}
D(\Q,E)&=D_0(\Q,E)+D_0(\Q,E) \Sigma(\Q,E) D(\Q,E)\nn\\
&=\frac{1}{D_0(\Q,E)^{-1}-\Sigma(\Q,E)},
\end{align}
where $E$ and $\Q$ are, respectively, the energy and momentum of the photon. In the absence of the active medium, we have the bare photon propagator
\begin{equation}\label{eq:bare_photon_propagator}
D_0(\Q,E)\equiv D_0(E-\epsilon_\Q^c)=\frac1{E-\omega-\epsilon^c_\Q + i 0},
\end{equation}
where the imaginary infinitesimal $+i0$ shifts the poles into the lower half of the complex plane and ensures that the photon propagates forward in time. Note that we can also incorporate the decay of photons from the microcavity by setting the imaginary part equal to the decay rate. However, we neglect this decay rate since it is much smaller than the Rabi coupling in the regime of strong light-matter coupling, and it therefore only has a negligible effect on the scattering. In other words, we consider a closed quantum system. In the following, we will always implicitly assume that the energy carries an infinitesimal positive imaginary part, that is, we will be working with retarded Green's functions.

As shown in Fig.~\ref{fig:diagram_dressphoton}(b), the photon self-energy  consists of two terms: $\Sigma(\Q,E)=\Sigma^{(1)}(E-\epsilon^{X}_\Q)+\Sigma^{(2)}(E-\epsilon^{X}_\Q)$, with $\epsilon^{X}_\Q=Q^2/2m_X$ the exciton kinetic energy and $m_X=m_e+m_h$ the exciton mass. 
These contain all possible processes that involve the excitation of an electron-hole pair, and they thus only depend on the energy in the 
exciton center-of-mass frame. Hence, the $\Q$ dependence simply appears as a shift in the energy.

Within the model \eqref{eq:Hamiltonian}, we have  
\begin{subequations}
\begin{align}
\Sigma^{(1)}(E)&=g^2 \sum_{\k}^\Lambda \frac{1}{E -\bar{\epsilon}_{\k}}\equiv g^2\,\Pi(E),\label{eq:self-energy I}\\
\Sigma^{(2)}(E)&=g^2\, \Pi^2(E)\, \mathcal{T}_0(E),
\label{eq:self-energy II}
\end{align}
\end{subequations}
where $\Pi(E)$ is  
the polarization bubble diagram corresponding to an electron-hole pair. 
Since $\Pi(E) \sim \ln \Lambda$ and the bare coupling $g\sim 1/\ln \Lambda$, the term $\Sigma^{(1)}(E)$ vanishes when the momentum cutoff $\Lambda \to \infty$. 
On the other hand, we have $g^2\Pi^2(E)=\Omega^2/Z_X$ from Eq.~\eqref{eq:g_renormalization} and the fact that $V_0 \Pi(E) \to -1$ as $\Lambda \to \infty$. 

The electron-hole $T$ matrix $\mathcal{T}_0(E)$ (or exciton propagator) is 
shown in Fig.~\ref{fig:diagram_dressphoton}(c), and corresponds to
\begin{equation}\label{eq:exciton_bare}
    \mathcal{T}_0(E)
    =\frac{1}{-V_0^{-1}-\Pi(E)}.
\end{equation}
We can renormalize the bare $V_0$ using Eq.~\eqref{eq:1overV0} to finally obtain the standard expression in Eq.~\eqref{eq:2d T matrix_copy_2}, see Appendix~\ref{app:fermion_loop}.

\begin{figure}
	\centering
	\includegraphics[width=\linewidth]{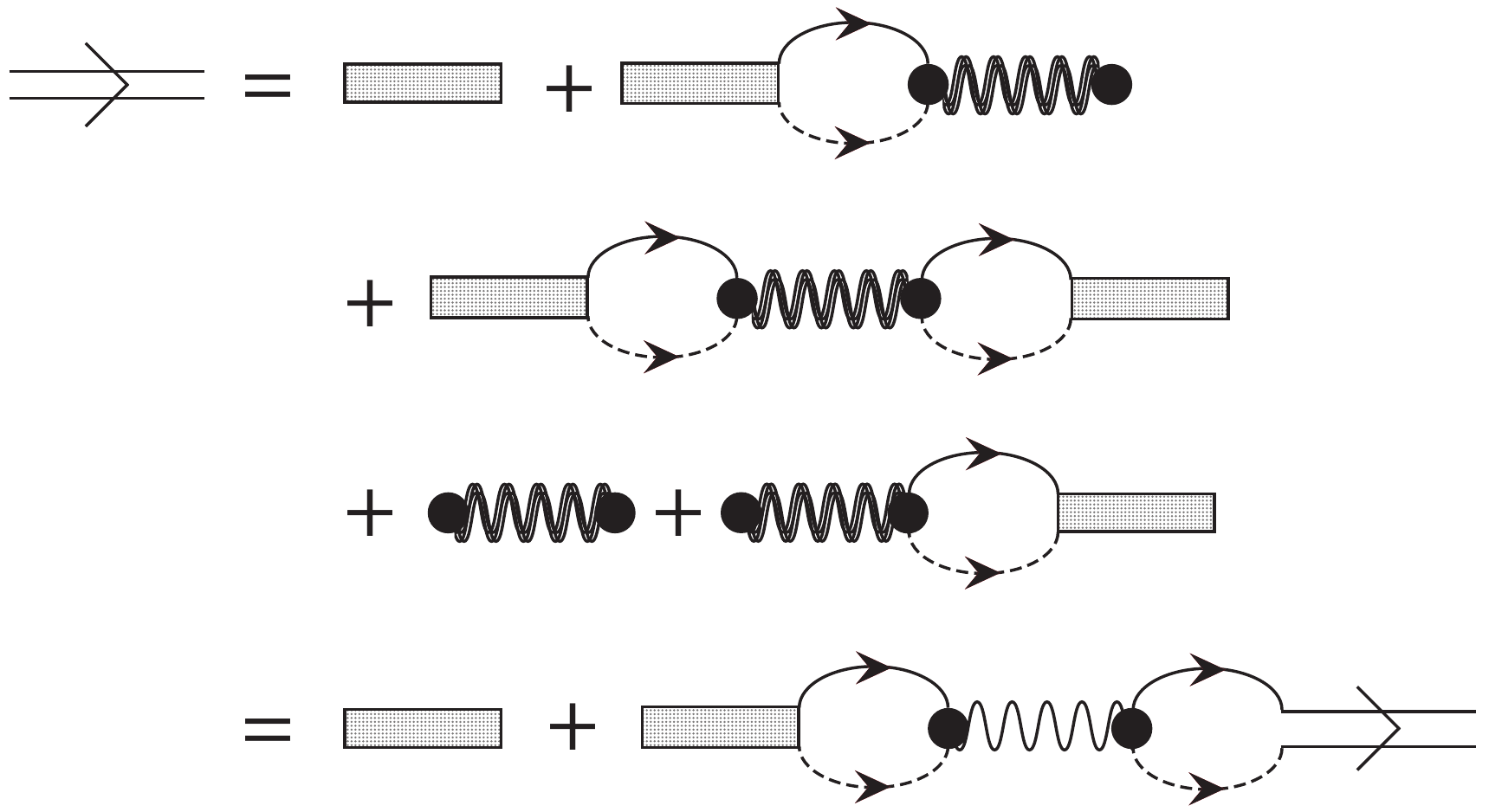}
	\caption{Polariton propagator (double line with an arrow) given by repeated interactions between electron-hole pairs and dressed photons --- see Eq.~\eqref{eq:TfromT0}. The symbols are the same as in Fig.~\ref{fig:diagram_dressphoton}.}	
    \label{fig:diagram_polariton}
\end{figure}

Combining all of these expressions yields the explicit form for 
the dressed photon propagator: 
\begin{equation}\label{eq:dressed_photon_explicit}
D (\Q,E)=\frac{1}{E-\omega-\epsilon^c_\Q+\frac{\Omega^2}{\eb}\,{\left[\ln\left(\frac{\epsilon^X_\Q-E
}{\eb}\right)\right]}^{-1}} ,
\end{equation}
where $\omega$ can be related to the physical photon-exciton detuning $\delta$ via Eq.~\eqref{eq:finite_detuning}. To simplify the notation, we have not explicitly written the imaginary infinitesimal on the energy. 
The poles of the photon propagator contain information about the entire polariton spectrum. Specifically, the denominator of Eq.~\eqref{eq:dressed_photon_explicit} is zero when we have
\begin{equation}\label{eq:Ep_dispersion_Q_dia}
    \left(\omega+\epsilon^c_\Q-E\right)\ln\left[\frac{\epsilon^X_\Q-E}{\eb}\right]=\frac{\Omega^2}{\eb}.
\end{equation}
Indeed, for $\Q =0$, this exactly corresponds to Eq.~\eqref{eq:Ep_dipserion_Q=0} from the operator approach.
For general $\Q$, we denote the polariton energy solutions of Eq.~\eqref{eq:Ep_dispersion_Q_dia} as $E_\pm(\Q)$.

In addition to the dressed photon propagator, we have what we call the ``polariton propagator'' $\mathcal{T}(\Q,E)$ shown in Fig.~\ref{fig:diagram_polariton}, which instead focuses on the behavior of the matter component of the polariton.
Like the dressed photon, 
it contains all the possible scattering processes involving an electron-hole pair or photon.
In the absence of light-matter coupling, it reduces to $\mathcal{T}_0(E-\epsilon^{X}_\Q)$, 
the electron-hole $T$ matrix at center-of-mass momentum $\Q$. 
Suppressing the dependence on energy and momentum,
Fig.~\ref{fig:diagram_polariton} 
reads:
\begin{align} \nn 
\mathcal{T}&= \mathcal{T}_0+g^2 \left(\mathcal{T}_0  \Pi D+\mathcal{T}_0\Pi D\Pi\mathcal{T}_0 + D
+ D\Pi \mathcal{T}_0 \right)\\ \nn 
&= \mathcal{T}_0+\mathcal{T}_0^2\,D\,g^2\Pi^2 \\ 
& =\frac{1}{\mathcal{T}_0^{-1}-D_0\, g^2\Pi^2}. \label{eq:TfromT0}
\end{align}
Here, we again use $g\sim 1/\ln\Lambda$ and $\Pi\sim \ln\Lambda$ 
to remove terms that vanish in the limit $\Lambda \to \infty$, 
while in the last line, we use the relation $D = (D_0^{-1} -g^2\Pi^2\mathcal{T}_0)^{-1}$. 
Thus, the polariton propagator finally reads:
\begin{equation}\label{eq:D2d full}
    \mathcal{T}(\Q,E)=\frac{2\pi/m_r}{-\ln\left(\frac{\epsilon^X_\Q-E}{\eb}\right)-\frac{\Omega^2}{\eb}\left(E-\omega-\epsilon^c_\Q\right)^{-1}}.
\end{equation}
It is also straightforward to show that the polariton and photon propagators are related via
\begin{align}
    \mathcal{T}(\Q,E)=\frac{\mathcal{T}_0(E-\epsilon^X_\Q)}{D_0(E-\epsilon_\Q^c)}D(\Q,E).
\end{align}

The polariton propagator has precisely the same poles as the dressed photon propagator in Eq.~\eqref{eq:dressed_photon_explicit}. However, the poles of the photon propagator have residues corresponding to the photon fractions of the polaritons: 
\begin{align}
  |C_\pm(\Q)|^2=\left(1+\frac{Z_X/Z_\pm(\Q)}{\Omega^2D_0^2(\Q, E_\pm(\Q))}\right)^{-1},
  \label{eq:photonfraction}
\end{align}
while the residues for  the poles of 
the polariton propagator instead correspond to the matter component,
\begin{align}
 Z_\pm(\Q) |X_\pm(\Q)|^2 \equiv Z_\pm(\Q) \left[1- |C_\pm(\Q)|^2\right] ,
 \label{eq:Zpm_copy_1}
\end{align}
with the electron-hole normalization factors
\begin{align}
    Z_\pm(\Q) = \frac{2 \pi|E_\pm(\Q) - \epsilon^X_\Q| }{m_r} .
\end{align}
At $\Q=0$, the Hopfield coefficients 
and the normalization factors 
reduce to the corresponding expressions in Eqs.~\eqref{eq:Hopfields} and \eqref{eq:Zpm}.

\section{Polariton-electron interactions in the Born approximation}\label{sec:e-P scattering zero k}

Due to the complexity of the Coulomb interaction, an exact solution of few-body scattering processes such as polariton-electron and polariton-polariton scattering in strongly coupled light matter systems remains elusive. Instead, the interactions are typically approximated by considering their corresponding excitonic counterparts multiplied by appropriate powers of the excitonic Hopfield coefficient:
\begin{align}
    g_{eP}\simeq |X_-|^2g_{eX}, \qquad g_{PP}\simeq |X_-|^4g_{XX},
    \label{eq:Born_assumption}
\end{align}
where $g_{eP}$ $(g_{eX})$ is the polariton-electron (exciton-electron) interaction constant, while 
$g_{PP}$ $(g_{XX})$ is the polariton-polariton (exciton-exciton) interaction constant. Note that we have specialized to the lower polariton here, but similar expressions are used for the upper polariton with the simple replacement $X_-\to X_+$. The standard approach is to calculate the interaction strengths within the Born approximation~\cite{TassonePRB99,RamonPRB02,Malpuech2002,SnokePRB2010}, with effects due to very strong light-matter coupling included via the calculation of additional matrix elements~\cite{TassonePRB99,RochatPRB00,Combescot_2007,COMBESCOTPhysRep2008, GlazovPRB2009}.
The polariton-polariton interaction strength has also been calculated in $T$-matrix approaches that treat the exciton as an inert object, first focusing on the case of cross-circularly polarized polaritons~\cite{WoutersPRB07} and recently for both the singlet and triplet scattering configurations in multilayer systems~\cite{Bleu2020}.

In this paper, we provide the first exact microscopic calculation of polariton-electron scattering, within a simplified model of contact interactions between charges. It is therefore imperative to compare our results with the standard Born approximation. Therefore, in this section we will describe the results of the Born approximation within our model. We will furthermore argue that the Born approximation provides a strict upper bound on the interaction energy shift (in the limit of zero momentum) when there are no lower-lying bound states.

\subsection{Operator approach}

The analytic expression for the polariton wave function obtained in the previous section (see also the finite-momentum generalization in Appendix~\ref{app:polaritonQ}) allows us to straightforwardly evaluate the Born approximation for polariton-electron scattering. For simplicity, we consider the frame where the center-of-mass momentum is zero, since in that case the different partial wave components of the scattering separate.
Since the electron has a relatively flat dispersion, our results are likely to be insensitive to the actual electron momentum and will thus be dominated by the polariton momentum.

The Born approximation consists in approximating the scattering $T$ matrix by the first term in the Born series, which corresponds to evaluating the matrix element of the polariton-electron interaction potential. Taking the incoming (outgoing) polariton and electron to have momenta $\pm\p_1$ ($\pm\p_2$), this matrix element takes the form
\begin{align}
    \label{eq:TB}
    &{T}^{B}(\p_1,\p_2) \nn \\ &\quad=
     \bra{0}e_{-\ptwo} P_{\ptwo}[H-E_-(\p_1)-\epsilon^e_{\p_1}]P_{\pone}^\dagger e_{-\pone}^\dagger\ket{0},
\end{align}
where the operator $P^\dag_\p$ creates a polariton with momentum $\p$, see Eq.~\eqref{eq:Psi_Q} in Appendix~\ref{app:polaritonQ}. We assume energy and momentum conservation such that the scattering is elastic, i.e., we take $|\p_1|=|\p_2|=p$. Here, the subtraction of the energy of the non-interacting particles from the Hamiltonian, Eq.~\eqref{eq:Hamiltonian}, ensures that we only evaluate the polariton-electron interaction potential. Using the exact polariton wave function from Appendix~\ref{app:polaritonQ}, we find
\begin{align}
\label{eq:T_matrix_Born}
    &{T}^{B}(\p_1,\p_2) \nn \\
&=-\left(\psi_{-,m_e \pone/m_X+\ptwo}^{(\pone)}\right)^2\left[E_-(\pone)-\epsilon^h_{\p_1+\p_2}-\epsilon^e_{\p_2} \right]\nn \\[0.2em]
& = \frac{Z_-(\p_1)|X_-(\p_1)|^2}{-E_-(\pone)+\epsilon^h_{\p_1+\p_2}+\epsilon^e_{\p_2}},
\end{align}
where the sign in the middle line arises from the exchange of identical fermions. As anticipated, we see that the exciton fraction $|X_-(\p_1)|^2$ naturally emerges from the normalization of the polariton wave function.

\begin{figure}
	\centering
	\includegraphics[width=\linewidth]{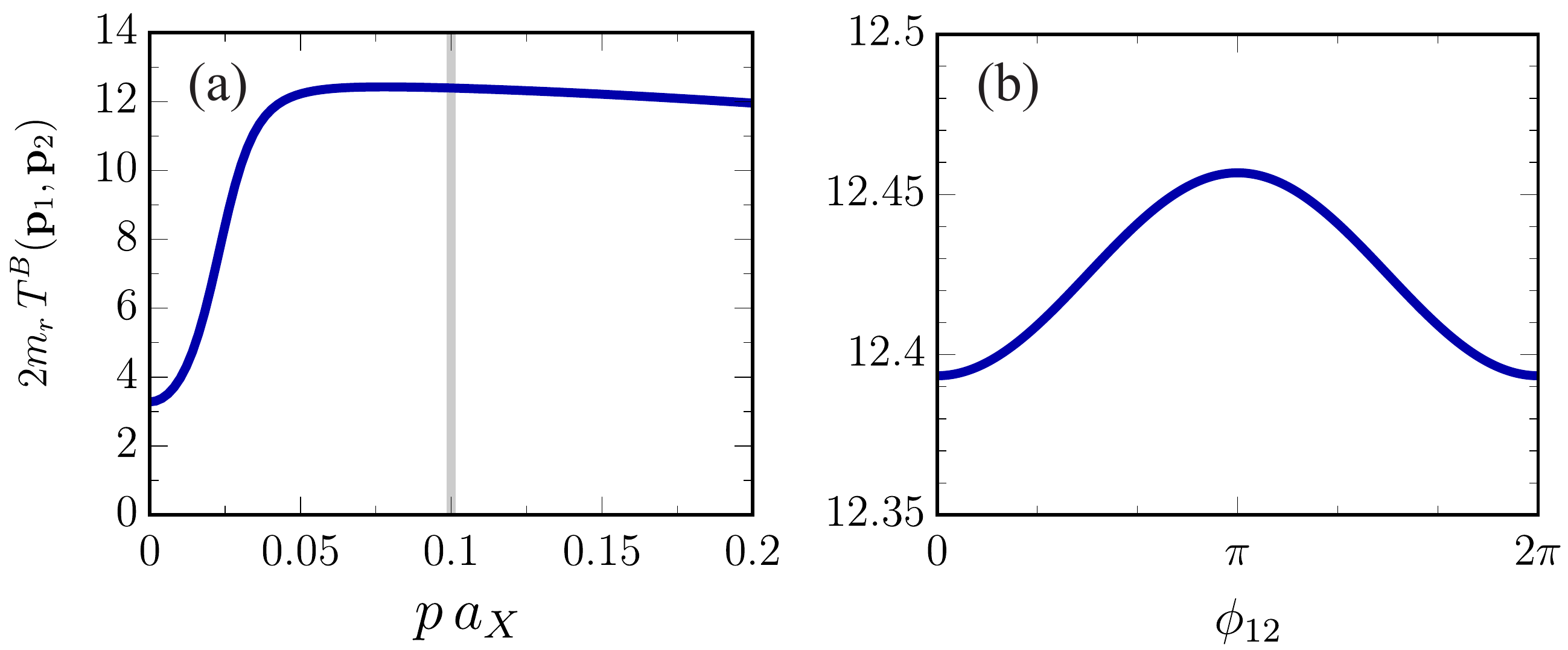}
	\caption{Born approximation for elastic polariton-electron scattering. (a) Momentum dependence at a fixed angle $\phi_{12}=0$ between $\p_1$ and $\p_2$, where we take $|\p_{1}|=|\p_{2}|=p$. Gray vertical line indicates the selected momentum in (b). (b) Angular dependence for a fixed $p=0.1 a_X^{-1}$. The results are shown for parameters relevant to a single GaAs quantum well, i.e., $\Omega/\eb =0.2$, $m_e=0.067 m_0$, $m_h=0.45 m_0$ \cite{GaAsEffectiveMass}, and $m_c=10^{-4} m_0$, with $m_0$ the free electron mass, at negative detuning $\delta/\eb =-0.2$.}
	\label{fig:Op_Born_Da_Born_Compare}
\end{figure}

In Fig.~\ref{fig:Op_Born_Da_Born_Compare} we illustrate the Born approximation for elastic polariton-electron scattering. The exciton fraction quickly increases at low momentum, which dominates the behavior at momenta up to the polariton dispersion inflection point above which the exciton fraction is very close to unity. Apart from the Hopfield coefficient $|X_-(\p_1)|^2$, there is also a momentum and angular dependence through the kinetic energies and the wave function normalization; however we find that this dependence is generally quite weak. Above the inflection point, it leads to a slow decline for increasing momentum.

Specializing to zero momentum, we may define a low-energy polariton-electron interaction constant $g_{eP}^B$, which takes the form
\begin{align}
    g_{eP}^B\equiv T^B(\0,\0)=\frac{Z_-|X_-|^2}{-E_-}=\frac{2\pi|X_-|^2}{m_r}.
    \label{eq:gepB}
\end{align}
This is precisely the exciton fraction multiplied by the exciton-electron interaction, $g_{eX}^B = 4\pi \eb a_X^2=2\pi/m_r$. Note that this is quite close to the corresponding result obtained in the case of Coulomb electronic interactions~\cite{Ramon2003}: Using the formalism developed in Ref.~\cite{JesperPRR19}, we find $g_{eX}^{B,\mathrm{Coulomb}}=(2\pi/m_r)(2-3\pi/8)\simeq 0.8g_{eX}^B$. We emphasize that the polariton-electron Born approximation in Eq.~\eqref{eq:gepB} only depends on the light-matter coupling through the exciton fraction, while it is independent of the absolute strength of the light-matter coupling, $\Omega/\eb$.

\subsection{Diagrammatic approach}

\begin{figure*}[t]
	\centering
	\includegraphics[width=\linewidth]{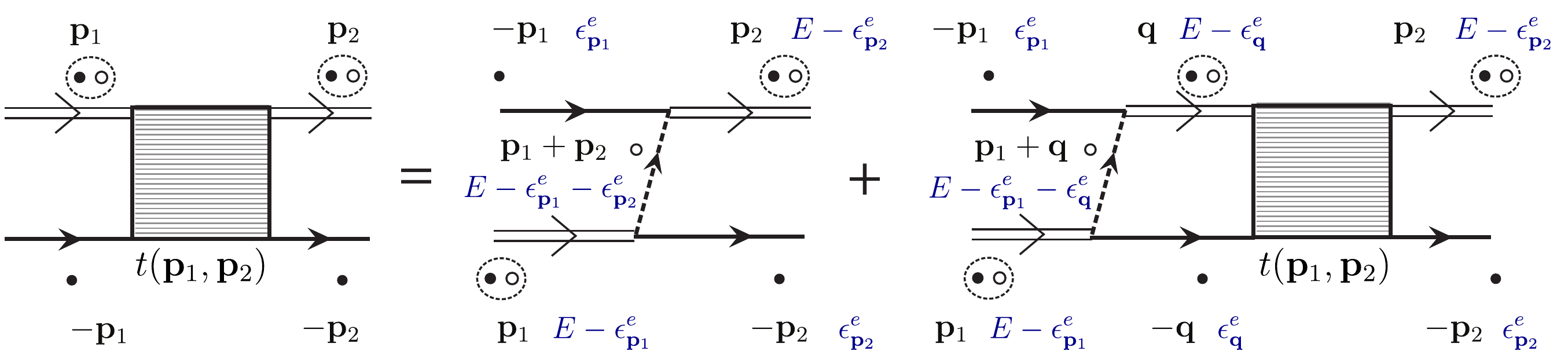}
	\caption{Diagrammatic representation of the (unnormalized) polariton-electron scattering matrix $t$ (shaded rectangle). 
	Black, white, and dashed circles represent electrons, holes, and polaritons, respectively, and black and blue text are the corresponding momenta and energies of the intermediate propagators. 
	}
	\label{fig:e_P_scattering_t_matrix}
\end{figure*}

The polariton-electron scattering process can also be calculated diagrammatically, as shown in Fig.~\ref{fig:e_P_scattering_t_matrix}. In this case, the Born approximation corresponds to the first diagram on the right hand side, 
which involves the exchange of a hole with momentum $\pone+\ptwo$~\cite{VudtiwatEPL13}. This gives
\begin{align}\label{eq:geP_int_Born}
&T^{B}(\p_1,\p_2) \nn \\ &\quad =
-Z_-(\p_1)|X_-(\p_1)|^2
G_h(\p_1+\p_2,E_-(\p_1)-\epsilon_{\p_2}),
\end{align}
where the minus sign is a consequence of fermionic statistics and 
\begin{align}
G_h(\p,E)=\frac1{E-\epsilon^h_\p+i0}    
\end{align} 
is the free hole Green's function. 
Each external polariton line in the diagram contributes the square root of the residue of the propagator $\cal{T}$ at the pole, 
and the normalization of the incoming and outgoing lines is the same since $|\p_1|=|\p_2|$. The residue is precisely $\operatorname{Res}\left[\mathcal{T}(\p, E_-(\p)\right]=Z_-(\p)|X_-(\p)|^2$ --- see Eq.~\eqref{eq:Zpm_copy_1}. Using the definition of the hole propagator, and comparing Eq.~\eqref{eq:T_matrix_Born} with Eq.~\eqref{eq:geP_int_Born} we thus see that the Born approximation within the two approaches coincides, as it should.

\subsection{
Upper bound on the interaction energy shift}\label{sec:Tep_zero_momentum_compare}

We now argue that the Born approximation is conceptually important 
since it provides an upper bound on the polariton-electron interaction constant $g_{eP}$. 
We start by noting that the ground-state energy for an electron and a lower polariton in an area $A$ can be written as
\begin{align} \label{eq:ePenergy}
    E_{eP} = \frac{\expval{H}{\Psi_{eP}}}{\braket{\Psi_{eP}}} = E_- + \frac{g_{eP}}{A}, 
\end{align}
where $\ket{\Psi_{eP}}$ is the exact interacting polariton-electron state. Here we have assumed that there are no lower-energy polariton-electron bound states so that $E_{eP}$ consists of the non-interacting kinetic energy plus a two-body interaction term. We momentarily keep the area $A$ explicit so that we can keep track of powers of $a_X^2/A$. 
In particular, note that $\braket{\Psi_{eP}} = 1 + O(a_X^2/A)$, where the last term arises from the composite nature of the polariton.

Rearranging Eq.~\eqref{eq:ePenergy} and keeping only terms up to order $a_X^2/A$ then gives
\begin{align} \nn
    \frac{g_{eP}}{A} &= \bra{\Psi_{eP}}(H-E_-)\ket{\Psi_{eP}} \\
    & \leq \bra{0}e_{\0} P_{\0}(H-E_-)P_{\0}^\dagger  e_{\0}^\dagger\ket{0} ,
\end{align}
where the second line follows from the fact that the non-interacting state $P_{\0}^\dagger e_{-\0}^\dagger\ket{0}$ can be viewed as a variational approximation to the exact interacting state $\ket{\Psi_{eP}}$. Thus, we finally arrive at
\begin{align}
    g_{eP} \leq g^{B}_{eP} .
\end{align}
Hence, the Born approximation serves as a strict upper bound on the interaction energy shift. This observation is quite general and is independent of the details of the underlying interactions, e.g., it also applies to polariton-polariton scattering and to more realistic interactions between charged particles. The only requirement is that there are no lower energy bound states (such as trions) and that the Hamiltonian faithfully reproduces the low-energy physics. Therefore, any diverging interaction strength obtained in the absence of a bound state must be an artefact of the approximation~\cite{Bleu2020}.

\section{Polariton-electron interactions}
\label{sec:e-P_scattering}

We now present our full diagrammatic calculation of the polariton-electron interaction strength, which is exact within the Hamiltonian~\eqref{eq:Hamiltonian}. Our approach follows similar calculations originally introduced in the context of neutron-deuteron scattering by Skorniakov and Ter-Martirosian~\cite{Skorniakov1957}, and later applied to cold atomic gases~\cite{Petrov2002,BrodskyPRA06,LevinsenPRA06,PetrovLesHouches2010}. Subsequently, the theory of the effective three-body problem at finite momentum was developed in a series of papers~\cite{BedaquePRC1998,Levinsen2009,Levinsen2011,Ngampruetikorn2013}, and this has already been successfully applied to describe the strong atom-dimer attraction observed in mass-imbalanced ultracold Fermi gases~\cite{Jag2014}.

The exact scattering of an electron and a polariton can be represented as an infinite sum of terms where the two electrons involved in the process exchange a hole, and the Born approximation corresponds to keeping only the first term in this series. While it is not possible to simply calculate each term in this sum separately and then sum them up, the key observation that enables an exact solution of the polariton-electron scattering is that the sum satisfies an integral equation as illustrated in Fig.~\ref{fig:e_P_scattering_t_matrix}. Indeed, iterating the right hand side of this equation generates all possible terms where the electrons exchange a hole. Diagrammatically, the equation for the polariton-electron $T$ matrix takes the exact same form as the exciton-electron $T$ matrix, the only difference being the replacement of the exciton with the polariton propagator, and the associated change in dispersion.

To compute the $T$ matrix, we take the incoming and outgoing electrons to have momentum $-\p_1$ and $-\p_2$ and energy $\epsilon_{\p_1}^e$ and $\epsilon_{\p_2}^e$, respectively, while the polaritons have momenta $\p_1$ and $\p_2$ and energies $E-\epsilon^e_{\p_1}$ and $E-\epsilon^e_{\p_2}$, respectively, with $E$ the total collision energy. The equation for the (unnormalized) polariton-electron $T$ matrix illustrated in Fig.~\ref{fig:e_P_scattering_t_matrix} 
takes the explicit form~\cite{BedaquePRC1998,Ngampruetikorn2013}
\begin{align}
&t (\p_1,\p_2)=-G_h(\p_1+\p_2,E-\epsilon^e_{\p_1}-\epsilon^e_{\p_2})\nn\\
&-\sum_\q G_h(\p_1+\q,E-\epsilon^e_{\p_1}-\epsilon^e_{\q}) \mathcal{T} (\q,E-\epsilon^e_{\q}) t(\q,\p_2),\label{eq:ePtMatrix}
\end{align}
where the minus signs on the right hand side follows from the exchange of identical electrons~\footnote{Note that the electron propagator $G_e(-\q,-q_0)=1/(-q_0-\epsilon_\q+i0)$ in the last term in Fig.~\ref{fig:e_P_scattering_t_matrix} has a simple pole in the upper half of the complex $q_0$ plane, while all other propagators have their non-analytic structure only in the lower half plane. This allows us to perform the integral over $q_0$ by closing the contour in the upper half plane, effectively removing that propagator and setting the electron dispersion to its on-shell value $\epsilon^e_{\q}=\q^2/(2 m_e)$ in the other propagators. For more details, see, e.g., Ref.~\cite{BrodskyPRA06}.}. We see that this is an integral equation in the first momentum index of $t$, and we can therefore only set $|\p_1|=|\p_2|$ at the end of calculation. In order to have on-shell scattering, we take the total energy to be $E=E_-(\p_2)+\epsilon_{\p_2}$. The normalization 
is the same as in Eq.~\eqref{eq:geP_int_Born}, and therefore the normalized and on-shell $T$ matrix takes the form
\begin{align}
    T(\p_1,\p_2)=Z_-(\p_1)|X_-(\p_1)|^2t 
    (\p_1,\p_2).
    \label{eq:Tfromt}
\end{align}
This quantifies the strength of scattering between electrons and polaritons.

\subsection{Scattering of slow particles}

It is of particular interest to consider the scattering of polaritons at vanishing momentum due to the realization of polariton Bose-Einstein condensation~\cite{BEC06}. First, it is important to clarify precisely what we mean by this limit. It is well-known~\cite{AdhikariAJP86,LevinsenBook15} that the scattering $T$ matrix of 2D quantum particles with short range interactions (such as the exciton-electron interaction between a charge and an induced dipole) must approach zero as the collision energy vanishes. However, in the case of polariton-electron scattering, this only becomes relevant for exceedingly small momenta $p$ below those where the typical scale of exciton-electron scattering $\sim1/m_e$ is comparable to $1/[m_c\ln(\eb/\epsilon_\p^c)]$~\cite{Bleu2020}. Because of the very small photon mass, this momentum scale corresponds to a length scale much larger than the size of the universe, and this effect can therefore be discarded. Thus, we can safely take the limit $p\to0$ in our numerical solutions, since we will never encounter such an extreme scale. We emphasize that this is a dramatic consequence of the lack of Galilean invariance in the polariton system due to the strong coupling of excitons and photons. 

For scattering at small momenta, we set $\ptwo=0$ and $\pone=\p$ in Eq.~\eqref{eq:ePtMatrix} and perform the integration over the angle between $\p$ and $\q$ assuming $E<0$. The remaining integral equation is then solved using Gauss-Legendre quadrature; for more details, see Appendix~\ref{app:ePfiniteQ}. As a result, we obtain the exact polariton-electron interaction constant
\begin{align}\label{eq:Tep}
g_{eP}=T(0,0)=Z_-|X_-|^2t(0,0),    
\end{align}
where we use the same normalization as in Eq.~\eqref{eq:Tfromt}. Note the similarities between this expression and that in Eq.~\eqref{eq:gepB} within the Born approximation.

\begin{figure}
	\centering
	\includegraphics[width=\linewidth]{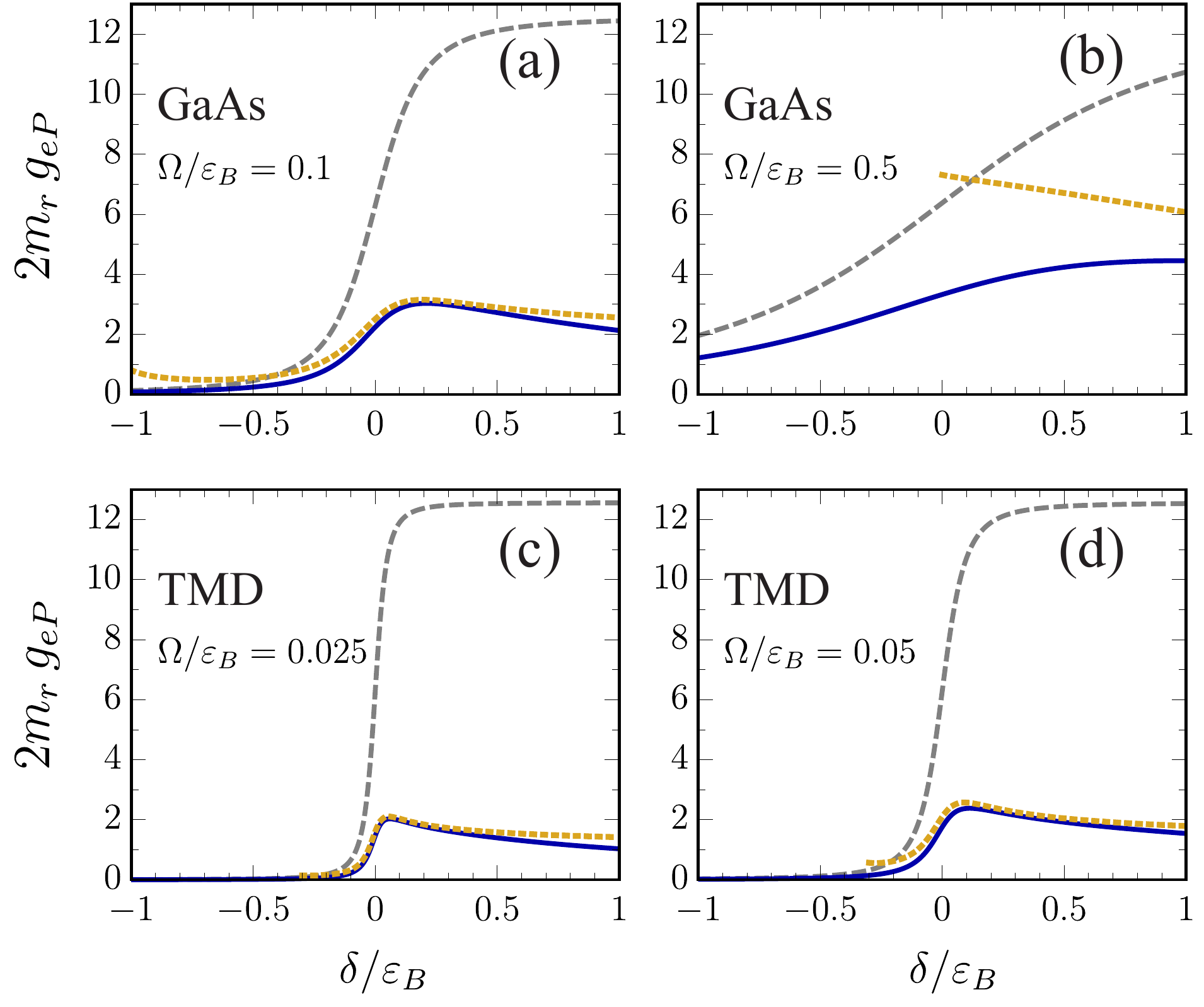}
	\caption{Comparison between the polariton-electron interaction constant $g_{eP}$ within our exact diagrammatic calculation (blue solid line), the Born approximation (gray dashed line), and the approximation Eq.~\eqref{eq:gepfromgex} (yellow dotted line). Top row: GaAs quantum well microcavity systems with $m_c=10^{-4} m_0$, 
	$m_e=0.067 m_0$, $m_h=0.45 m_0$ \cite{GaAsEffectiveMass,GaAsLargeRabi}, and $\varepsilon_{eX}/\eb\simeq 1.18$ obtained from the low-energy exciton-electron $T$ matrix.
	Bottom row: TMD monolayer microcavity systems with $m_e=m_h= m_0$ and $\varepsilon_{eX}/\eb=0.47$~\cite{atomdimer};
	the parameters correspond approximately to the case of a MoSe$_2$~ \cite{dufferwiel2015exciton}, MoS$_2$~\cite{liu2015strong}, or WSe$_2$ \cite{lundt2016room,He2014} monolayer in (c) or to a WS$_2$~\cite{Flatten2016} monolayer in (d).
}
	\label{fig:Born_Com_Zero_k}
\end{figure}

Figure~\ref{fig:Born_Com_Zero_k} shows a comparison of the polariton-electron interaction constant obtained from different methods for parameters relevant to both TMDs and GaAs quantum wells.
The exact diagrammatic calculation, $g_{eP}$, can be used as a standard to gauge the correctness of other calculations. We see that generally the interaction strength increases as the polariton goes from being photonic at negative detuning to more excitonic at positive detuning, in agreement with the expectation from the Born approximation, Eq.~\eqref{eq:gepB}. 
However, we find that the exact calculation features a peak at small positive detuning when the ratio $\Omega/\eb$ is not too large, which is a qualitative feature that is missed by the Born approximation.
Furthermore, the Born approximation 
generically overestimates the interaction strength, and this effect is most dramatic for relatively small values of the ratio $\Omega/\eb$ such as in TMDs. For larger $\Omega/\eb$, the exact result approaches $g_{eP}^B$ and thus we expect the Born approximation to be more accurate for GaAs quantum wells, which is consistent with recent measurements of the polariton-polariton interaction strength~\cite{EliPRB2019}. 
Note that $\Omega/\eb \gtrsim 0.5$ applies to the case of multiple GaAs quantum wells and thus a full description requires us to take account of 
the non-trivial effects of multiple layers~\cite{Bleu2020}.

We can understand these results by noting that at momenta above the inflection point, the polariton propagator in Eq.~\eqref{eq:D2d full} quickly approaches the exciton propagator, ${\cal T}_0$, in Eq.~\eqref{eq:2d T matrix_copy_2}. Taking the limit ${\cal T}\to{\cal T}_0$, we observe that Eq.~\eqref{eq:ePtMatrix} has the exact same functional form as in exciton-electron scattering~\cite{Ngampruetikorn2013}, the only difference being the change in collision energy due to the light-matter coupling. This motivates the interpretation of polariton-electron scattering as an off-shell exciton-electron scattering process, similarly to recent results on polariton-polariton interactions~\cite{Bleu2020}. To make this explicit, we consider the universal form of the low-energy exciton-electron scattering $T$ matrix
\begin{align} \label{eq:Tex}
    T_{eX}(E)\simeq\frac{2\pi}{m_{eX}}\frac1{\ln[-\varepsilon_{eX}/(E+\eb)]},
\end{align}
where we define $m_{eX}=m_em_X/(m_e+m_X)$ as the reduced exciton-electron mass. Unlike the light-matter coupled system, the exciton-electron scattering only depends on the collision energy, since the exciton-electron system is Galilean invariant. Equation~\eqref{eq:Tex} is valid at collision energies $|E+\eb|\ll\varepsilon_{eX}$, where the energy scale $\varepsilon_{eX}\sim\eb$ depends on the precise details of the electronic interactions and on the electron-hole mass ratio~\footnote{In general, there are finite range corrections to Eq.~\eqref{eq:Tex}, which add a term $\sim m_{eX}r_e^2(E+\eb)$ in the denominator~\cite{LevinsenBook15}. In the absence of other scales, the effective range $r_e\sim a_X$ and therefore such terms are not important when $|E+\eb|\ll\varepsilon_{eX}$.}. We thus estimate the polariton-electron scattering in the limit of low momentum by the off-shell exciton-electron expression
\begin{align}
    g_{eP}\simeq |X_-|^2T_{eX}(E_-),
    \label{eq:gepfromgex}
\end{align}
which is valid when $|E_-+\eb|\ll\eb$.

\begin{figure*}[ht]
	\centering
	\includegraphics[width=\linewidth]{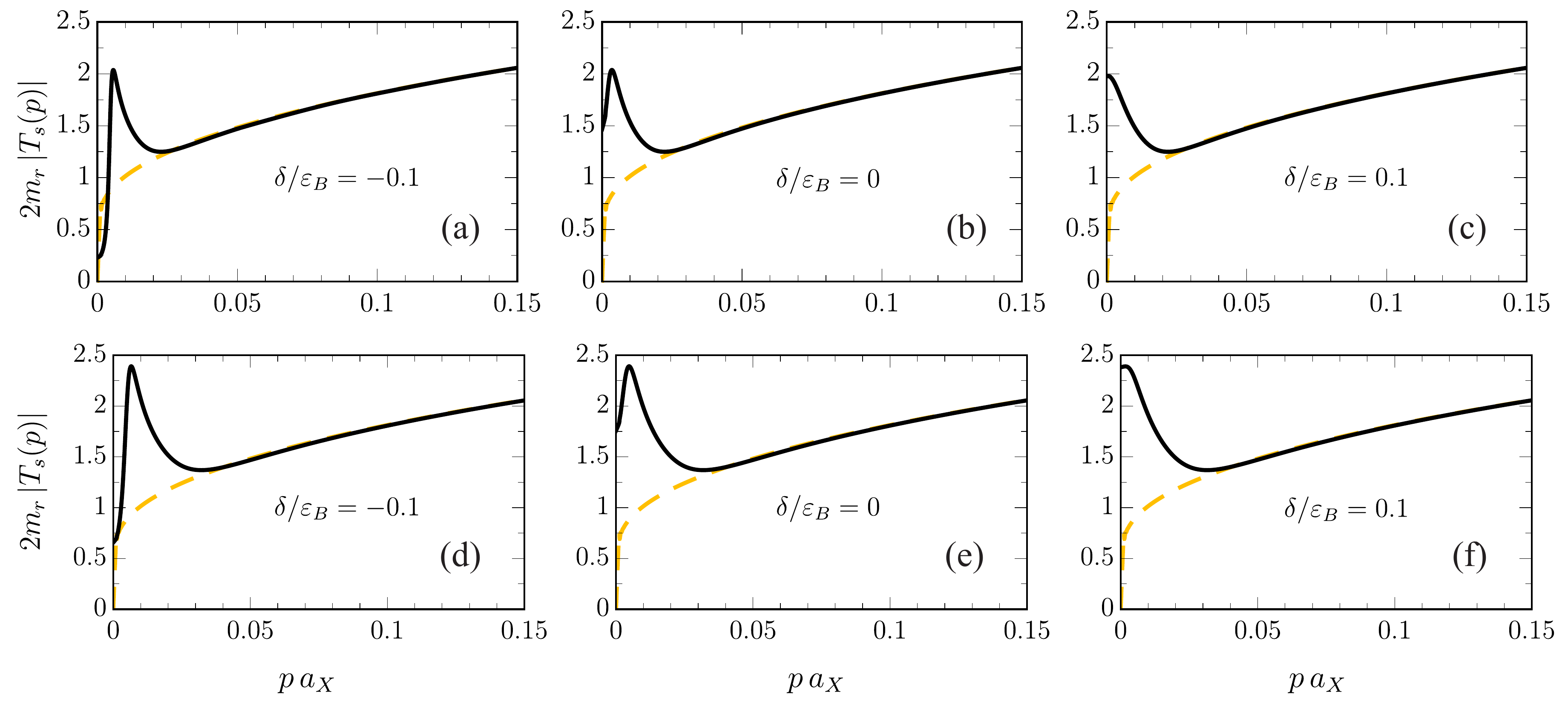}
	\caption{Polariton-electron scattering $T$ matrix at finite momentum for parameters corresponding to TMD monolayers. We show our exact diagrammatic calculation as a black solid line, while the yellow dashed line is the corresponding exciton-electron scattering $T$ matrix. Here, $m_e=m_h=m_0$ and $m_c=10^{-4} m_0$, and in the top row $\Omega/\eb=0.025$ corresponding approximately to the case of a MoSe$_2$~\cite{dufferwiel2015exciton}, MoS$_2$~\cite{liu2015strong}, or WSe$_2$~\cite{lundt2016room,He2014} monolayer, while in the bottom row $\Omega/\eb=0.05$ corresponding to a WS$_2$~\cite{Flatten2016} monolayer.}
	\label{fig:e-P_tmd}
\end{figure*}

Equation~\eqref{eq:gepfromgex} yields the yellow dashed line in Fig.~\ref{fig:Born_Com_Zero_k}, which is seen to closely follow the exact polariton-electron scattering result for sufficiently low Rabi coupling, reproducing all qualitative features. In particular, it works exceptionally well in the case of TMD monolayers, which supports the approximation for polariton-polariton interactions introduced in Ref.~\cite{Bleu2020}. Note that Eq.~\eqref{eq:gepfromgex} predicts a resonance when the (negative) collision energy is comparable to $\varepsilon_{eX}$, such as in GaAs with a very large Rabi coupling in panel (b). However, this resonance is spurious and can be cured by including higher order terms in the phase shift, and it is also absent in our exact calculation. 

While Eq.~\eqref{eq:gepfromgex} is conceptually similar to the Born approximation, Eq.~\eqref{eq:gepB}, we see that the fact that the exciton-electron scattering should be considered off-shell leads to a qualitative difference in the behavior as a function of detuning. We stress that there is no \textit{a priori} reason why this conclusion should not generalize beyond contact electronic interactions to Coulomb or Keldysh~\cite{Keldysh1979} type potentials in, e.g., GaAs or TMDs: At large electron-exciton separation $r$, both of these electronic potentials will result in an effective exciton-electron potential that decays like $-1/r^4$ corresponding to the interaction between a charge and an induced dipole. Such a potential qualifies as short ranged~\cite{landau2013quantum}, and therefore the low-energy exciton-electron scattering necessarily follows the form in Eq.~\eqref{eq:Tex}. Furthermore, this approximation relies only on a single quantity, the low-energy scale $\varepsilon_{eX}$, which can be fitted in experiment or calculated in exact solutions of the exciton-electron three-body problem such as in the recent work by Fey \textit{et al.}~\cite{Fey2020}. Finally, we note that since $\varepsilon_{eX}$ only appears under a logarithm, our results are not very sensitive to realistic variations in this parameter, and they may therefore be expected to be quite accurate even at a quantitative level.

\subsection{Polariton-electron interactions at finite momentum}
\label{sec:e-P_scattering_finiteQ}

We now turn to the interaction of an electron and a polariton at finite momentum which, as we shall show, can be strongly enhanced compared with their exciton-electron counterpart. As previously noted, we will restrict our attention to the case of zero center-of-mass momentum, which allows us to perform a partial wave decomposition of the $T$ matrix as follows~\cite{AdhikariAJP86}:
\begin{align}
    t(\p_1,\p_2)&=\sum_{l=0}^\infty (2-\delta_{l0})\cos(l \phi_{12})t_{l}(p_1,p_2),
    \label{eq:angular decompose 2D_t}
\end{align}
where $l$ is the angular momentum quantum number and $\phi_{12}$ is the angle between $\p_1$ and $\p_2$. Obviously, the scattering in the limit of zero momentum discussed in the previous subsection corresponds to $s$-wave, $l=0$. This decomposition allows us to solve the integral equation as a function of a single parameter, the magnitude of the incident relative momentum. The main challenge then is that the electron-hole $T$ matrix in Eq.~\eqref{eq:ePtMatrix} has a simple pole when $q=p_2$~\cite{BedaquePRC1998}, and we discuss how to deal with this in Appendix~\ref{app:ePfiniteQ}. Once we have solved for the partial wave amplitudes $t_l$, we obtain the normalized scattering $T$ matrix in the $l$'th partial wave via
\begin{align}\label{eq:Tep_exact}
    T_{l}(p)=Z_-(\p)|X_-(\p)|^2t_{l} (p),
\end{align}
similarly to Eq.~\eqref{eq:Tfromt}. In this work, we focus on $s$-wave scattering, which we denote by $T_s(p)$. However, we note that higher partial waves (in particular $p$-wave) can become important in the scattering of polaritons with heavy holes~\cite{Levinsen2009,Ngampruetikorn2013} due to the presence of hole-hole-electron trion bound states which have previously been discussed in the context of quantum wells~\cite{Sergeev2001} and TMD monolayers~\cite{Ganchev2015,Courtade2017} or strongly screened electronic interactions~\cite{Pricoupenko2010,atomdimer}.

\begin{figure*}[t]
	\centering
	\includegraphics[width=\linewidth]{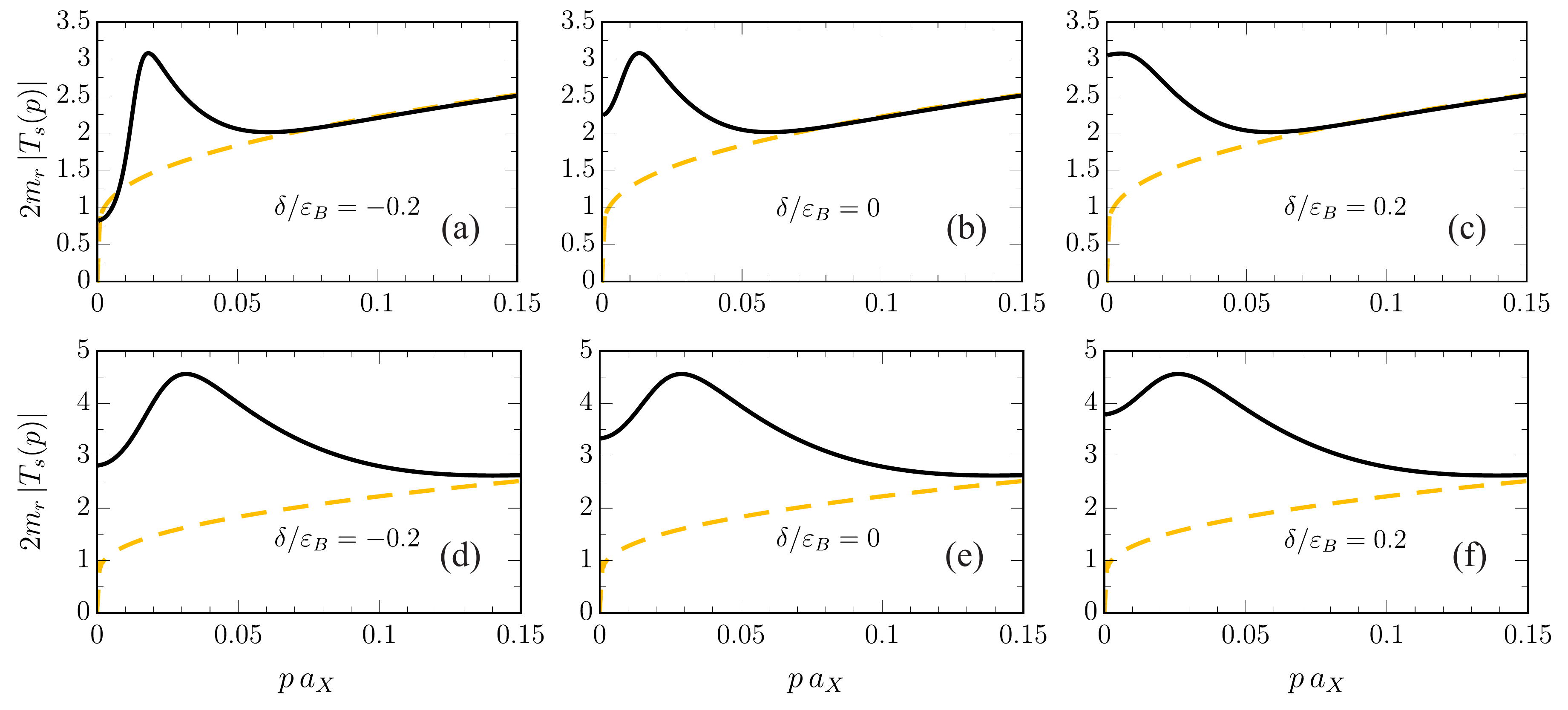}
	\caption{Polariton-electron scattering $T$ matrix at finite momentum for parameters relevant to GaAs semiconductor microcavities.
We show our exact diagrammatic calculation (black solid line) and the corresponding exciton-electron scattering $T$ matrix (yellow dashed line) for $m_e=0.067 m_0$, $m_h=0.45 m_0$ \cite{GaAsEffectiveMass}, and $m_c=10^{-4} m_0$. In the top row we have $\Omega/\eb=0.1$ while in the bottom row $\Omega/\eb=0.5$ \cite{GaAsLargeRabi}.}
	\label{fig:e-P_gaas}
\end{figure*}

Figures~\ref{fig:e-P_tmd} and \ref{fig:e-P_gaas} show our results for the polariton-electron scattering at finite relative momentum for the case of a TMD monolayer and for a GaAs quantum well, respectively. A striking aspect of our results is the presence of a resonance-like peak at momenta close to the polariton inflection point. The peak is most pronounced for zero and negative detuning and it is enhanced with increased Rabi coupling. By contrast, we find that the peak is completely absent in the case of exciton-electron scattering (also shown in the figures), where we obtain the corresponding $T$ matrix using Eq.~\eqref{eq:Tex} at collision energy $E+\eb=p^2/2m_{eX}$. At a qualitative level, the results for GaAs and a TMD monolayer embedded in a microcavity are similar, although the generally larger ratios of Rabi coupling to exciton binding energy in GaAs means that the relative enhancement of interactions in GaAs microcavity is larger. For relative momenta above the polariton inflection point, we see that the polariton-electron $T$ matrix quickly reduces to the exciton-electron $T$ matrix, as we would expect since in that limit the exciton fraction approaches unity.

\begin{figure}
\centering
	\includegraphics[width=\linewidth]{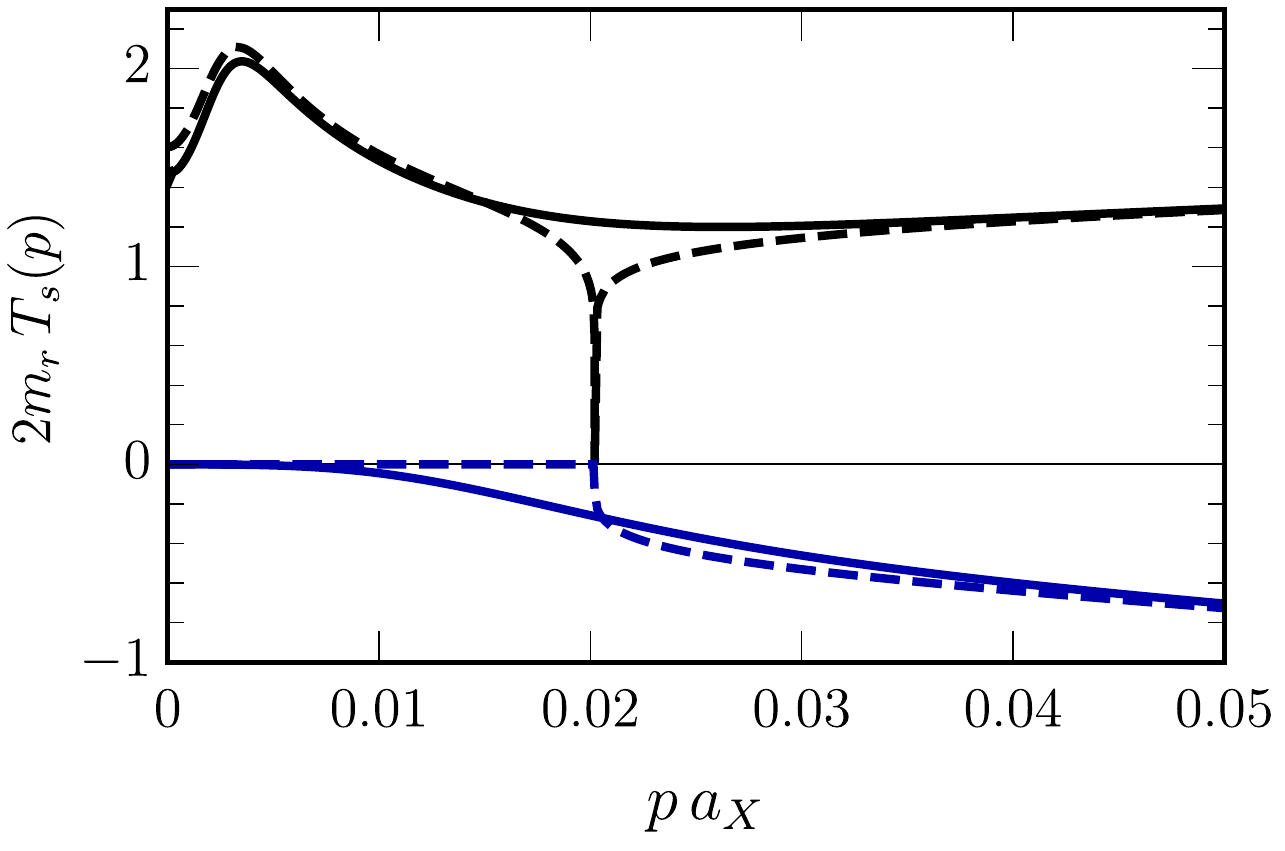}
\caption{Comparison between the real and imaginary parts of the electron-polariton scattering $T$ matrix. Black and blue solid lines correspond to the real and imaginary part of $T_s(p)$ in Eq.~\eqref{eq:Tep_exact}, respectively. Black and blue dashed lines are the real and imaginary parts of the off-shell exciton approximation in Eq.~\eqref{eq:Tep_approximated}, respectively. For this comparison, we take the parameters $m_e=m_h=m_0$, $m_c=10^{-4} m_0$, $\delta/\eb=0$, and $\Omega/\eb=0.025$.} 
 	\label{fig:Tfinitecomparison}
\end{figure}

As in the case of scattering at low momenta, we can understand the resonance-like feature in the polariton-electron scattering at finite momentum in terms of off-shell exciton-electron scattering. Using a similar reasoning to that which led to Eq.~\eqref{eq:gepfromgex}, we approximate
\begin{align}
    T_s(p)\simeq |X_-(\p)|^2T_{eX}(E_-(\p)+\epsilon_\p^e).
    \label{eq:Tep_approximated}
\end{align}
We illustrate this in Fig.~\ref{fig:Tfinitecomparison} for the case of the TMDs with a smaller ratio of Rabi coupling to exciton binding energy (MoSe$_2$, WSe$_2$, or MoS$_2$). We see that this approximation works extremely well for almost all momenta, the only difference being a narrow region around the momentum where the collision energy $E+\eb$ vanishes. By analyzing the two contributions in Eq.~\eqref{eq:Tep_approximated}, we conclude that the resonance-like feature is due to a competition between the exciton fraction --- which quickly approaches unity when the momentum approaches the polariton inflection point --- and the fact that scattering is suppressed when the collision energy approaches zero. Figure~\ref{fig:Tfinitecomparison} also shows that the polariton-electron $T$ matrix is nearly purely real at the peak, which we can also understand from Eq.~\eqref{eq:Tep_approximated} since exciton-electron scattering only features an imaginary part for positive collision energy.

Crucially, our results in Figures~\ref{fig:e-P_tmd}-\ref{fig:Tfinitecomparison} imply that the polariton-electron $T$ matrix is essentially never smaller than that of an electron and an exciton at a given momentum, the only exception being at a very large negative detuning. This is in spite of the suppression due to the reduced exciton fraction in the polariton, and it demonstrates that the strong energy shift of the scattering process due to the light-matter coupling is the dominant effect. Our result can thus be understood as an effective photon-mediated enhancement of exciton-electron scattering, which is a major qualitative difference compared with the commonly applied assumption in the Born approximation, Eq.~\eqref{eq:Born_assumption}. As we have derived this effect in a fully microscopic model, and since it agrees with the universal low-energy behavior, Eq.~\eqref{eq:Tep_approximated}, we expect that this conclusion carries over to more realistic interactions between charged particles.

Given the general nature of our arguments, the enhancement of interactions between polaritons and electrons is also expected to carry over to other geometries, such as systems where a 2D electron gas is in proximity to a 2D exciton-polariton condensate. Here, it would be interesting to investigate whether the enhanced interactions could lead to stronger drag effects~\cite{OlegPRB2010}, or to more robust superconductivity~\cite{LaussyPRL10,CotlePRB16}.

\section{Conclusions and outlook}
\label{sec:conclusion}

We have presented a microscopic approach to exciton-polaritons in a 2D semiconductor embedded in a microcavity. By treating the semiconductor interactions as strongly screened, we have found an analytic expression for the polariton propagator, which serves as a starting point for further few- and many-body calculations. We have applied our microscopic description of polaritons to the calculation of the scattering of electrons and polaritons in a manner that is exact within our approximation. Remarkably, we have shown that polariton-electron scattering is nearly always stronger than exciton-electron scattering at a given momentum, in contrast to the prevailing belief based on the Born approximation. In particular, we have shown that the interactions can be strongly enhanced up to and beyond the inflection point of the polariton dispersion, giving rise to a resonance-like feature. While this qualitatively new and unexpected behavior may appear counter-intuitive, we have argued that the enhancement of scattering is due to the strong light-matter coupling that shifts the collision energy compared with exciton-electron scattering, and hence this is a generic effect which is independent of our approximation of screened interactions.

Our approach can be directly applied to a large class of other few-body problems in light-matter coupled systems. Of particular interest is the case of spin-polarized electrons interacting with an exciton-polariton of opposite circular polarization. In this case, the exciton-electron system features a trion bound state which leads to resonantly enhanced interactions, and this has recently enabled the observation of polaron-polaritons~\cite{SidlerNatPhys16,Chervy2020}. Furthermore, using the four-body techniques developed in the context of cold atomic gases Refs.~\cite{Petrov2003,Petrov2005,BrodskyPRA06,LevinsenPRA06}, our approach can be straightforwardly generalized to the calculation of polariton-polariton and polariton-exciton interactions. This would, for instance, allow a fully microscopic description of the scattering resonances due to coupling to a biexciton bound state, as observed in Refs.~\cite{TakemuraNPhys2014,Takemura2017,Navadeh-Toupchi2019} and also analyzed in Refs.~\cite{Levinsen2019,Bastarrachea-Magnani2020}. Finally, the microscopic theory of exciton-polaritons for the case of Coulomb interactions developed in Ref.~\cite{JesperPRR19} also allows a natural diagrammatic representation, which can be applied to few-body scattering problems. Hence our approach holds promise as a fully microscopic theory in which to obtain key quantities in exciton-polariton systems. 


\acknowledgements

We thank Olivier Bleu, Dmitry Efimkin, Eliezer Estrecho, Emma Laird, David Neilson, Elena Ostrovskaya, and Maciej Pieczarka for useful discussions, and we acknowledge support from the Australian Research Council Centre of Excellence in Future Low-Energy Electronics Technologies (CE170100039). JL is furthermore supported through the Australian Research Council Future Fellowship FT160100244.


\appendix

\section{Polariton at finite momentum}
\label{app:polaritonQ}

In this Appendix, we apply the operator approach discussed in Sec.~\ref{sec:operator} to a polariton at finite momentum $\Q$. 
The polariton creation operator $P^\dag_\Q$ is defined as:
\begin{equation}\label{eq:Psi_Q}
P_{\Q}^\dagger\ket{0}\equiv \ket{\Psi_\Q} =\sum_\k \psi_\k^{(\Q)}\, e^\dag_{\Qe+\k} h^\dag_{\Qh-\k}\ket{0}+\gamma^{(\Q)} c^\dag_\Q \ket{0},
\end{equation}
where $\mathbf{Q}_{e,h}\equiv m_{e,h}\,\Q/m_X$ with $m_X=m_e+m_h$ the exciton mass. The \sch equation can be obtained by projecting $(E-\hat{H})\ket{\Psi_\Q}=0$ onto the electron-hole and photon parts of Eq.~\eqref{eq:Psi_Q}, which gives:
\begin{subequations}
\label{eq:sch_contact_Q}
\begin{align}
    \left(E-\epsilon^X_\Q-\bar{\epsilon}_\k\right)\psi^{(\Q)}_\k&=-V_0 \sum_{\kP} \psi^{(\Q)}_{\kP}+g\gamma^{(\Q)},\label{eq:app_A_Sch_Q_1}\\
\left(E-\omega-\epsilon^c_\Q\right)\gamma^{(\Q)}&=g\sum_\k \varphi^{(\Q)}_\k.\label{eq:app_A_Sch_Q_2}
\end{align}
\end{subequations}
The momentum $\Q$ serves as an external parameter in Eq.~\eqref{eq:sch_contact_Q}
and thus the renormalization procedure is the same as in Sec.~\ref{sec:operator}, as we now demonstrate.

We first consider the case of small exciton-photon Rabi coupling where the polariton energy is close to the exciton energy, i.e., $E=-\eb+\epsilon^X_\Q +\Delta E$. In the limit $\Delta E\to 0$, the electron-hole part of the polariton wave function $\psi_\k^{(\Q)}$ can be approximated as proportional to the exciton wave function, 
$\psi_\k^{(\Q)}\simeq\beta^{(\Q)}\, \phi_{X\k}$, where $\beta^{(\Q)}$ is a complex number. Within this approximation, Eq.~\eqref{eq:sch_contact_Q} takes the form:
\begin{subequations}
\begin{align}
    (E+\eb-\epsilon^X_\Q )\beta^{(\Q)}&=\gamma^{(\Q)}\,g\sum_\k \phi_{X\k},   \label{eq:two-level_sch_1_app_Q}\\
    (E-\omega-\epsilon^c_\Q)\gamma^{(\Q)}&=\beta^{(\Q)}\,g\sum_\k \phi_{X\k}. \label{eq:two-level_sch_2_app_Q}
\end{align}
\end{subequations}
Written in matrix form we have
\begin{align}
    \begin{bmatrix}
-\eb+\epsilon^X_\Q  & \Omega \\[0.2em]
\Omega & \omega+ \epsilon^c_\Q
\end{bmatrix}\begin{bmatrix}
\beta^{(\Q)} \\[0.2em]
\gamma^{(\Q)} 
\end{bmatrix}=E \begin{bmatrix}
\beta^{(\Q)} \\[0.2em]
\gamma^{(\Q)} 
\end{bmatrix} \label{eq:matrix_eq_two_level_Q},
\end{align}
where we have identified the off-diagonal term $g\sum_\k \phi_{X\k}$ as the experimentally measurable Rabi coupling $\Omega$ introduced in Eq.~\eqref{eq:g_renormalization0}. Equation~\eqref{eq:matrix_eq_two_level_Q} yields the spectrum of two coupled oscillators as
\begin{align}\label{eq:app_two_level_dispersion}
E^{\rm osc}_\pm =\frac{1}{2}\left[\epsilon^X_\Q+\delta+\epsilon_\Q^c 
   \pm\sqrt{(\delta+\epsilon_\Q^c-\epsilon^X_\Q )^2+4\Omega^2}\,\right]
   -\eb,
\end{align}
with the Hopfield coefficients
\begin{subequations}
\begin{align}
    \abs{C_\pm^{\rm osc}(\Q)}^2&=\frac{1}{2}\left(1\mp\frac{\epsilon^X_\Q-\delta-\epsilon_\Q^c}{\sqrt{(\epsilon^X_\Q-\delta-\epsilon_\Q^c)^2+4\Omega^2}}\right),\\
    \abs{X_\pm^{\rm osc}(\Q)}^2&=\frac{1}{2}\left(1\pm\frac{\epsilon^X_\Q-\delta-\epsilon_\Q^c}{\sqrt{(\epsilon^X_\Q-\delta-\epsilon_\Q^c)^2+4\Omega^2}}\right).
\end{align}
\end{subequations}

After the renormalization, we follow a similar derivation as that carried out in Eqs.~\eqref{eq:f_and_gamma} and \eqref{eq:complicated} to find that the exact polariton dispersion $E_\pm(\Q)$ satisfies 
\begin{equation}\label{eq:app_Ep_dispersion_Q}
\left[\omega+\epsilon^c_\Q-E_\pm(\Q)\right]\ln\left[\frac{-E_\pm(\Q)+\epsilon^X_\Q}{\eb }\right]=\frac{\Omega^2}{\eb }.
\end{equation}

With the polariton dispersion established, we can return to Eq.~\eqref{eq:sch_contact_Q} to solve for the electron-hole wave function. Since the right hand side of Eq.~\eqref{eq:app_A_Sch_Q_1} is independent of $\k$, we immediately know that the electron-hole
wave function $\psi_{\pm,\k}^{(\Q)}$ has the same functional form as in Eq.~\eqref{eq:ehwave}:
\begin{align}
    \psi_{\pm,\k}^{(\Q)}=\frac{\sqrt{Z_\pm{(\Q)}}\sqrt{1-|\gamma_\pm^{(\Q)}|^2}}{-E_\pm(\Q)+\epsilon^X_\Q+\ok},
    \label{eq:ehwave_Q}
\end{align}
with
\begin{align}
    Z_\pm{(\Q)}&=\left[\sum_\k\frac{1}{(E_\pm(\Q)-\epsilon^X_\Q -\bar{\epsilon}_{\k})^2} \right]^{-1}\nn\\[0.2em]
    &=\frac{2\pi\left|E_\pm(\Q)-\epsilon^X_\Q\right|}{m_r}.
    \label{eq:Zpm_Q}
\end{align}
We can also extract the Hopfield coefficients in Eq.~\eqref{eq:Hopfields} in the same manner as in Sec.~\ref{sec:operator}. This straightforwardly yields the same expressions as in the diagrammatic section~\ref{sec:Diagrammatic formulation}, and allows us to obtain an analytic expression for the electron-hole wave function in terms of experimentally measurable parameters.

\section{Electron-hole \textit{T} matrix}\label{app:fermion_loop}

In this Appendix, we derive the expression for the exciton propagator, Eq.~\eqref{eq:2d T matrix_copy_2}. Starting from the electron-hole $T$ matrix diagram shown in Fig.~\ref{fig:diagram_dressphoton}(c), the exciton propagator at zero momentum reads:
\begin{equation}\label{eq:T_0_app_copy}
    \mathcal{T}_0(E)
    =\frac{1}{-V_0^{-1}-\Pi(E)},
\end{equation}
where $V_0^{-1}$ has been related to an integral by Eq.~\eqref{eq:1overV0},
\begin{align}\label{eq:eq:1overV0_copy}
    \frac1{V_0}=\sum_\k^\Lambda \frac1{\eb+\ok}.
\end{align}
The bubble integral $\Pi(E)$ in Eq.~\eqref{eq:T_0_app_copy} can be worked out explicitly:
\begin{align}
\Pi(E)=&\sum_\k^\Lambda \frac{1}{E-\ok}\nn\\
=&\sum_{\k}^{\Lambda\to\infty} \left(\frac{1}{E-\ok}+\frac{1}{\eb +\ok}\right)-\sum_\k^\Lambda\frac{1}{\eb +\ok}\nn\\
=&\frac{m_r}{2\pi}\ln\left(\frac{-E}{\eb }\right)-\sum_\k^\Lambda\frac{1}{\eb +\ok},\label{eq:APP_Pi(E)}
\end{align}
where we can take the limit $\Lambda\to\infty$ in the difference in the second line, since this is ultraviolet convergent, while the last sum diverges as $\sim\ln\Lambda$ when $\Lambda\to\infty$.
Inserting Eqs.~\eqref{eq:APP_Pi(E)} and \eqref{eq:eq:1overV0_copy} into Eq.~\eqref{eq:T_0_app_copy} we obtain the expression for $\mathcal{T}_0(E)$:
\begin{equation}
\mathcal{T}_0(E)=\frac{2\pi/m_r}{-\ln\left[E/\eb+i0\right]+i\pi}.
\end{equation}

\section{Partial wave decomposition and numerical solution of the polariton-electron scattering integral equation}
\label{app:ePfiniteQ}

We now outline how we numerically investigate the key equation of polariton-electron scattering, namely Eq.~\eqref{eq:ePtMatrix}, which we rewrite here in integral form by taking the continuum limit
\begin{align}
t (\p_1,\p_2)=&-G_h(\p_1+\p_2,E-\epsilon^e_{\p_1}-\epsilon^e_{\p_2})\nn\\
&\hspace{-10mm}-\int_0^\infty \frac{q\,d q}{2\pi} \int_0^{2\pi} \frac{d \phi}{2\pi} G_h(\p_1+\q,E-\epsilon^e_{\p_1}-\epsilon^e_{\q})\nn\\
&\qquad\qquad\quad\times \mathcal{T} (\q,E-\epsilon^e_{\q}) t (\q,\p_2),\label{eq:ePtMatrix_2D_integral form}
\end{align}
where we again take the total energy $E=E_-(\p_2)+\epsilon_{\p_2}$. This represents an integral equation in the first momentum of the function $t(\p_1,\p_2)$, while it only depends parametrically on the second. There are two main challenges to deal with when solving this equation. First, the integral equation depends on the angle between incident and outgoing relative momenta. This angular dependence increases the numerical complexity, which we resolve by performing a partial wave decomposition. Second, the polariton propagator has a simple pole, which we address by using a principal value prescription.

Let us first discuss the partial wave decomposition. As in Eq.~\eqref{eq:angular decompose 2D_t}, we define
\begin{align}
    t(\p_1,\p_2)&=\sum_{l=0}^\infty (2-\delta_{l0})\cos(l \phi_{12})t_{l}(p_1,p_2),
    \label{eq:twave}
\end{align}    
with $\phi_{12}$ the angle between $\p_1$ and $\p_2$. This can be inverted to find
\begin{align}
t_{l}(p_1,p_2)&= \int_0^{2\pi}\frac{d\phi_{12}}{2\pi} \cos(l \phi_{12})t(\p_1,\p_2).
\end{align}
Similarly, we define the partial wave decomposition of the hole Green's function
\begin{align}
    &G_h(\p_1+\p_2,E-\epsilon^e_{\p_1}-\epsilon^e_{\p_2})\nn \\ &\qquad=\sum_{l=0}^\infty (2-\delta_{l0})\cos(l \phi_{12})g_l(p_1,p_2),
    \label{eq:gwave}
\end{align}
and 
\begin{align}
    &g_l(p_1,p_2)\nn \\ &\qquad =\int_0^{2\pi}\frac{d\phi_{12}}{2\pi} \cos(l \phi_{12})G_h(\p_1+\p_2,E-\epsilon^e_{\p_1}-\epsilon^e_{\p_2}).
\end{align}

Now let us define the angles of the vectors $\p_1$, $\p_2$, and $\q$ in Eq.~\eqref{eq:ePtMatrix_2D_integral form} as $\phi_1$, $\phi_2$, $\phi_q$, respectively. Then we can take $\phi\equiv\phi_{q2}=\phi_q-\phi_2$ in the integral, while the angle between $\p_1$ and $\q$ is $\phi_{1q}\equiv\phi_1-\phi_q=\phi_{12}-\phi_{q2}$ with $\phi_{12}=\phi_1-\phi_2$. Inserting the expansions \eqref{eq:twave} and \eqref{eq:gwave} in Eq.~\eqref{eq:ePtMatrix_2D_integral form} and focussing only on the angular integral, we have
\begin{align}
    &\int_0^{2\pi}\frac{d\phi_{q2}}{2\pi}(2-\delta_{l_h0})(2-\delta_{l_t0})\cos(l_h\phi_{1q})\cos(l_t\phi_{q2})\nn \\ 
    &\quad=\delta_{l_hl_t}(2-\delta_{l_t0})\cos(l_t\phi_{12}),
\end{align}
where $l_h$ and $l_t$ are the partial wave quantum numbers of the hole propagator and the polariton-electron $T$ matrix inside the integral, respectively. We see that this only depends on the angle $\phi_{12}$ which is the same that we have on the left hand side of Eq.~\eqref{eq:ePtMatrix_2D_integral form}. Therefore, applying the operator $\int_0^{2\pi}\frac{d\phi_{12}}{2\pi}\cos(l\phi_{12})[\cdot]$ to the equation, we find that we must have $l_h=l_t=l$, and we arrive at the equation
\begin{align}
&t_l(p_1,p_2)=-g_l(p_1,p_2)
\nn\\ &\qquad
-\int_0^\infty \frac{q\,dq}{2\pi}g_l(p_1,q)\mathcal{T}(q,E-\epsilon^e_q)t_l(q,p_2).
\label{eq:2D t matrix equation}
\end{align}

The polariton propagator in Eq.~\eqref{eq:2D t matrix equation} has a simple pole at $q=p_2+i0$, since $\mathcal{T}(q,E-\epsilon^e_q)=\mathcal{T}(q,E_-(p_2)+\epsilon^e_{p_2}-\epsilon^e_q)$ by definition diverges as $q\to p_2$. To circumvent this pole, we use the fact that
\begin{align}
    \frac1{q-k-i0}=\frac{\cal P}{q-k}+i\pi\delta(q-k),
\end{align}
where ${\cal P}$ denotes the principal part. Using this in Eq.~\eqref{eq:2D t matrix equation}, we find
\begin{align}
&t_l(p_1,p_2)=-[1-i\pi\alpha(p_2)t_l(p_2,p_2)]g_l(p_1,p_2)
\nn\\ &\qquad
-{\cal P}\int_0^\infty \frac{q\,dq}{2\pi}g_l(p_1,q)\mathcal{T}(q,E-\epsilon^e_q)t_l(q,p_2),
\label{eq:2D t matrix equation_v2}
\end{align}
where
\begin{align}
    \alpha(p)&=-\lim_{q\to p}\frac{q(q-p)}{2\pi}{\cal T}(q,E_-(p)+\epsilon_p^e-\epsilon_q^e)\\
    & = \frac{p}{2\pi}\frac{|X_-(p)|^2Z_(p)}{\partial(E_-(p)+\epsilon^e_p)/\partial p}.
\end{align}
The prefactor in the first term of Eq.~\eqref{eq:2D t matrix equation_v2} is independent of $p_1$ and depends only on $p_2$. Since the multiplication by a constant prefactor in the inhomogenous term of an integral equation does not change its structure, we can instead solve the auxilliary equation
\begin{align}
&K_l(p_1,p_2)=-g_l(p_1,p_2)
\nn\\ &\qquad
-{\cal P}\int_0^\infty \frac{q\,dq}{2\pi}g_l(p_1,q)\mathcal{T}(q,E-\epsilon^e_q)K_l(q,p_2),
\label{eq:2D t matrix equation_v3}
\end{align}
where the pole is explicitly excluded using the principal value prescription~\cite{BedaquePRC1998}. We then relate the on-shell unnormalized $t$ matrix to $K$ via
\begin{align}
    t_l(k,k)=[1-i\pi\alpha(k)t_l(k,k)]K_l(k,k),
\end{align}
or, in other words, 
\begin{align}
    t_l(k,k)=\frac1{K_l^{-1}(k,k)+i\pi\alpha(k)}.
\end{align}
We solve Eq.~\eqref{eq:2D t matrix equation_v3} using Gauss-Legendre quadrature.

In the case of scattering at low momentum, we can further simplify the problem, since the simple pole in this case disappears. We then set $p_2=0$ and $p_1=p$ in Eq.~\eqref{eq:2D t matrix equation}, which can then be written as
\begin{align}
&t_0(p,0)=-\frac{1}{E-p^2/2m_r}\nn\\
&\quad+\int_0^\infty \frac{q\,dq}{2\pi} \frac{\mathcal{T}(q, E-q^2/2m_e) 
t_0(q,0)}{\sqrt{(E-p^2/2m_r-q^2/2m_r)^2-(p\, q/m_h)^2}}.\label{eq:STM_p_equation}
\end{align}

\bibliography{e_P_scattering_refs}

\begin{thebibliography}{76}%
\makeatletter
\providecommand \@ifxundefined [1]{%
 \@ifx{#1\undefined}
}%
\providecommand \@ifnum [1]{%
 \ifnum #1\expandafter \@firstoftwo
 \else \expandafter \@secondoftwo
 \fi
}%
\providecommand \@ifx [1]{%
 \ifx #1\expandafter \@firstoftwo
 \else \expandafter \@secondoftwo
 \fi
}%
\providecommand \natexlab [1]{#1}%
\providecommand \enquote  [1]{``#1''}%
\providecommand \bibnamefont  [1]{#1}%
\providecommand \bibfnamefont [1]{#1}%
\providecommand \citenamefont [1]{#1}%
\providecommand \href@noop [0]{\@secondoftwo}%
\providecommand \href [0]{\begingroup \@sanitize@url \@href}%
\providecommand \@href[1]{\@@startlink{#1}\@@href}%
\providecommand \@@href[1]{\endgroup#1\@@endlink}%
\providecommand \@sanitize@url [0]{\catcode `\\12\catcode `\$12\catcode
  `\&12\catcode `\#12\catcode `\^12\catcode `\_12\catcode `\%12\relax}%
\providecommand \@@startlink[1]{}%
\providecommand \@@endlink[0]{}%
\providecommand \url  [0]{\begingroup\@sanitize@url \@url }%
\providecommand \@url [1]{\endgroup\@href {#1}{\urlprefix }}%
\providecommand \urlprefix  [0]{URL }%
\providecommand \Eprint [0]{\href }%
\providecommand \doibase [0]{http://dx.doi.org/}%
\providecommand \selectlanguage [0]{\@gobble}%
\providecommand \bibinfo  [0]{\@secondoftwo}%
\providecommand \bibfield  [0]{\@secondoftwo}%
\providecommand \translation [1]{[#1]}%
\providecommand \BibitemOpen [0]{}%
\providecommand \bibitemStop [0]{}%
\providecommand \bibitemNoStop [0]{.\EOS\space}%
\providecommand \EOS [0]{\spacefactor3000\relax}%
\providecommand \BibitemShut  [1]{\csname bibitem#1\endcsname}%
\let\auto@bib@innerbib\@empty
\bibitem [{\citenamefont {Deng}\ \emph {et~al.}(2010)\citenamefont {Deng},
  \citenamefont {Haug},\ and\ \citenamefont {Yamamoto}}]{DengRevMod10}%
  \BibitemOpen
  \bibfield  {author} {\bibinfo {author} {\bibfnamefont {H.}~\bibnamefont
  {Deng}}, \bibinfo {author} {\bibfnamefont {H.}~\bibnamefont {Haug}}, \ and\
  \bibinfo {author} {\bibfnamefont {Y.}~\bibnamefont {Yamamoto}},\ }\bibfield
  {title} {\bibinfo {title} {\emph {Exciton-polariton Bose-{E}instein
  condensation}},\ }\href {\doibase 10.1103/RevModPhys.82.1489} {\bibfield
  {journal} {\bibinfo  {journal} {Rev. Mod. Phys.}\ }\textbf {\bibinfo {volume}
  {82}},\ \bibinfo {pages} {1489} (\bibinfo {year} {2010})}\BibitemShut
  {NoStop}%
\bibitem [{\citenamefont {Carusotto}\ and\ \citenamefont
  {Ciuti}(2013)}]{CiutiRevMod13}%
  \BibitemOpen
  \bibfield  {author} {\bibinfo {author} {\bibfnamefont {I.}~\bibnamefont
  {Carusotto}}\ and\ \bibinfo {author} {\bibfnamefont {C.}~\bibnamefont
  {Ciuti}},\ }\bibfield  {title} {\bibinfo {title} {\emph {Quantum fluids of
  light}},\ }\href {\doibase 10.1103/RevModPhys.85.299} {\bibfield  {journal}
  {\bibinfo  {journal} {Rev. Mod. Phys.}\ }\textbf {\bibinfo {volume} {85}},\
  \bibinfo {pages} {299} (\bibinfo {year} {2013})}\BibitemShut {NoStop}%
\bibitem [{\citenamefont {Byrnes}\ \emph {et~al.}(2014)\citenamefont {Byrnes},
  \citenamefont {Kim},\ and\ \citenamefont {Yamamoto}}]{ByrnesNatPhys14}%
  \BibitemOpen
  \bibfield  {author} {\bibinfo {author} {\bibfnamefont {T.}~\bibnamefont
  {Byrnes}}, \bibinfo {author} {\bibfnamefont {N.~Y.}\ \bibnamefont {Kim}}, \
  and\ \bibinfo {author} {\bibfnamefont {Y.}~\bibnamefont {Yamamoto}},\
  }\bibfield  {title} {\bibinfo {title} {\emph {Exciton-polariton
  condensates}},\ }\href {\doibase 10.1038/nphys3143} {\bibfield  {journal}
  {\bibinfo  {journal} {Nat. Phys.}\ }\textbf {\bibinfo {volume} {10}},\
  \bibinfo {pages} {803} (\bibinfo {year} {2014})}\BibitemShut {NoStop}%
\bibitem [{\citenamefont {Amo}\ \emph {et~al.}(2009)\citenamefont {Amo},
  \citenamefont {Lefr{\`e}re}, \citenamefont {Pigeon}, \citenamefont {Adrados},
  \citenamefont {Ciuti}, \citenamefont {Carusotto}, \citenamefont {Houdr{\'e}},
  \citenamefont {Giacobino},\ and\ \citenamefont {Bramati}}]{Amo2009}%
  \BibitemOpen
  \bibfield  {author} {\bibinfo {author} {\bibfnamefont {A.}~\bibnamefont
  {Amo}}, \bibinfo {author} {\bibfnamefont {J.}~\bibnamefont {Lefr{\`e}re}},
  \bibinfo {author} {\bibfnamefont {S.}~\bibnamefont {Pigeon}}, \bibinfo
  {author} {\bibfnamefont {C.}~\bibnamefont {Adrados}}, \bibinfo {author}
  {\bibfnamefont {C.}~\bibnamefont {Ciuti}}, \bibinfo {author} {\bibfnamefont
  {I.}~\bibnamefont {Carusotto}}, \bibinfo {author} {\bibfnamefont
  {R.}~\bibnamefont {Houdr{\'e}}}, \bibinfo {author} {\bibfnamefont
  {E.}~\bibnamefont {Giacobino}}, \ and\ \bibinfo {author} {\bibfnamefont
  {A.}~\bibnamefont {Bramati}},\ }\bibfield  {title} {\bibinfo {title} {\emph
  {Superfluidity of polaritons in semiconductor microcavities}},\ }\href
  {https://doi.org/10.1038/nphys1364} {\bibfield  {journal} {\bibinfo
  {journal} {Nature Physics}\ }\textbf {\bibinfo {volume} {5}},\ \bibinfo
  {pages} {805} (\bibinfo {year} {2009})}\BibitemShut {NoStop}%
\bibitem [{\citenamefont {Sanvitto}\ \emph {et~al.}(2010)\citenamefont
  {Sanvitto}, \citenamefont {Marchetti}, \citenamefont {Szyma{\'n}ska},
  \citenamefont {Tosi}, \citenamefont {Baudisch}, \citenamefont {Laussy},
  \citenamefont {Krizhanovskii}, \citenamefont {Skolnick}, \citenamefont
  {Marrucci}, \citenamefont {Lema{\^\i}tre}, \citenamefont {Bloch},
  \citenamefont {Tejedor},\ and\ \citenamefont {Vi{\~n}a}}]{Sanvitto2010}%
  \BibitemOpen
  \bibfield  {author} {\bibinfo {author} {\bibfnamefont {D.}~\bibnamefont
  {Sanvitto}}, \bibinfo {author} {\bibfnamefont {F.~M.}\ \bibnamefont
  {Marchetti}}, \bibinfo {author} {\bibfnamefont {M.~H.}\ \bibnamefont
  {Szyma{\'n}ska}}, \bibinfo {author} {\bibfnamefont {G.}~\bibnamefont {Tosi}},
  \bibinfo {author} {\bibfnamefont {M.}~\bibnamefont {Baudisch}}, \bibinfo
  {author} {\bibfnamefont {F.~P.}\ \bibnamefont {Laussy}}, \bibinfo {author}
  {\bibfnamefont {D.~N.}\ \bibnamefont {Krizhanovskii}}, \bibinfo {author}
  {\bibfnamefont {M.~S.}\ \bibnamefont {Skolnick}}, \bibinfo {author}
  {\bibfnamefont {L.}~\bibnamefont {Marrucci}}, \bibinfo {author}
  {\bibfnamefont {A.}~\bibnamefont {Lema{\^\i}tre}}, \bibinfo {author}
  {\bibfnamefont {J.}~\bibnamefont {Bloch}}, \bibinfo {author} {\bibfnamefont
  {C.}~\bibnamefont {Tejedor}}, \ and\ \bibinfo {author} {\bibfnamefont
  {L.}~\bibnamefont {Vi{\~n}a}},\ }\bibfield  {title} {\bibinfo {title} {\emph
  {Persistent currents and quantized vortices in a polariton superfluid}},\
  }\href {https://doi.org/10.1038/nphys1668} {\bibfield  {journal} {\bibinfo
  {journal} {Nature Physics}\ }\textbf {\bibinfo {volume} {6}},\ \bibinfo
  {pages} {527} (\bibinfo {year} {2010})}\BibitemShut {NoStop}%
\bibitem [{\citenamefont {Amo}\ \emph {et~al.}(2010)\citenamefont {Amo},
  \citenamefont {Liew}, \citenamefont {Adrados}, \citenamefont {Houdr{\'e}},
  \citenamefont {Giacobino}, \citenamefont {Kavokin},\ and\ \citenamefont
  {Bramati}}]{Amo2010}%
  \BibitemOpen
  \bibfield  {author} {\bibinfo {author} {\bibfnamefont {A.}~\bibnamefont
  {Amo}}, \bibinfo {author} {\bibfnamefont {T.~C.~H.}\ \bibnamefont {Liew}},
  \bibinfo {author} {\bibfnamefont {C.}~\bibnamefont {Adrados}}, \bibinfo
  {author} {\bibfnamefont {R.}~\bibnamefont {Houdr{\'e}}}, \bibinfo {author}
  {\bibfnamefont {E.}~\bibnamefont {Giacobino}}, \bibinfo {author}
  {\bibfnamefont {A.~V.}\ \bibnamefont {Kavokin}}, \ and\ \bibinfo {author}
  {\bibfnamefont {A.}~\bibnamefont {Bramati}},\ }\bibfield  {title} {\bibinfo
  {title} {\emph {Exciton-polariton spin switches}},\ }\href
  {https://doi.org/10.1038/nphoton.2010.79} {\bibfield  {journal} {\bibinfo
  {journal} {Nature Photonics}\ }\textbf {\bibinfo {volume} {4}},\ \bibinfo
  {pages} {361} (\bibinfo {year} {2010})}\BibitemShut {NoStop}%
\bibitem [{\citenamefont {Mu{\~n}oz-Matutano}\ \emph
  {et~al.}(2019)\citenamefont {Mu{\~n}oz-Matutano}, \citenamefont {Wood},
  \citenamefont {Johnsson}, \citenamefont {Vidal}, \citenamefont {Baragiola},
  \citenamefont {Reinhard}, \citenamefont {Lema{\^\i}tre}, \citenamefont
  {Bloch}, \citenamefont {Amo}, \citenamefont {Nogues}, \citenamefont {Besga},
  \citenamefont {Richard},\ and\ \citenamefont {Volz}}]{Munoz2019}%
  \BibitemOpen
  \bibfield  {author} {\bibinfo {author} {\bibfnamefont {G.}~\bibnamefont
  {Mu{\~n}oz-Matutano}}, \bibinfo {author} {\bibfnamefont {A.}~\bibnamefont
  {Wood}}, \bibinfo {author} {\bibfnamefont {M.}~\bibnamefont {Johnsson}},
  \bibinfo {author} {\bibfnamefont {X.}~\bibnamefont {Vidal}}, \bibinfo
  {author} {\bibfnamefont {B.~Q.}\ \bibnamefont {Baragiola}}, \bibinfo {author}
  {\bibfnamefont {A.}~\bibnamefont {Reinhard}}, \bibinfo {author}
  {\bibfnamefont {A.}~\bibnamefont {Lema{\^\i}tre}}, \bibinfo {author}
  {\bibfnamefont {J.}~\bibnamefont {Bloch}}, \bibinfo {author} {\bibfnamefont
  {A.}~\bibnamefont {Amo}}, \bibinfo {author} {\bibfnamefont {G.}~\bibnamefont
  {Nogues}}, \bibinfo {author} {\bibfnamefont {B.}~\bibnamefont {Besga}},
  \bibinfo {author} {\bibfnamefont {M.}~\bibnamefont {Richard}}, \ and\
  \bibinfo {author} {\bibfnamefont {T.}~\bibnamefont {Volz}},\ }\bibfield
  {title} {\bibinfo {title} {\emph {Emergence of quantum correlations from
  interacting fibre-cavity polaritons}},\ }\href {\doibase
  10.1038/s41563-019-0281-z} {\bibfield  {journal} {\bibinfo  {journal} {Nature
  Materials}\ }\textbf {\bibinfo {volume} {18}},\ \bibinfo {pages} {213}
  (\bibinfo {year} {2019})}\BibitemShut {NoStop}%
\bibitem [{\citenamefont {Delteil}\ \emph {et~al.}(2019)\citenamefont
  {Delteil}, \citenamefont {Fink}, \citenamefont {Schade}, \citenamefont
  {H{\"o}fling}, \citenamefont {Schneider},\ and\ \citenamefont {{\.I}mamo{\u
  g}lu}}]{Delteil2019}%
  \BibitemOpen
  \bibfield  {author} {\bibinfo {author} {\bibfnamefont {A.}~\bibnamefont
  {Delteil}}, \bibinfo {author} {\bibfnamefont {T.}~\bibnamefont {Fink}},
  \bibinfo {author} {\bibfnamefont {A.}~\bibnamefont {Schade}}, \bibinfo
  {author} {\bibfnamefont {S.}~\bibnamefont {H{\"o}fling}}, \bibinfo {author}
  {\bibfnamefont {C.}~\bibnamefont {Schneider}}, \ and\ \bibinfo {author}
  {\bibfnamefont {A.}~\bibnamefont {{\.I}mamo{\u g}lu}},\ }\bibfield  {title}
  {\bibinfo {title} {\emph {Towards polariton blockade of confined
  exciton--polaritons}},\ }\href {\doibase 10.1038/s41563-019-0282-y}
  {\bibfield  {journal} {\bibinfo  {journal} {Nature Materials}\ }\textbf
  {\bibinfo {volume} {18}},\ \bibinfo {pages} {219} (\bibinfo {year}
  {2019})}\BibitemShut {NoStop}%
\bibitem [{\citenamefont {Takemura}\ \emph {et~al.}(2014)\citenamefont
  {Takemura}, \citenamefont {Trebaol}, \citenamefont {Wouters}, \citenamefont
  {Portella-Oberli},\ and\ \citenamefont {Deveaud}}]{TakemuraNPhys2014}%
  \BibitemOpen
  \bibfield  {author} {\bibinfo {author} {\bibfnamefont {N.}~\bibnamefont
  {Takemura}}, \bibinfo {author} {\bibfnamefont {S.}~\bibnamefont {Trebaol}},
  \bibinfo {author} {\bibfnamefont {M.}~\bibnamefont {Wouters}}, \bibinfo
  {author} {\bibfnamefont {M.~T.}\ \bibnamefont {Portella-Oberli}}, \ and\
  \bibinfo {author} {\bibfnamefont {B.}~\bibnamefont {Deveaud}},\ }\bibfield
  {title} {\bibinfo {title} {\emph {Polaritonic Feshbach resonance}},\ }\href
  {https://doi.org/10.1038/nphys2999} {\bibfield  {journal} {\bibinfo
  {journal} {Nature Physics}\ }\textbf {\bibinfo {volume} {10}},\ \bibinfo
  {pages} {500} (\bibinfo {year} {2014})}\BibitemShut {NoStop}%
\bibitem [{\citenamefont {Tan}\ \emph {et~al.}(2020)\citenamefont {Tan},
  \citenamefont {Cotlet}, \citenamefont {Bergschneider}, \citenamefont
  {Schmidt}, \citenamefont {Back}, \citenamefont {Shimazaki}, \citenamefont
  {Kroner},\ and\ \citenamefont {\ifmmode \dot{I}\else
  \.{I}\fi{}mamo\ifmmode~\breve{g}\else \u{g}\fi{}lu}}]{Tan2020}%
  \BibitemOpen
  \bibfield  {author} {\bibinfo {author} {\bibfnamefont {L.~B.}\ \bibnamefont
  {Tan}}, \bibinfo {author} {\bibfnamefont {O.}~\bibnamefont {Cotlet}},
  \bibinfo {author} {\bibfnamefont {A.}~\bibnamefont {Bergschneider}}, \bibinfo
  {author} {\bibfnamefont {R.}~\bibnamefont {Schmidt}}, \bibinfo {author}
  {\bibfnamefont {P.}~\bibnamefont {Back}}, \bibinfo {author} {\bibfnamefont
  {Y.}~\bibnamefont {Shimazaki}}, \bibinfo {author} {\bibfnamefont
  {M.}~\bibnamefont {Kroner}}, \ and\ \bibinfo {author} {\bibfnamefont
  {A.}~\bibnamefont {\ifmmode \dot{I}\else
  \.{I}\fi{}mamo\ifmmode~\breve{g}\else \u{g}\fi{}lu}},\ }\bibfield  {title}
  {\bibinfo {title} {\emph {Interacting Polaron-Polaritons}},\ }\href {\doibase
  10.1103/PhysRevX.10.021011} {\bibfield  {journal} {\bibinfo  {journal} {Phys.
  Rev. X}\ }\textbf {\bibinfo {volume} {10}},\ \bibinfo {pages} {021011}
  (\bibinfo {year} {2020})}\BibitemShut {NoStop}%
\bibitem [{\citenamefont {Tassone}\ and\ \citenamefont
  {Yamamoto}(1999)}]{TassonePRB99}%
  \BibitemOpen
  \bibfield  {author} {\bibinfo {author} {\bibfnamefont {F.}~\bibnamefont
  {Tassone}}\ and\ \bibinfo {author} {\bibfnamefont {Y.}~\bibnamefont
  {Yamamoto}},\ }\bibfield  {title} {\bibinfo {title} {\emph {Exciton-exciton
  scattering dynamics in a semiconductor microcavity and stimulated scattering
  into polaritons}},\ }\href {\doibase 10.1103/PhysRevB.59.10830} {\bibfield
  {journal} {\bibinfo  {journal} {Phys. Rev. B}\ }\textbf {\bibinfo {volume}
  {59}},\ \bibinfo {pages} {10830} (\bibinfo {year} {1999})}\BibitemShut
  {NoStop}%
\bibitem [{\citenamefont {Ramon}\ \emph {et~al.}(2002)\citenamefont {Ramon},
  \citenamefont {Rapaport}, \citenamefont {Qarry}, \citenamefont {Cohen},
  \citenamefont {Mann}, \citenamefont {Ron},\ and\ \citenamefont
  {Pfeiffer}}]{RamonPRB02}%
  \BibitemOpen
  \bibfield  {author} {\bibinfo {author} {\bibfnamefont {G.}~\bibnamefont
  {Ramon}}, \bibinfo {author} {\bibfnamefont {R.}~\bibnamefont {Rapaport}},
  \bibinfo {author} {\bibfnamefont {A.}~\bibnamefont {Qarry}}, \bibinfo
  {author} {\bibfnamefont {E.}~\bibnamefont {Cohen}}, \bibinfo {author}
  {\bibfnamefont {A.}~\bibnamefont {Mann}}, \bibinfo {author} {\bibfnamefont
  {A.}~\bibnamefont {Ron}}, \ and\ \bibinfo {author} {\bibfnamefont {L.~N.}\
  \bibnamefont {Pfeiffer}},\ }\bibfield  {title} {\bibinfo {title} {\emph
  {Scattering of polaritons by a two-dimensional electron gas in a
  semiconductor microcavity}},\ }\href {\doibase 10.1103/PhysRevB.65.085323}
  {\bibfield  {journal} {\bibinfo  {journal} {Phys. Rev. B}\ }\textbf {\bibinfo
  {volume} {65}},\ \bibinfo {pages} {085323} (\bibinfo {year}
  {2002})}\BibitemShut {NoStop}%
\bibitem [{\citenamefont {Malpuech}\ \emph {et~al.}(2002)\citenamefont
  {Malpuech}, \citenamefont {Kavokin}, \citenamefont {Di~Carlo},\ and\
  \citenamefont {Baumberg}}]{Malpuech2002}%
  \BibitemOpen
  \bibfield  {author} {\bibinfo {author} {\bibfnamefont {G.}~\bibnamefont
  {Malpuech}}, \bibinfo {author} {\bibfnamefont {A.}~\bibnamefont {Kavokin}},
  \bibinfo {author} {\bibfnamefont {A.}~\bibnamefont {Di~Carlo}}, \ and\
  \bibinfo {author} {\bibfnamefont {J.~J.}\ \bibnamefont {Baumberg}},\
  }\bibfield  {title} {\bibinfo {title} {\emph {Polariton lasing by
  exciton-electron scattering in semiconductor microcavities}},\ }\href
  {\doibase 10.1103/PhysRevB.65.153310} {\bibfield  {journal} {\bibinfo
  {journal} {Phys. Rev. B}\ }\textbf {\bibinfo {volume} {65}},\ \bibinfo
  {pages} {153310} (\bibinfo {year} {2002})}\BibitemShut {NoStop}%
\bibitem [{\citenamefont {Hartwell}\ and\ \citenamefont
  {Snoke}(2010)}]{SnokePRB2010}%
  \BibitemOpen
  \bibfield  {author} {\bibinfo {author} {\bibfnamefont {V.~E.}\ \bibnamefont
  {Hartwell}}\ and\ \bibinfo {author} {\bibfnamefont {D.~W.}\ \bibnamefont
  {Snoke}},\ }\bibfield  {title} {\bibinfo {title} {\emph {Numerical
  simulations of the polariton kinetic energy distribution in GaAs quantum-well
  microcavity structures}},\ }\href {\doibase 10.1103/PhysRevB.82.075307}
  {\bibfield  {journal} {\bibinfo  {journal} {Phys. Rev. B}\ }\textbf {\bibinfo
  {volume} {82}},\ \bibinfo {pages} {075307} (\bibinfo {year}
  {2010})}\BibitemShut {NoStop}%
\bibitem [{\citenamefont {Shahnazaryan}\ \emph {et~al.}(2017)\citenamefont
  {Shahnazaryan}, \citenamefont {Iorsh}, \citenamefont {Shelykh},\ and\
  \citenamefont {Kyriienko}}]{Shahnazaryan2017}%
  \BibitemOpen
  \bibfield  {author} {\bibinfo {author} {\bibfnamefont {V.}~\bibnamefont
  {Shahnazaryan}}, \bibinfo {author} {\bibfnamefont {I.}~\bibnamefont {Iorsh}},
  \bibinfo {author} {\bibfnamefont {I.~A.}\ \bibnamefont {Shelykh}}, \ and\
  \bibinfo {author} {\bibfnamefont {O.}~\bibnamefont {Kyriienko}},\ }\bibfield
  {title} {\bibinfo {title} {\emph {Exciton-exciton interaction in
  transition-metal dichalcogenide monolayers}},\ }\href {\doibase
  10.1103/PhysRevB.96.115409} {\bibfield  {journal} {\bibinfo  {journal} {Phys.
  Rev. B}\ }\textbf {\bibinfo {volume} {96}},\ \bibinfo {pages} {115409}
  (\bibinfo {year} {2017})}\BibitemShut {NoStop}%
\bibitem [{\citenamefont {Rochat}\ \emph {et~al.}(2000)\citenamefont {Rochat},
  \citenamefont {Ciuti}, \citenamefont {Savona}, \citenamefont {Piermarocchi},
  \citenamefont {Quattropani},\ and\ \citenamefont
  {Schwendimann}}]{RochatPRB00}%
  \BibitemOpen
  \bibfield  {author} {\bibinfo {author} {\bibfnamefont {G.}~\bibnamefont
  {Rochat}}, \bibinfo {author} {\bibfnamefont {C.}~\bibnamefont {Ciuti}},
  \bibinfo {author} {\bibfnamefont {V.}~\bibnamefont {Savona}}, \bibinfo
  {author} {\bibfnamefont {C.}~\bibnamefont {Piermarocchi}}, \bibinfo {author}
  {\bibfnamefont {A.}~\bibnamefont {Quattropani}}, \ and\ \bibinfo {author}
  {\bibfnamefont {P.}~\bibnamefont {Schwendimann}},\ }\bibfield  {title}
  {\bibinfo {title} {\emph {Excitonic Bloch equations for a two-dimensional
  system of interacting excitons}},\ }\href {\doibase
  10.1103/PhysRevB.61.13856} {\bibfield  {journal} {\bibinfo  {journal} {Phys.
  Rev. B}\ }\textbf {\bibinfo {volume} {61}},\ \bibinfo {pages} {13856}
  (\bibinfo {year} {2000})}\BibitemShut {NoStop}%
\bibitem [{\citenamefont {Combescot}\ \emph {et~al.}(2007)\citenamefont
  {Combescot}, \citenamefont {Dupertuis},\ and\ \citenamefont
  {Betbeder-Matibet}}]{Combescot_2007}%
  \BibitemOpen
  \bibfield  {author} {\bibinfo {author} {\bibfnamefont {M.}~\bibnamefont
  {Combescot}}, \bibinfo {author} {\bibfnamefont {M.~A.}\ \bibnamefont
  {Dupertuis}}, \ and\ \bibinfo {author} {\bibfnamefont {O.}~\bibnamefont
  {Betbeder-Matibet}},\ }\bibfield  {title} {\bibinfo {title} {\emph
  {Polariton-polariton scattering: Exact results through a novel approach}},\
  }\href {\doibase 10.1209/0295-5075/79/17001} {\bibfield  {journal} {\bibinfo
  {journal} {Europhysics Letters ({EPL})}\ }\textbf {\bibinfo {volume} {79}},\
  \bibinfo {pages} {17001} (\bibinfo {year} {2007})}\BibitemShut {NoStop}%
\bibitem [{\citenamefont {Combescot}\ \emph {et~al.}(2008)\citenamefont
  {Combescot}, \citenamefont {Betbeder-Matibet},\ and\ \citenamefont
  {Dubin}}]{COMBESCOTPhysRep2008}%
  \BibitemOpen
  \bibfield  {author} {\bibinfo {author} {\bibfnamefont {M.}~\bibnamefont
  {Combescot}}, \bibinfo {author} {\bibfnamefont {O.}~\bibnamefont
  {Betbeder-Matibet}}, \ and\ \bibinfo {author} {\bibfnamefont
  {F.}~\bibnamefont {Dubin}},\ }\bibfield  {title} {\bibinfo {title} {\emph
  {The many-body physics of composite bosons}},\ }\href {\doibase
  https://doi.org/10.1016/j.physrep.2007.11.003} {\bibfield  {journal}
  {\bibinfo  {journal} {Physics Reports}\ }\textbf {\bibinfo {volume} {463}},\
  \bibinfo {pages} {215 } (\bibinfo {year} {2008})}\BibitemShut {NoStop}%
\bibitem [{\citenamefont {Glazov}\ \emph {et~al.}(2009)\citenamefont {Glazov},
  \citenamefont {Ouerdane}, \citenamefont {Pilozzi}, \citenamefont {Malpuech},
  \citenamefont {Kavokin},\ and\ \citenamefont {D'Andrea}}]{GlazovPRB2009}%
  \BibitemOpen
  \bibfield  {author} {\bibinfo {author} {\bibfnamefont {M.~M.}\ \bibnamefont
  {Glazov}}, \bibinfo {author} {\bibfnamefont {H.}~\bibnamefont {Ouerdane}},
  \bibinfo {author} {\bibfnamefont {L.}~\bibnamefont {Pilozzi}}, \bibinfo
  {author} {\bibfnamefont {G.}~\bibnamefont {Malpuech}}, \bibinfo {author}
  {\bibfnamefont {A.~V.}\ \bibnamefont {Kavokin}}, \ and\ \bibinfo {author}
  {\bibfnamefont {A.}~\bibnamefont {D'Andrea}},\ }\bibfield  {title} {\bibinfo
  {title} {\emph {Polariton-polariton scattering in microcavities: A
  microscopic theory}},\ }\href {\doibase 10.1103/PhysRevB.80.155306}
  {\bibfield  {journal} {\bibinfo  {journal} {Phys. Rev. B}\ }\textbf {\bibinfo
  {volume} {80}},\ \bibinfo {pages} {155306} (\bibinfo {year}
  {2009})}\BibitemShut {NoStop}%
\bibitem [{\citenamefont {Li}\ \emph {et~al.}(2020)\citenamefont {Li},
  \citenamefont {Bleu}, \citenamefont {Parish},\ and\ \citenamefont
  {Levinsen}}]{ePshort}%
  \BibitemOpen
  \bibfield  {author} {\bibinfo {author} {\bibfnamefont {G.}~\bibnamefont
  {Li}}, \bibinfo {author} {\bibfnamefont {O.}~\bibnamefont {Bleu}}, \bibinfo
  {author} {\bibfnamefont {M.~M.}\ \bibnamefont {Parish}}, \ and\ \bibinfo
  {author} {\bibfnamefont {J.}~\bibnamefont {Levinsen}},\ }\href@noop {}
  {\bibinfo {title} {\emph {Enhanced scattering between electrons and
  exciton-polaritons in a microcavity}}} (\bibinfo {year} {2020}),\ \Eprint
  {http://arxiv.org/abs/2008.09281} {arXiv:2008.09281} \BibitemShut {NoStop}%
\bibitem [{\citenamefont {Yamaguchi}\ \emph {et~al.}(2012)\citenamefont
  {Yamaguchi}, \citenamefont {Kamide}, \citenamefont {Ogawa},\ and\
  \citenamefont {Yamamoto}}]{Yamaguchi2012}%
  \BibitemOpen
  \bibfield  {author} {\bibinfo {author} {\bibfnamefont {M.}~\bibnamefont
  {Yamaguchi}}, \bibinfo {author} {\bibfnamefont {K.}~\bibnamefont {Kamide}},
  \bibinfo {author} {\bibfnamefont {T.}~\bibnamefont {Ogawa}}, \ and\ \bibinfo
  {author} {\bibfnamefont {Y.}~\bibnamefont {Yamamoto}},\ }\bibfield  {title}
  {\bibinfo {title} {\emph {{BEC}{\textendash}{BCS}-laser crossover in
  Coulomb-correlated electron{\textendash}hole{\textendash}photon systems}},\
  }\href {\doibase 10.1088/1367-2630/14/6/065001} {\bibfield  {journal}
  {\bibinfo  {journal} {New Journal of Physics}\ }\textbf {\bibinfo {volume}
  {14}},\ \bibinfo {pages} {065001} (\bibinfo {year} {2012})}\BibitemShut
  {NoStop}%
\bibitem [{\citenamefont {Hanai}\ \emph {et~al.}(2017)\citenamefont {Hanai},
  \citenamefont {Littlewood},\ and\ \citenamefont {Ohashi}}]{Hanai2017}%
  \BibitemOpen
  \bibfield  {author} {\bibinfo {author} {\bibfnamefont {R.}~\bibnamefont
  {Hanai}}, \bibinfo {author} {\bibfnamefont {P.~B.}\ \bibnamefont
  {Littlewood}}, \ and\ \bibinfo {author} {\bibfnamefont {Y.}~\bibnamefont
  {Ohashi}},\ }\bibfield  {title} {\bibinfo {title} {\emph {Dynamical
  instability of a driven-dissipative electron-hole condensate in the BCS-BEC
  crossover region}},\ }\href {\doibase 10.1103/PhysRevB.96.125206} {\bibfield
  {journal} {\bibinfo  {journal} {Phys. Rev. B}\ }\textbf {\bibinfo {volume}
  {96}},\ \bibinfo {pages} {125206} (\bibinfo {year} {2017})}\BibitemShut
  {NoStop}%
\bibitem [{\citenamefont {Hu}\ and\ \citenamefont
  {Liu}(2020)}]{Hu2019QuantumFI}%
  \BibitemOpen
  \bibfield  {author} {\bibinfo {author} {\bibfnamefont {H.}~\bibnamefont
  {Hu}}\ and\ \bibinfo {author} {\bibfnamefont {X.-J.}\ \bibnamefont {Liu}},\
  }\bibfield  {title} {\bibinfo {title} {\emph {Quantum fluctuations in a
  strongly interacting Bardeen-Cooper-Schrieffer polariton condensate at
  thermal equilibrium}},\ }\href {\doibase 10.1103/PhysRevA.101.011602}
  {\bibfield  {journal} {\bibinfo  {journal} {Phys. Rev. A}\ }\textbf {\bibinfo
  {volume} {101}},\ \bibinfo {pages} {011602} (\bibinfo {year}
  {2020})}\BibitemShut {NoStop}%
\bibitem [{\citenamefont {Levinsen}\ \emph
  {et~al.}(2019{\natexlab{a}})\citenamefont {Levinsen}, \citenamefont {Li},\
  and\ \citenamefont {Parish}}]{JesperPRR19}%
  \BibitemOpen
  \bibfield  {author} {\bibinfo {author} {\bibfnamefont {J.}~\bibnamefont
  {Levinsen}}, \bibinfo {author} {\bibfnamefont {G.}~\bibnamefont {Li}}, \ and\
  \bibinfo {author} {\bibfnamefont {M.~M.}\ \bibnamefont {Parish}},\ }\bibfield
   {title} {\bibinfo {title} {\emph {Microscopic description of
  exciton-polaritons in microcavities}},\ }\href {\doibase
  10.1103/PhysRevResearch.1.033120} {\bibfield  {journal} {\bibinfo  {journal}
  {Phys. Rev. Research}\ }\textbf {\bibinfo {volume} {1}},\ \bibinfo {pages}
  {033120} (\bibinfo {year} {2019}{\natexlab{a}})}\BibitemShut {NoStop}%
\bibitem [{\citenamefont {Adhikari}(1986)}]{AdhikariAJP86}%
  \BibitemOpen
  \bibfield  {author} {\bibinfo {author} {\bibfnamefont {S.~K.}\ \bibnamefont
  {Adhikari}},\ }\bibfield  {title} {\bibinfo {title} {\emph {Quantum
  scattering in two dimensions}},\ }\href {\doibase 10.1119/1.14623} {\bibfield
   {journal} {\bibinfo  {journal} {American Journal of Physics}\ }\textbf
  {\bibinfo {volume} {54}},\ \bibinfo {pages} {362} (\bibinfo {year}
  {1986})}\BibitemShut {NoStop}%
\bibitem [{\citenamefont {Kasprzak}\ \emph {et~al.}(2006)\citenamefont
  {Kasprzak}, \citenamefont {Richard}, \citenamefont {Kundermann},
  \citenamefont {Baas}, \citenamefont {Jeambrun}, \citenamefont {Keeling},
  \citenamefont {Marchetti}, \citenamefont {Szyma{\'n}ska}, \citenamefont
  {Andr{\'e}}, \citenamefont {Staehli}, \citenamefont {Savona}, \citenamefont
  {Littlewood}, \citenamefont {Deveaud},\ and\ \citenamefont {Dang}}]{BEC06}%
  \BibitemOpen
  \bibfield  {author} {\bibinfo {author} {\bibfnamefont {J.}~\bibnamefont
  {Kasprzak}}, \bibinfo {author} {\bibfnamefont {M.}~\bibnamefont {Richard}},
  \bibinfo {author} {\bibfnamefont {S.}~\bibnamefont {Kundermann}}, \bibinfo
  {author} {\bibfnamefont {A.}~\bibnamefont {Baas}}, \bibinfo {author}
  {\bibfnamefont {P.}~\bibnamefont {Jeambrun}}, \bibinfo {author}
  {\bibfnamefont {J.~M.~J.}\ \bibnamefont {Keeling}}, \bibinfo {author}
  {\bibfnamefont {F.~M.}\ \bibnamefont {Marchetti}}, \bibinfo {author}
  {\bibfnamefont {M.~H.}\ \bibnamefont {Szyma{\'n}ska}}, \bibinfo {author}
  {\bibfnamefont {R.}~\bibnamefont {Andr{\'e}}}, \bibinfo {author}
  {\bibfnamefont {J.~L.}\ \bibnamefont {Staehli}}, \bibinfo {author}
  {\bibfnamefont {V.}~\bibnamefont {Savona}}, \bibinfo {author} {\bibfnamefont
  {P.~B.}\ \bibnamefont {Littlewood}}, \bibinfo {author} {\bibfnamefont
  {B.}~\bibnamefont {Deveaud}}, \ and\ \bibinfo {author} {\bibfnamefont
  {L.~S.}\ \bibnamefont {Dang}},\ }\bibfield  {title} {\bibinfo {title} {\emph
  {Bose--Einstein condensation of exciton polaritons}},\ }\href {\doibase
  10.1038/nature05131} {\bibfield  {journal} {\bibinfo  {journal} {Nature}\
  }\textbf {\bibinfo {volume} {443}},\ \bibinfo {pages} {409} (\bibinfo {year}
  {2006})}\BibitemShut {NoStop}%
\bibitem [{\citenamefont {Landau}\ and\ \citenamefont
  {Lifshitz}(2013)}]{landau2013quantum}%
  \BibitemOpen
  \bibfield  {author} {\bibinfo {author} {\bibfnamefont {L.~D.}\ \bibnamefont
  {Landau}}\ and\ \bibinfo {author} {\bibfnamefont {E.~M.}\ \bibnamefont
  {Lifshitz}},\ }\href@noop {} {\emph {\bibinfo {title} {Quantum mechanics:
  non-relativistic theory}}},\ Vol.~\bibinfo {volume} {3}\ (\bibinfo
  {publisher} {Elsevier},\ \bibinfo {year} {2013})\BibitemShut {NoStop}%
\bibitem [{\citenamefont {Levinsen}\ and\ \citenamefont
  {Parish}(2015)}]{LevinsenBook15}%
  \BibitemOpen
  \bibfield  {author} {\bibinfo {author} {\bibfnamefont {J.}~\bibnamefont
  {Levinsen}}\ and\ \bibinfo {author} {\bibfnamefont {M.~M.}\ \bibnamefont
  {Parish}},\ }\bibfield  {title} {\bibinfo {title} {\emph {{Strongly
  interacting two-dimensional Fermi gases}}},\ }\href {\doibase
  10.1142/9789814667746_0001} {\bibfield  {journal} {\bibinfo  {journal} {Annu.
  Rev. Cold Atoms Mol.}\ }\textbf {\bibinfo {volume} {3}},\ \bibinfo {pages}
  {1} (\bibinfo {year} {2015})}\BibitemShut {NoStop}%
\bibitem [{Note1()}]{Note1}%
  \BibitemOpen
  \bibinfo {note} {If $\Omega \gtrsim \varepsilon _B$ and/or $\delta \gtrsim
  \varepsilon _B$, the upper polariton enters the continuum of unbound
  electron-hole states. In that case, one must analytically continue the energy
  slightly into the complex plane, $E\to E+i0$}\BibitemShut {NoStop}%
\bibitem [{Note2()}]{Note2}%
  \BibitemOpen
  \bibinfo {note} {The variational approach developed by Khurgin~\cite
  {Khurgin2001} approximated the functional form of the electron-hole wave
  function as unchanged in the presence of light-matter coupling. While this is
  an approximation in the Coulomb case studied in that work~\cite
  {JesperPRR19}, we see that it is exact for strongly screened electron-hole
  interactions.}\BibitemShut {Stop}%
\bibitem [{\citenamefont {Hopfield}(1958)}]{HopfieldPR1958}%
  \BibitemOpen
  \bibfield  {author} {\bibinfo {author} {\bibfnamefont {J.~J.}\ \bibnamefont
  {Hopfield}},\ }\bibfield  {title} {\bibinfo {title} {\emph {Theory of the
  Contribution of Excitons to the Complex Dielectric Constant of Crystals}},\
  }\href {\doibase 10.1103/PhysRev.112.1555} {\bibfield  {journal} {\bibinfo
  {journal} {Phys. Rev.}\ }\textbf {\bibinfo {volume} {112}},\ \bibinfo {pages}
  {1555} (\bibinfo {year} {1958})}\BibitemShut {NoStop}%
\bibitem [{\citenamefont {Werner}\ \emph {et~al.}(2009)\citenamefont {Werner},
  \citenamefont {Tarruell},\ and\ \citenamefont {Castin}}]{Werner2009}%
  \BibitemOpen
  \bibfield  {author} {\bibinfo {author} {\bibfnamefont {F.}~\bibnamefont
  {Werner}}, \bibinfo {author} {\bibfnamefont {L.}~\bibnamefont {Tarruell}}, \
  and\ \bibinfo {author} {\bibfnamefont {Y.}~\bibnamefont {Castin}},\
  }\bibfield  {title} {\bibinfo {title} {\emph {Number of closed-channel
  molecules in the BEC-BCS crossover}},\ }\href {\doibase
  10.1140/epjb/e2009-00040-8} {\bibfield  {journal} {\bibinfo  {journal} {The
  European Physical Journal B}\ }\textbf {\bibinfo {volume} {68}},\ \bibinfo
  {pages} {401} (\bibinfo {year} {2009})}\BibitemShut {NoStop}%
\bibitem [{\citenamefont {Fetter}\ and\ \citenamefont
  {Walecka}(2003)}]{FetterBook}%
  \BibitemOpen
  \bibfield  {author} {\bibinfo {author} {\bibfnamefont {A.~L.}\ \bibnamefont
  {Fetter}}\ and\ \bibinfo {author} {\bibfnamefont {J.~D.}\ \bibnamefont
  {Walecka}},\ }\href@noop {} {\emph {\bibinfo {title} {Quantum Theory of
  Many-Particle Systems}}}\ (\bibinfo  {publisher} {Dover},\ \bibinfo {address}
  {New York, USA},\ \bibinfo {year} {2003})\BibitemShut {NoStop}%
\bibitem [{\citenamefont {Wouters}(2007)}]{WoutersPRB07}%
  \BibitemOpen
  \bibfield  {author} {\bibinfo {author} {\bibfnamefont {M.}~\bibnamefont
  {Wouters}},\ }\bibfield  {title} {\bibinfo {title} {\emph {Resonant
  polariton-polariton scattering in semiconductor microcavities}},\ }\href
  {\doibase 10.1103/PhysRevB.76.045319} {\bibfield  {journal} {\bibinfo
  {journal} {Phys. Rev. B}\ }\textbf {\bibinfo {volume} {76}},\ \bibinfo
  {pages} {045319} (\bibinfo {year} {2007})}\BibitemShut {NoStop}%
\bibitem [{\citenamefont {Bleu}\ \emph {et~al.}(2020)\citenamefont {Bleu},
  \citenamefont {Li}, \citenamefont {Levinsen},\ and\ \citenamefont
  {Parish}}]{Bleu2020}%
  \BibitemOpen
  \bibfield  {author} {\bibinfo {author} {\bibfnamefont {O.}~\bibnamefont
  {Bleu}}, \bibinfo {author} {\bibfnamefont {G.}~\bibnamefont {Li}}, \bibinfo
  {author} {\bibfnamefont {J.}~\bibnamefont {Levinsen}}, \ and\ \bibinfo
  {author} {\bibfnamefont {M.~M.}\ \bibnamefont {Parish}},\ }\bibfield  {title}
  {\bibinfo {title} {\emph {Polariton interactions in microcavities with
  atomically thin semiconductor layers}},\ }\href {\doibase
  10.1103/PhysRevResearch.2.043185} {\bibfield  {journal} {\bibinfo  {journal}
  {Phys. Rev. Research}\ }\textbf {\bibinfo {volume} {2}},\ \bibinfo {pages}
  {043185} (\bibinfo {year} {2020})}\BibitemShut {NoStop}%
\bibitem [{\citenamefont {Sze}(2007)}]{GaAsEffectiveMass}%
  \BibitemOpen
  \bibfield  {author} {\bibinfo {author} {\bibfnamefont {S.~M.}\ \bibnamefont
  {Sze}},\ }\href@noop {} {\emph {\bibinfo {title} {Physics of semiconductor
  devices}}},\ \bibinfo {edition} {3rd}\ ed.\ (\bibinfo  {publisher} {Hoboken,
  NJ : Wiley-Interscience},\ \bibinfo {address} {Hoboken, NJ Hoboken, N.J.},\
  \bibinfo {year} {2007})\BibitemShut {NoStop}%
\bibitem [{\citenamefont {Ramon}\ \emph {et~al.}(2003)\citenamefont {Ramon},
  \citenamefont {Mann},\ and\ \citenamefont {Cohen}}]{Ramon2003}%
  \BibitemOpen
  \bibfield  {author} {\bibinfo {author} {\bibfnamefont {G.}~\bibnamefont
  {Ramon}}, \bibinfo {author} {\bibfnamefont {A.}~\bibnamefont {Mann}}, \ and\
  \bibinfo {author} {\bibfnamefont {E.}~\bibnamefont {Cohen}},\ }\bibfield
  {title} {\bibinfo {title} {\emph {Theory of neutral and charged exciton
  scattering with electrons in semiconductor quantum wells}},\ }\href {\doibase
  10.1103/PhysRevB.67.045323} {\bibfield  {journal} {\bibinfo  {journal} {Phys.
  Rev. B}\ }\textbf {\bibinfo {volume} {67}},\ \bibinfo {pages} {045323}
  (\bibinfo {year} {2003})}\BibitemShut {NoStop}%
\bibitem [{\citenamefont {Ngampruetikorn}\ \emph
  {et~al.}(2013{\natexlab{a}})\citenamefont {Ngampruetikorn}, \citenamefont
  {Parish},\ and\ \citenamefont {Levinsen}}]{VudtiwatEPL13}%
  \BibitemOpen
  \bibfield  {author} {\bibinfo {author} {\bibfnamefont {V.}~\bibnamefont
  {Ngampruetikorn}}, \bibinfo {author} {\bibfnamefont {M.~M.}\ \bibnamefont
  {Parish}}, \ and\ \bibinfo {author} {\bibfnamefont {J.}~\bibnamefont
  {Levinsen}},\ }\bibfield  {title} {\bibinfo {title} {\emph {Three-body
  problem in a two-dimensional Fermi gas}},\ }\href {\doibase
  10.1209/0295-5075/102/13001} {\bibfield  {journal} {\bibinfo  {journal}
  {{EPL} (Europhysics Letters)}\ }\textbf {\bibinfo {volume} {102}},\ \bibinfo
  {pages} {13001} (\bibinfo {year} {2013}{\natexlab{a}})}\BibitemShut {NoStop}%
\bibitem [{\citenamefont {Skorniakov}\ and\ \citenamefont
  {Ter-Martirosian}(1957)}]{Skorniakov1957}%
  \BibitemOpen
  \bibfield  {author} {\bibinfo {author} {\bibfnamefont {G.~V.}\ \bibnamefont
  {Skorniakov}}\ and\ \bibinfo {author} {\bibfnamefont {K.~A.}\ \bibnamefont
  {Ter-Martirosian}},\ }\bibfield  {title} {\bibinfo {title} {\emph {Three Body
  Problem for Short Range Forces. I. Scattering of Low Energy Neutrons by
  Deuterons}},\ }\href
  {http://www.jetp.ac.ru/cgi-bin/e/index/e/4/5/p648?a=list} {\bibfield
  {journal} {\bibinfo  {journal} {Sov. Phys. JETP}\ }\textbf {\bibinfo {volume}
  {4}},\ \bibinfo {pages} {648} (\bibinfo {year} {1957})}\BibitemShut {NoStop}%
\bibitem [{\citenamefont {Petrov}(2003)}]{Petrov2002}%
  \BibitemOpen
  \bibfield  {author} {\bibinfo {author} {\bibfnamefont {D.~S.}\ \bibnamefont
  {Petrov}},\ }\bibfield  {title} {\bibinfo {title} {\emph {Three-body problem
  in Fermi gases with short-range interparticle interaction}},\ }\href
  {\doibase 10.1103/PhysRevA.67.010703} {\bibfield  {journal} {\bibinfo
  {journal} {Phys. Rev. A}\ }\textbf {\bibinfo {volume} {67}},\ \bibinfo
  {pages} {010703} (\bibinfo {year} {2003})}\BibitemShut {NoStop}%
\bibitem [{\citenamefont {Brodsky}\ \emph {et~al.}(2006)\citenamefont
  {Brodsky}, \citenamefont {Kagan}, \citenamefont {Klaptsov}, \citenamefont
  {Combescot},\ and\ \citenamefont {Leyronas}}]{BrodskyPRA06}%
  \BibitemOpen
  \bibfield  {author} {\bibinfo {author} {\bibfnamefont {I.~V.}\ \bibnamefont
  {Brodsky}}, \bibinfo {author} {\bibfnamefont {M.~Y.}\ \bibnamefont {Kagan}},
  \bibinfo {author} {\bibfnamefont {A.~V.}\ \bibnamefont {Klaptsov}}, \bibinfo
  {author} {\bibfnamefont {R.}~\bibnamefont {Combescot}}, \ and\ \bibinfo
  {author} {\bibfnamefont {X.}~\bibnamefont {Leyronas}},\ }\bibfield  {title}
  {\bibinfo {title} {\emph {Exact diagrammatic approach for dimer-dimer
  scattering and bound states of three and four resonantly interacting
  particles}},\ }\href {\doibase 10.1103/PhysRevA.73.032724} {\bibfield
  {journal} {\bibinfo  {journal} {Phys. Rev. A}\ }\textbf {\bibinfo {volume}
  {73}},\ \bibinfo {pages} {032724} (\bibinfo {year} {2006})}\BibitemShut
  {NoStop}%
\bibitem [{\citenamefont {Levinsen}\ and\ \citenamefont
  {Gurarie}(2006)}]{LevinsenPRA06}%
  \BibitemOpen
  \bibfield  {author} {\bibinfo {author} {\bibfnamefont {J.}~\bibnamefont
  {Levinsen}}\ and\ \bibinfo {author} {\bibfnamefont {V.}~\bibnamefont
  {Gurarie}},\ }\bibfield  {title} {\bibinfo {title} {\emph {Properties of
  strongly paired fermionic condensates}},\ }\href {\doibase
  10.1103/PhysRevA.73.053607} {\bibfield  {journal} {\bibinfo  {journal} {Phys.
  Rev. A}\ }\textbf {\bibinfo {volume} {73}},\ \bibinfo {pages} {053607}
  (\bibinfo {year} {2006})}\BibitemShut {NoStop}%
\bibitem [{Pet()}]{PetrovLesHouches2010}%
  \BibitemOpen
  \href@noop {} {}\bibinfo {note} {D.~S.~Petrov, in {\it Proceedings of the Les
  Houches Summer Schools, Session 94}, edited by C. Salomon, G. V. Shlyapnikov,
  and L. F. Cugliandolo (Oxford University Press, Oxford, England, 2013)},\
  \Eprint {http://arxiv.org/abs/1206.5752} {arXiv:1206.5752} \BibitemShut
  {NoStop}%
\bibitem [{\citenamefont {Bedaque}\ \emph {et~al.}(1998)\citenamefont
  {Bedaque}, \citenamefont {Hammer},\ and\ \citenamefont {van
  Kolck}}]{BedaquePRC1998}%
  \BibitemOpen
  \bibfield  {author} {\bibinfo {author} {\bibfnamefont {P.~F.}\ \bibnamefont
  {Bedaque}}, \bibinfo {author} {\bibfnamefont {H.-W.}\ \bibnamefont {Hammer}},
  \ and\ \bibinfo {author} {\bibfnamefont {U.}~\bibnamefont {van Kolck}},\
  }\bibfield  {title} {\bibinfo {title} {\emph {Effective theory for
  neutron-deuteron scattering: Energy dependence}},\ }\href {\doibase
  10.1103/PhysRevC.58.R641} {\bibfield  {journal} {\bibinfo  {journal} {Phys.
  Rev. C}\ }\textbf {\bibinfo {volume} {58}},\ \bibinfo {pages} {R641}
  (\bibinfo {year} {1998})}\BibitemShut {NoStop}%
\bibitem [{\citenamefont {Levinsen}\ \emph {et~al.}(2009)\citenamefont
  {Levinsen}, \citenamefont {Tiecke}, \citenamefont {Walraven},\ and\
  \citenamefont {Petrov}}]{Levinsen2009}%
  \BibitemOpen
  \bibfield  {author} {\bibinfo {author} {\bibfnamefont {J.}~\bibnamefont
  {Levinsen}}, \bibinfo {author} {\bibfnamefont {T.~G.}\ \bibnamefont
  {Tiecke}}, \bibinfo {author} {\bibfnamefont {J.~T.~M.}\ \bibnamefont
  {Walraven}}, \ and\ \bibinfo {author} {\bibfnamefont {D.~S.}\ \bibnamefont
  {Petrov}},\ }\bibfield  {title} {\bibinfo {title} {\emph {Atom-Dimer
  Scattering and Long-Lived Trimers in Fermionic Mixtures}},\ }\href {\doibase
  10.1103/PhysRevLett.103.153202} {\bibfield  {journal} {\bibinfo  {journal}
  {Phys. Rev. Lett.}\ }\textbf {\bibinfo {volume} {103}},\ \bibinfo {pages}
  {153202} (\bibinfo {year} {2009})}\BibitemShut {NoStop}%
\bibitem [{\citenamefont {Levinsen}\ and\ \citenamefont
  {Petrov}(2011)}]{Levinsen2011}%
  \BibitemOpen
  \bibfield  {author} {\bibinfo {author} {\bibfnamefont {J.}~\bibnamefont
  {Levinsen}}\ and\ \bibinfo {author} {\bibfnamefont {D.~S.}\ \bibnamefont
  {Petrov}},\ }\bibfield  {title} {\bibinfo {title} {\emph {Atom-dimer and
  dimer-dimer scattering in fermionic mixtures near a narrow Feshbach
  resonance}},\ }\href {\doibase 10.1140/epjd/e2011-20071-x} {\bibfield
  {journal} {\bibinfo  {journal} {The European Physical Journal D}\ }\textbf
  {\bibinfo {volume} {65}},\ \bibinfo {pages} {67} (\bibinfo {year}
  {2011})}\BibitemShut {NoStop}%
\bibitem [{\citenamefont {Ngampruetikorn}\ \emph
  {et~al.}(2013{\natexlab{b}})\citenamefont {Ngampruetikorn}, \citenamefont
  {Parish},\ and\ \citenamefont {Levinsen}}]{Ngampruetikorn2013}%
  \BibitemOpen
  \bibfield  {author} {\bibinfo {author} {\bibfnamefont {V.}~\bibnamefont
  {Ngampruetikorn}}, \bibinfo {author} {\bibfnamefont {M.~M.}\ \bibnamefont
  {Parish}}, \ and\ \bibinfo {author} {\bibfnamefont {J.}~\bibnamefont
  {Levinsen}},\ }\bibfield  {title} {\bibinfo {title} {\emph {Three-body
  problem in a two-dimensional Fermi gas}},\ }\href {\doibase
  10.1209/0295-5075/102/13001} {\bibfield  {journal} {\bibinfo  {journal}
  {{EPL} (Europhysics Letters)}\ }\textbf {\bibinfo {volume} {102}},\ \bibinfo
  {pages} {13001} (\bibinfo {year} {2013}{\natexlab{b}})}\BibitemShut {NoStop}%
\bibitem [{\citenamefont {Jag}\ \emph {et~al.}(2014)\citenamefont {Jag},
  \citenamefont {Zaccanti}, \citenamefont {Cetina}, \citenamefont {Lous},
  \citenamefont {Schreck}, \citenamefont {Grimm}, \citenamefont {Petrov},\ and\
  \citenamefont {Levinsen}}]{Jag2014}%
  \BibitemOpen
  \bibfield  {author} {\bibinfo {author} {\bibfnamefont {M.}~\bibnamefont
  {Jag}}, \bibinfo {author} {\bibfnamefont {M.}~\bibnamefont {Zaccanti}},
  \bibinfo {author} {\bibfnamefont {M.}~\bibnamefont {Cetina}}, \bibinfo
  {author} {\bibfnamefont {R.~S.}\ \bibnamefont {Lous}}, \bibinfo {author}
  {\bibfnamefont {F.}~\bibnamefont {Schreck}}, \bibinfo {author} {\bibfnamefont
  {R.}~\bibnamefont {Grimm}}, \bibinfo {author} {\bibfnamefont {D.~S.}\
  \bibnamefont {Petrov}}, \ and\ \bibinfo {author} {\bibfnamefont
  {J.}~\bibnamefont {Levinsen}},\ }\bibfield  {title} {\bibinfo {title} {\emph
  {Observation of a Strong Atom-Dimer Attraction in a Mass-Imbalanced
  Fermi-Fermi Mixture}},\ }\href {\doibase 10.1103/PhysRevLett.112.075302}
  {\bibfield  {journal} {\bibinfo  {journal} {Phys. Rev. Lett.}\ }\textbf
  {\bibinfo {volume} {112}},\ \bibinfo {pages} {075302} (\bibinfo {year}
  {2014})}\BibitemShut {NoStop}%
\bibitem [{Note3()}]{Note3}%
  \BibitemOpen
  \bibinfo {note} {Note that the electron propagator $G_e(-\protect \mathbf
  {q},-q_0)=1/(-q_0-\epsilon _\protect \mathbf {q}+i0)$ in the last term in
  Fig.~\ref {fig:e_P_scattering_t_matrix} has a simple pole in the upper half
  of the complex $q_0$ plane, while all other propagators have their
  non-analytic structure only in the lower half plane. This allows us to
  perform the integral over $q_0$ by closing the contour in the upper half
  plane, effectively removing that propagator and setting the electron
  dispersion to its on-shell value $\epsilon ^e_{\protect \mathbf {q}}=\protect
  \mathbf {q}^2/(2 m_e)$ in the other propagators. For more details, see, e.g.,
  Ref.~\cite {BrodskyPRA06}.}\BibitemShut {Stop}%
\bibitem [{\citenamefont {Brodbeck}\ \emph {et~al.}(2017)\citenamefont
  {Brodbeck}, \citenamefont {De~Liberato}, \citenamefont {Amthor},
  \citenamefont {Klaas}, \citenamefont {Kamp}, \citenamefont {Worschech},
  \citenamefont {Schneider},\ and\ \citenamefont {H\"ofling}}]{GaAsLargeRabi}%
  \BibitemOpen
  \bibfield  {author} {\bibinfo {author} {\bibfnamefont {S.}~\bibnamefont
  {Brodbeck}}, \bibinfo {author} {\bibfnamefont {S.}~\bibnamefont
  {De~Liberato}}, \bibinfo {author} {\bibfnamefont {M.}~\bibnamefont {Amthor}},
  \bibinfo {author} {\bibfnamefont {M.}~\bibnamefont {Klaas}}, \bibinfo
  {author} {\bibfnamefont {M.}~\bibnamefont {Kamp}}, \bibinfo {author}
  {\bibfnamefont {L.}~\bibnamefont {Worschech}}, \bibinfo {author}
  {\bibfnamefont {C.}~\bibnamefont {Schneider}}, \ and\ \bibinfo {author}
  {\bibfnamefont {S.}~\bibnamefont {H\"ofling}},\ }\bibfield  {title} {\bibinfo
  {title} {\emph {Experimental Verification of the Very Strong Coupling Regime
  in a GaAs Quantum Well Microcavity}},\ }\href {\doibase
  10.1103/PhysRevLett.119.027401} {\bibfield  {journal} {\bibinfo  {journal}
  {Phys. Rev. Lett.}\ }\textbf {\bibinfo {volume} {119}},\ \bibinfo {pages}
  {027401} (\bibinfo {year} {2017})}\BibitemShut {NoStop}%
\bibitem [{ato()}]{atomdimer}%
  \BibitemOpen
  \href@noop {} {}\bibinfo {note} {Reference~\cite{Ngampruetikorn2013}
  determined the scattering phase shift of atoms and diatomic molecules in a
  two-component Fermi gas with equal mass fermions, and in particular
  identified the low-energy energy scale
  $\varepsilon_{eX}\simeq1/[2m_{eX}(1.27a_X)^2]$. In the absence of coupling to
  light, this scenario is equivalent to exciton-electron scattering given our
  approximation of contact interactions. Likewise, Ref.~\cite{Pricoupenko2010}
  determined the bound three-body states in the same problem.}\BibitemShut
  {Stop}%
\bibitem [{\citenamefont {Dufferwiel}\ \emph {et~al.}(2015)\citenamefont
  {Dufferwiel}, \citenamefont {Schwarz}, \citenamefont {Withers}, \citenamefont
  {Trichet}, \citenamefont {Li}, \citenamefont {Sich}, \citenamefont {Del
  Pozo-Zamudio}, \citenamefont {Clark}, \citenamefont {Nalitov}, \citenamefont
  {Solnyshkov}, \citenamefont {Malpuech}, \citenamefont {Novoselov},
  \citenamefont {Smith}, \citenamefont {Skolnick}, \citenamefont
  {Krizhanovskii},\ and\ \citenamefont {Tartakovskii}}]{dufferwiel2015exciton}%
  \BibitemOpen
  \bibfield  {author} {\bibinfo {author} {\bibfnamefont {S.}~\bibnamefont
  {Dufferwiel}}, \bibinfo {author} {\bibfnamefont {S.}~\bibnamefont {Schwarz}},
  \bibinfo {author} {\bibfnamefont {F.}~\bibnamefont {Withers}}, \bibinfo
  {author} {\bibfnamefont {A.~A.~P.}\ \bibnamefont {Trichet}}, \bibinfo
  {author} {\bibfnamefont {F.}~\bibnamefont {Li}}, \bibinfo {author}
  {\bibfnamefont {M.}~\bibnamefont {Sich}}, \bibinfo {author} {\bibfnamefont
  {O.}~\bibnamefont {Del Pozo-Zamudio}}, \bibinfo {author} {\bibfnamefont
  {C.}~\bibnamefont {Clark}}, \bibinfo {author} {\bibfnamefont
  {A.}~\bibnamefont {Nalitov}}, \bibinfo {author} {\bibfnamefont {D.~D.}\
  \bibnamefont {Solnyshkov}}, \bibinfo {author} {\bibfnamefont
  {G.}~\bibnamefont {Malpuech}}, \bibinfo {author} {\bibfnamefont {K.~S.}\
  \bibnamefont {Novoselov}}, \bibinfo {author} {\bibfnamefont {J.~M.}\
  \bibnamefont {Smith}}, \bibinfo {author} {\bibfnamefont {M.~S.}\ \bibnamefont
  {Skolnick}}, \bibinfo {author} {\bibfnamefont {D.~N.}\ \bibnamefont
  {Krizhanovskii}}, \ and\ \bibinfo {author} {\bibfnamefont {A.~I.}\
  \bibnamefont {Tartakovskii}},\ }\bibfield  {title} {\bibinfo {title} {\emph
  {Exciton-polaritons in van der Waals heterostructures embedded in tunable
  microcavities}},\ }\href {https://doi.org/10.1038/ncomms9579} {\bibfield
  {journal} {\bibinfo  {journal} {Nature Communications}\ }\textbf {\bibinfo
  {volume} {6}},\ \bibinfo {pages} {8579} (\bibinfo {year} {2015})}\BibitemShut
  {NoStop}%
\bibitem [{\citenamefont {Liu}\ \emph {et~al.}(2014)\citenamefont {Liu},
  \citenamefont {Galfsky}, \citenamefont {Sun}, \citenamefont {Xia},
  \citenamefont {Lin}, \citenamefont {Lee}, \citenamefont {K{\'e}na-Cohen},\
  and\ \citenamefont {Menon}}]{liu2015strong}%
  \BibitemOpen
  \bibfield  {author} {\bibinfo {author} {\bibfnamefont {X.}~\bibnamefont
  {Liu}}, \bibinfo {author} {\bibfnamefont {T.}~\bibnamefont {Galfsky}},
  \bibinfo {author} {\bibfnamefont {Z.}~\bibnamefont {Sun}}, \bibinfo {author}
  {\bibfnamefont {F.}~\bibnamefont {Xia}}, \bibinfo {author} {\bibfnamefont
  {E.-c.}\ \bibnamefont {Lin}}, \bibinfo {author} {\bibfnamefont {Y.-H.}\
  \bibnamefont {Lee}}, \bibinfo {author} {\bibfnamefont {S.}~\bibnamefont
  {K{\'e}na-Cohen}}, \ and\ \bibinfo {author} {\bibfnamefont {V.~M.}\
  \bibnamefont {Menon}},\ }\bibfield  {title} {\bibinfo {title} {\emph {Strong
  light--matter coupling in two-dimensional atomic crystals}},\ }\href
  {\doibase 10.1038/nphoton.2014.304} {\bibfield  {journal} {\bibinfo
  {journal} {Nature Photonics}\ }\textbf {\bibinfo {volume} {9}},\ \bibinfo
  {pages} {30} (\bibinfo {year} {2014})}\BibitemShut {NoStop}%
\bibitem [{\citenamefont {Lundt}\ \emph {et~al.}(2016)\citenamefont {Lundt},
  \citenamefont {Klembt}, \citenamefont {Cherotchenko}, \citenamefont
  {Betzold}, \citenamefont {Iff}, \citenamefont {Nalitov}, \citenamefont
  {Klaas}, \citenamefont {Dietrich}, \citenamefont {Kavokin}, \citenamefont
  {H{\"o}fling},\ and\ \citenamefont {Schneider}}]{lundt2016room}%
  \BibitemOpen
  \bibfield  {author} {\bibinfo {author} {\bibfnamefont {N.}~\bibnamefont
  {Lundt}}, \bibinfo {author} {\bibfnamefont {S.}~\bibnamefont {Klembt}},
  \bibinfo {author} {\bibfnamefont {E.}~\bibnamefont {Cherotchenko}}, \bibinfo
  {author} {\bibfnamefont {S.}~\bibnamefont {Betzold}}, \bibinfo {author}
  {\bibfnamefont {O.}~\bibnamefont {Iff}}, \bibinfo {author} {\bibfnamefont
  {A.~V.}\ \bibnamefont {Nalitov}}, \bibinfo {author} {\bibfnamefont
  {M.}~\bibnamefont {Klaas}}, \bibinfo {author} {\bibfnamefont {C.~P.}\
  \bibnamefont {Dietrich}}, \bibinfo {author} {\bibfnamefont {A.~V.}\
  \bibnamefont {Kavokin}}, \bibinfo {author} {\bibfnamefont {S.}~\bibnamefont
  {H{\"o}fling}}, \ and\ \bibinfo {author} {\bibfnamefont {C.}~\bibnamefont
  {Schneider}},\ }\bibfield  {title} {\bibinfo {title} {\emph {Room-temperature
  Tamm-plasmon exciton-polaritons with a WSe2 monolayer}},\ }\href
  {https://doi.org/10.1038/ncomms13328} {\bibfield  {journal} {\bibinfo
  {journal} {Nature Communications}\ }\textbf {\bibinfo {volume} {7}},\
  \bibinfo {pages} {13328} (\bibinfo {year} {2016})}\BibitemShut {NoStop}%
\bibitem [{\citenamefont {He}\ \emph {et~al.}(2014)\citenamefont {He},
  \citenamefont {Kumar}, \citenamefont {Zhao}, \citenamefont {Wang},
  \citenamefont {Mak}, \citenamefont {Zhao},\ and\ \citenamefont
  {Shan}}]{He2014}%
  \BibitemOpen
  \bibfield  {author} {\bibinfo {author} {\bibfnamefont {K.}~\bibnamefont
  {He}}, \bibinfo {author} {\bibfnamefont {N.}~\bibnamefont {Kumar}}, \bibinfo
  {author} {\bibfnamefont {L.}~\bibnamefont {Zhao}}, \bibinfo {author}
  {\bibfnamefont {Z.}~\bibnamefont {Wang}}, \bibinfo {author} {\bibfnamefont
  {K.~F.}\ \bibnamefont {Mak}}, \bibinfo {author} {\bibfnamefont
  {H.}~\bibnamefont {Zhao}}, \ and\ \bibinfo {author} {\bibfnamefont
  {J.}~\bibnamefont {Shan}},\ }\bibfield  {title} {\bibinfo {title} {\emph
  {Tightly Bound Excitons in Monolayer ${\mathrm{WSe}}_{2}$}},\ }\href
  {\doibase 10.1103/PhysRevLett.113.026803} {\bibfield  {journal} {\bibinfo
  {journal} {Phys. Rev. Lett.}\ }\textbf {\bibinfo {volume} {113}},\ \bibinfo
  {pages} {026803} (\bibinfo {year} {2014})}\BibitemShut {NoStop}%
\bibitem [{\citenamefont {Flatten}\ \emph {et~al.}(2016)\citenamefont
  {Flatten}, \citenamefont {He}, \citenamefont {Coles}, \citenamefont
  {Trichet}, \citenamefont {Powell}, \citenamefont {Taylor}, \citenamefont
  {Warner},\ and\ \citenamefont {Smith}}]{Flatten2016}%
  \BibitemOpen
  \bibfield  {author} {\bibinfo {author} {\bibfnamefont {L.~C.}\ \bibnamefont
  {Flatten}}, \bibinfo {author} {\bibfnamefont {Z.}~\bibnamefont {He}},
  \bibinfo {author} {\bibfnamefont {D.~M.}\ \bibnamefont {Coles}}, \bibinfo
  {author} {\bibfnamefont {A.~A.~P.}\ \bibnamefont {Trichet}}, \bibinfo
  {author} {\bibfnamefont {A.~W.}\ \bibnamefont {Powell}}, \bibinfo {author}
  {\bibfnamefont {R.~A.}\ \bibnamefont {Taylor}}, \bibinfo {author}
  {\bibfnamefont {J.~H.}\ \bibnamefont {Warner}}, \ and\ \bibinfo {author}
  {\bibfnamefont {J.~M.}\ \bibnamefont {Smith}},\ }\bibfield  {title} {\bibinfo
  {title} {\emph {Room-temperature exciton-polaritons with two-dimensional
  WS$_2$}},\ }\href {https://doi.org/10.1038/srep33134} {\bibfield  {journal}
  {\bibinfo  {journal} {Scientific Reports}\ }\textbf {\bibinfo {volume} {6}},\
  \bibinfo {pages} {33134} (\bibinfo {year} {2016})}\BibitemShut {NoStop}%
\bibitem [{\citenamefont {Estrecho}\ \emph {et~al.}(2019)\citenamefont
  {Estrecho}, \citenamefont {Gao}, \citenamefont {Bobrovska}, \citenamefont
  {Comber-Todd}, \citenamefont {Fraser}, \citenamefont {Steger}, \citenamefont
  {West}, \citenamefont {Pfeiffer}, \citenamefont {Levinsen}, \citenamefont
  {Parish}, \citenamefont {Liew}, \citenamefont {Matuszewski}, \citenamefont
  {Snoke}, \citenamefont {Truscott},\ and\ \citenamefont
  {Ostrovskaya}}]{EliPRB2019}%
  \BibitemOpen
  \bibfield  {author} {\bibinfo {author} {\bibfnamefont {E.}~\bibnamefont
  {Estrecho}}, \bibinfo {author} {\bibfnamefont {T.}~\bibnamefont {Gao}},
  \bibinfo {author} {\bibfnamefont {N.}~\bibnamefont {Bobrovska}}, \bibinfo
  {author} {\bibfnamefont {D.}~\bibnamefont {Comber-Todd}}, \bibinfo {author}
  {\bibfnamefont {M.~D.}\ \bibnamefont {Fraser}}, \bibinfo {author}
  {\bibfnamefont {M.}~\bibnamefont {Steger}}, \bibinfo {author} {\bibfnamefont
  {K.}~\bibnamefont {West}}, \bibinfo {author} {\bibfnamefont {L.~N.}\
  \bibnamefont {Pfeiffer}}, \bibinfo {author} {\bibfnamefont {J.}~\bibnamefont
  {Levinsen}}, \bibinfo {author} {\bibfnamefont {M.~M.}\ \bibnamefont
  {Parish}}, \bibinfo {author} {\bibfnamefont {T.~C.~H.}\ \bibnamefont {Liew}},
  \bibinfo {author} {\bibfnamefont {M.}~\bibnamefont {Matuszewski}}, \bibinfo
  {author} {\bibfnamefont {D.~W.}\ \bibnamefont {Snoke}}, \bibinfo {author}
  {\bibfnamefont {A.~G.}\ \bibnamefont {Truscott}}, \ and\ \bibinfo {author}
  {\bibfnamefont {E.~A.}\ \bibnamefont {Ostrovskaya}},\ }\bibfield  {title}
  {\bibinfo {title} {\emph {Direct measurement of polariton-polariton
  interaction strength in the Thomas-Fermi regime of exciton-polariton
  condensation}},\ }\href {\doibase 10.1103/PhysRevB.100.035306} {\bibfield
  {journal} {\bibinfo  {journal} {Phys. Rev. B}\ }\textbf {\bibinfo {volume}
  {100}},\ \bibinfo {pages} {035306} (\bibinfo {year} {2019})}\BibitemShut
  {NoStop}%
\bibitem [{Note4()}]{Note4}%
  \BibitemOpen
  \bibinfo {note} {{\protect \color {blue}In general, there are finite range
  corrections to Eq.~\protect \textup {\hbox {\mathsurround \z@ \protect
  \normalfont (\ignorespaces \ref {eq:Tex}\unskip \@@italiccorr )}}, which add
  a term $\sim m_{eX}r_e^2(E+\varepsilon _B)$ in the denominator~\cite
  {LevinsenBook15}. In the absence of other scales, the effective range
  $r_e\sim a_X$ and therefore such terms are not important when $|E+\varepsilon
  _B|\ll \varepsilon _{eX}$.}}\BibitemShut {Stop}%
\bibitem [{\citenamefont {Keldysh}(1979)}]{Keldysh1979}%
  \BibitemOpen
  \bibfield  {author} {\bibinfo {author} {\bibfnamefont {L.~V.}\ \bibnamefont
  {Keldysh}},\ }\bibfield  {title} {\bibinfo {title} {\emph {Coulomb
  interaction in thin semiconductor and semimetal films}},\ }\href@noop {}
  {\bibfield  {journal} {\bibinfo  {journal} {Soviet Journal of Experimental
  and Theoretical Physics Letters}\ }\textbf {\bibinfo {volume} {29}},\
  \bibinfo {pages} {658} (\bibinfo {year} {1979})}\BibitemShut {NoStop}%
\bibitem [{\citenamefont {Fey}\ \emph {et~al.}(2020)\citenamefont {Fey},
  \citenamefont {Schmelcher}, \citenamefont {Imamoglu},\ and\ \citenamefont
  {Schmidt}}]{Fey2020}%
  \BibitemOpen
  \bibfield  {author} {\bibinfo {author} {\bibfnamefont {C.}~\bibnamefont
  {Fey}}, \bibinfo {author} {\bibfnamefont {P.}~\bibnamefont {Schmelcher}},
  \bibinfo {author} {\bibfnamefont {A.}~\bibnamefont {Imamoglu}}, \ and\
  \bibinfo {author} {\bibfnamefont {R.}~\bibnamefont {Schmidt}},\ }\bibfield
  {title} {\bibinfo {title} {\emph {Theory of exciton-electron scattering in
  atomically thin semiconductors}},\ }\href {\doibase
  10.1103/PhysRevB.101.195417} {\bibfield  {journal} {\bibinfo  {journal}
  {Phys. Rev. B}\ }\textbf {\bibinfo {volume} {101}},\ \bibinfo {pages}
  {195417} (\bibinfo {year} {2020})}\BibitemShut {NoStop}%
\bibitem [{\citenamefont {Sergeev}\ and\ \citenamefont
  {Suris}(2001)}]{Sergeev2001}%
  \BibitemOpen
  \bibfield  {author} {\bibinfo {author} {\bibfnamefont {R.}~\bibnamefont
  {Sergeev}}\ and\ \bibinfo {author} {\bibfnamefont {R.}~\bibnamefont
  {Suris}},\ }\bibfield  {title} {\bibinfo {title} {\emph {The Triplet State of
  X+ Trion in 2D Quantum Wells}},\ }\href {\doibase
  10.1002/1521-3951(200110)227:2<387::AID-PSSB387>3.0.CO;2-0} {\bibfield
  {journal} {\bibinfo  {journal} {physica status solidi (b)}\ }\textbf
  {\bibinfo {volume} {227}},\ \bibinfo {pages} {387} (\bibinfo {year}
  {2001})}\BibitemShut {NoStop}%
\bibitem [{\citenamefont {Ganchev}\ \emph {et~al.}(2015)\citenamefont
  {Ganchev}, \citenamefont {Drummond}, \citenamefont {Aleiner},\ and\
  \citenamefont {Fal'ko}}]{Ganchev2015}%
  \BibitemOpen
  \bibfield  {author} {\bibinfo {author} {\bibfnamefont {B.}~\bibnamefont
  {Ganchev}}, \bibinfo {author} {\bibfnamefont {N.}~\bibnamefont {Drummond}},
  \bibinfo {author} {\bibfnamefont {I.}~\bibnamefont {Aleiner}}, \ and\
  \bibinfo {author} {\bibfnamefont {V.}~\bibnamefont {Fal'ko}},\ }\bibfield
  {title} {\bibinfo {title} {\emph {Three-Particle Complexes in Two-Dimensional
  Semiconductors}},\ }\href {\doibase 10.1103/PhysRevLett.114.107401}
  {\bibfield  {journal} {\bibinfo  {journal} {Phys. Rev. Lett.}\ }\textbf
  {\bibinfo {volume} {114}},\ \bibinfo {pages} {107401} (\bibinfo {year}
  {2015})}\BibitemShut {NoStop}%
\bibitem [{\citenamefont {Courtade}\ \emph {et~al.}(2017)\citenamefont
  {Courtade}, \citenamefont {Semina}, \citenamefont {Manca}, \citenamefont
  {Glazov}, \citenamefont {Robert}, \citenamefont {Cadiz}, \citenamefont
  {Wang}, \citenamefont {Taniguchi}, \citenamefont {Watanabe}, \citenamefont
  {Pierre}, \citenamefont {Escoffier}, \citenamefont {Ivchenko}, \citenamefont
  {Renucci}, \citenamefont {Marie}, \citenamefont {Amand},\ and\ \citenamefont
  {Urbaszek}}]{Courtade2017}%
  \BibitemOpen
  \bibfield  {author} {\bibinfo {author} {\bibfnamefont {E.}~\bibnamefont
  {Courtade}}, \bibinfo {author} {\bibfnamefont {M.}~\bibnamefont {Semina}},
  \bibinfo {author} {\bibfnamefont {M.}~\bibnamefont {Manca}}, \bibinfo
  {author} {\bibfnamefont {M.~M.}\ \bibnamefont {Glazov}}, \bibinfo {author}
  {\bibfnamefont {C.}~\bibnamefont {Robert}}, \bibinfo {author} {\bibfnamefont
  {F.}~\bibnamefont {Cadiz}}, \bibinfo {author} {\bibfnamefont
  {G.}~\bibnamefont {Wang}}, \bibinfo {author} {\bibfnamefont {T.}~\bibnamefont
  {Taniguchi}}, \bibinfo {author} {\bibfnamefont {K.}~\bibnamefont {Watanabe}},
  \bibinfo {author} {\bibfnamefont {M.}~\bibnamefont {Pierre}}, \bibinfo
  {author} {\bibfnamefont {W.}~\bibnamefont {Escoffier}}, \bibinfo {author}
  {\bibfnamefont {E.~L.}\ \bibnamefont {Ivchenko}}, \bibinfo {author}
  {\bibfnamefont {P.}~\bibnamefont {Renucci}}, \bibinfo {author} {\bibfnamefont
  {X.}~\bibnamefont {Marie}}, \bibinfo {author} {\bibfnamefont
  {T.}~\bibnamefont {Amand}}, \ and\ \bibinfo {author} {\bibfnamefont
  {B.}~\bibnamefont {Urbaszek}},\ }\bibfield  {title} {\bibinfo {title} {\emph
  {Charged excitons in monolayer ${\mathrm{WSe}}_{2}$: Experiment and
  theory}},\ }\href {\doibase 10.1103/PhysRevB.96.085302} {\bibfield  {journal}
  {\bibinfo  {journal} {Phys. Rev. B}\ }\textbf {\bibinfo {volume} {96}},\
  \bibinfo {pages} {085302} (\bibinfo {year} {2017})}\BibitemShut {NoStop}%
\bibitem [{\citenamefont {Pricoupenko}\ and\ \citenamefont
  {Pedri}(2010)}]{Pricoupenko2010}%
  \BibitemOpen
  \bibfield  {author} {\bibinfo {author} {\bibfnamefont {L.}~\bibnamefont
  {Pricoupenko}}\ and\ \bibinfo {author} {\bibfnamefont {P.}~\bibnamefont
  {Pedri}},\ }\bibfield  {title} {\bibinfo {title} {\emph {Universal
  ($1+2$)-body bound states in planar atomic waveguides}},\ }\href {\doibase
  10.1103/PhysRevA.82.033625} {\bibfield  {journal} {\bibinfo  {journal} {Phys.
  Rev. A}\ }\textbf {\bibinfo {volume} {82}},\ \bibinfo {pages} {033625}
  (\bibinfo {year} {2010})}\BibitemShut {NoStop}%
\bibitem [{\citenamefont {Berman}\ \emph {et~al.}(2010)\citenamefont {Berman},
  \citenamefont {Kezerashvili},\ and\ \citenamefont {Lozovik}}]{OlegPRB2010}%
  \BibitemOpen
  \bibfield  {author} {\bibinfo {author} {\bibfnamefont {O.~L.}\ \bibnamefont
  {Berman}}, \bibinfo {author} {\bibfnamefont {R.~Y.}\ \bibnamefont
  {Kezerashvili}}, \ and\ \bibinfo {author} {\bibfnamefont {Y.~E.}\
  \bibnamefont {Lozovik}},\ }\bibfield  {title} {\bibinfo {title} {\emph {Drag
  effects in a system of electrons and microcavity polaritons}},\ }\href
  {\doibase 10.1103/PhysRevB.82.125307} {\bibfield  {journal} {\bibinfo
  {journal} {Phys. Rev. B}\ }\textbf {\bibinfo {volume} {82}},\ \bibinfo
  {pages} {125307} (\bibinfo {year} {2010})}\BibitemShut {NoStop}%
\bibitem [{\citenamefont {Laussy}\ \emph {et~al.}(2010)\citenamefont {Laussy},
  \citenamefont {Kavokin},\ and\ \citenamefont {Shelykh}}]{LaussyPRL10}%
  \BibitemOpen
  \bibfield  {author} {\bibinfo {author} {\bibfnamefont {F.~P.}\ \bibnamefont
  {Laussy}}, \bibinfo {author} {\bibfnamefont {A.~V.}\ \bibnamefont {Kavokin}},
  \ and\ \bibinfo {author} {\bibfnamefont {I.~A.}\ \bibnamefont {Shelykh}},\
  }\bibfield  {title} {\bibinfo {title} {\emph {Exciton-Polariton Mediated
  Superconductivity}},\ }\href {\doibase 10.1103/PhysRevLett.104.106402}
  {\bibfield  {journal} {\bibinfo  {journal} {Phys. Rev. Lett.}\ }\textbf
  {\bibinfo {volume} {104}},\ \bibinfo {pages} {106402} (\bibinfo {year}
  {2010})}\BibitemShut {NoStop}%
\bibitem [{\citenamefont {Cotle\ifmmode~\mbox{\c{t}}\else \c{t}\fi{}}\ \emph
  {et~al.}(2016)\citenamefont {Cotle\ifmmode~\mbox{\c{t}}\else \c{t}\fi{}},
  \citenamefont {Zeytino\ifmmode~\check{g}\else \v{g}\fi{}lu}, \citenamefont
  {Sigrist}, \citenamefont {Demler},\ and\ \citenamefont
  {Imamo\ifmmode~\check{g}\else \v{g}\fi{}lu}}]{CotlePRB16}%
  \BibitemOpen
  \bibfield  {author} {\bibinfo {author} {\bibfnamefont {O.}~\bibnamefont
  {Cotle\ifmmode~\mbox{\c{t}}\else \c{t}\fi{}}}, \bibinfo {author}
  {\bibfnamefont {S.}~\bibnamefont {Zeytino\ifmmode~\check{g}\else
  \v{g}\fi{}lu}}, \bibinfo {author} {\bibfnamefont {M.}~\bibnamefont
  {Sigrist}}, \bibinfo {author} {\bibfnamefont {E.}~\bibnamefont {Demler}}, \
  and\ \bibinfo {author} {\bibfnamefont {A.}~\bibnamefont
  {Imamo\ifmmode~\check{g}\else \v{g}\fi{}lu}},\ }\bibfield  {title} {\bibinfo
  {title} {\emph {Superconductivity and other collective phenomena in a hybrid
  Bose-Fermi mixture formed by a polariton condensate and an electron system in
  two dimensions}},\ }\href {\doibase 10.1103/PhysRevB.93.054510} {\bibfield
  {journal} {\bibinfo  {journal} {Phys. Rev. B}\ }\textbf {\bibinfo {volume}
  {93}},\ \bibinfo {pages} {054510} (\bibinfo {year} {2016})}\BibitemShut
  {NoStop}%
\bibitem [{\citenamefont {Sidler}\ \emph {et~al.}(2017)\citenamefont {Sidler},
  \citenamefont {Back}, \citenamefont {Cotlet}, \citenamefont {Srivastava},
  \citenamefont {Fink}, \citenamefont {Kroner}, \citenamefont {Demler},\ and\
  \citenamefont {Imamoglu}}]{SidlerNatPhys16}%
  \BibitemOpen
  \bibfield  {author} {\bibinfo {author} {\bibfnamefont {M.}~\bibnamefont
  {Sidler}}, \bibinfo {author} {\bibfnamefont {P.}~\bibnamefont {Back}},
  \bibinfo {author} {\bibfnamefont {O.}~\bibnamefont {Cotlet}}, \bibinfo
  {author} {\bibfnamefont {A.}~\bibnamefont {Srivastava}}, \bibinfo {author}
  {\bibfnamefont {T.}~\bibnamefont {Fink}}, \bibinfo {author} {\bibfnamefont
  {M.}~\bibnamefont {Kroner}}, \bibinfo {author} {\bibfnamefont
  {E.}~\bibnamefont {Demler}}, \ and\ \bibinfo {author} {\bibfnamefont
  {A.}~\bibnamefont {Imamoglu}},\ }\bibfield  {title} {\bibinfo {title} {\emph
  {Fermi polaron-polaritons in charge-tunable atomically thin
  semiconductors}},\ }\href {https://doi.org/10.1038/nphys3949} {\bibfield
  {journal} {\bibinfo  {journal} {Nature Physics}\ }\textbf {\bibinfo {volume}
  {13}},\ \bibinfo {pages} {255} (\bibinfo {year} {2017})}\BibitemShut
  {NoStop}%
\bibitem [{\citenamefont {Chervy}\ \emph {et~al.}(2020)\citenamefont {Chervy},
  \citenamefont {Kn\"uppel}, \citenamefont {Abbaspour}, \citenamefont
  {Lupatini}, \citenamefont {F\"alt}, \citenamefont {Wegscheider},
  \citenamefont {Kroner},\ and\ \citenamefont {Imamo\ifmmode~\check{g}\else
  \v{g}\fi{}lu}}]{Chervy2020}%
  \BibitemOpen
  \bibfield  {author} {\bibinfo {author} {\bibfnamefont {T.}~\bibnamefont
  {Chervy}}, \bibinfo {author} {\bibfnamefont {P.}~\bibnamefont {Kn\"uppel}},
  \bibinfo {author} {\bibfnamefont {H.}~\bibnamefont {Abbaspour}}, \bibinfo
  {author} {\bibfnamefont {M.}~\bibnamefont {Lupatini}}, \bibinfo {author}
  {\bibfnamefont {S.}~\bibnamefont {F\"alt}}, \bibinfo {author} {\bibfnamefont
  {W.}~\bibnamefont {Wegscheider}}, \bibinfo {author} {\bibfnamefont
  {M.}~\bibnamefont {Kroner}}, \ and\ \bibinfo {author} {\bibfnamefont
  {A.}~\bibnamefont {Imamo\ifmmode~\check{g}\else \v{g}\fi{}lu}},\ }\bibfield
  {title} {\bibinfo {title} {\emph {Accelerating Polaritons with External
  Electric and Magnetic Fields}},\ }\href {\doibase 10.1103/PhysRevX.10.011040}
  {\bibfield  {journal} {\bibinfo  {journal} {Phys. Rev. X}\ }\textbf {\bibinfo
  {volume} {10}},\ \bibinfo {pages} {011040} (\bibinfo {year}
  {2020})}\BibitemShut {NoStop}%
\bibitem [{\citenamefont {Petrov}\ \emph {et~al.}(2003)\citenamefont {Petrov},
  \citenamefont {Baranov},\ and\ \citenamefont {Shlyapnikov}}]{Petrov2003}%
  \BibitemOpen
  \bibfield  {author} {\bibinfo {author} {\bibfnamefont {D.~S.}\ \bibnamefont
  {Petrov}}, \bibinfo {author} {\bibfnamefont {M.~A.}\ \bibnamefont {Baranov}},
  \ and\ \bibinfo {author} {\bibfnamefont {G.~V.}\ \bibnamefont
  {Shlyapnikov}},\ }\bibfield  {title} {\bibinfo {title} {\emph {Superfluid
  transition in quasi-two-dimensional Fermi gases}},\ }\href {\doibase
  10.1103/PhysRevA.67.031601} {\bibfield  {journal} {\bibinfo  {journal} {Phys.
  Rev. A}\ }\textbf {\bibinfo {volume} {67}},\ \bibinfo {pages} {031601}
  (\bibinfo {year} {2003})}\BibitemShut {NoStop}%
\bibitem [{\citenamefont {Petrov}\ \emph {et~al.}(2005)\citenamefont {Petrov},
  \citenamefont {Salomon},\ and\ \citenamefont {Shlyapnikov}}]{Petrov2005}%
  \BibitemOpen
  \bibfield  {author} {\bibinfo {author} {\bibfnamefont {D.~S.}\ \bibnamefont
  {Petrov}}, \bibinfo {author} {\bibfnamefont {C.}~\bibnamefont {Salomon}}, \
  and\ \bibinfo {author} {\bibfnamefont {G.~V.}\ \bibnamefont {Shlyapnikov}},\
  }\bibfield  {title} {\bibinfo {title} {\emph {Diatomic molecules in ultracold
  Fermi gases{\textemdash}novel composite bosons}},\ }\href {\doibase
  10.1088/0953-4075/38/9/014} {\bibfield  {journal} {\bibinfo  {journal}
  {Journal of Physics B: Atomic, Molecular and Optical Physics}\ }\textbf
  {\bibinfo {volume} {38}},\ \bibinfo {pages} {S645} (\bibinfo {year}
  {2005})}\BibitemShut {NoStop}%
\bibitem [{\citenamefont {Takemura}\ \emph {et~al.}(2017)\citenamefont
  {Takemura}, \citenamefont {Anderson}, \citenamefont {Navadeh-Toupchi},
  \citenamefont {Oberli}, \citenamefont {Portella-Oberli},\ and\ \citenamefont
  {Deveaud}}]{Takemura2017}%
  \BibitemOpen
  \bibfield  {author} {\bibinfo {author} {\bibfnamefont {N.}~\bibnamefont
  {Takemura}}, \bibinfo {author} {\bibfnamefont {M.~D.}\ \bibnamefont
  {Anderson}}, \bibinfo {author} {\bibfnamefont {M.}~\bibnamefont
  {Navadeh-Toupchi}}, \bibinfo {author} {\bibfnamefont {D.~Y.}\ \bibnamefont
  {Oberli}}, \bibinfo {author} {\bibfnamefont {M.~T.}\ \bibnamefont
  {Portella-Oberli}}, \ and\ \bibinfo {author} {\bibfnamefont {B.}~\bibnamefont
  {Deveaud}},\ }\bibfield  {title} {\bibinfo {title} {\emph {Spin anisotropic
  interactions of lower polaritons in the vicinity of polaritonic Feshbach
  resonance}},\ }\href {\doibase 10.1103/PhysRevB.95.205303} {\bibfield
  {journal} {\bibinfo  {journal} {Phys. Rev. B}\ }\textbf {\bibinfo {volume}
  {95}},\ \bibinfo {pages} {205303} (\bibinfo {year} {2017})}\BibitemShut
  {NoStop}%
\bibitem [{\citenamefont {Navadeh-Toupchi}\ \emph {et~al.}(2019)\citenamefont
  {Navadeh-Toupchi}, \citenamefont {Takemura}, \citenamefont {Anderson},
  \citenamefont {Oberli},\ and\ \citenamefont
  {Portella-Oberli}}]{Navadeh-Toupchi2019}%
  \BibitemOpen
  \bibfield  {author} {\bibinfo {author} {\bibfnamefont {M.}~\bibnamefont
  {Navadeh-Toupchi}}, \bibinfo {author} {\bibfnamefont {N.}~\bibnamefont
  {Takemura}}, \bibinfo {author} {\bibfnamefont {M.~D.}\ \bibnamefont
  {Anderson}}, \bibinfo {author} {\bibfnamefont {D.~Y.}\ \bibnamefont
  {Oberli}}, \ and\ \bibinfo {author} {\bibfnamefont {M.~T.}\ \bibnamefont
  {Portella-Oberli}},\ }\bibfield  {title} {\bibinfo {title} {\emph
  {Polaritonic Cross Feshbach Resonance}},\ }\href {\doibase
  10.1103/PhysRevLett.122.047402} {\bibfield  {journal} {\bibinfo  {journal}
  {Phys. Rev. Lett.}\ }\textbf {\bibinfo {volume} {122}},\ \bibinfo {pages}
  {047402} (\bibinfo {year} {2019})}\BibitemShut {NoStop}%
\bibitem [{\citenamefont {Levinsen}\ \emph
  {et~al.}(2019{\natexlab{b}})\citenamefont {Levinsen}, \citenamefont
  {Marchetti}, \citenamefont {Keeling},\ and\ \citenamefont
  {Parish}}]{Levinsen2019}%
  \BibitemOpen
  \bibfield  {author} {\bibinfo {author} {\bibfnamefont {J.}~\bibnamefont
  {Levinsen}}, \bibinfo {author} {\bibfnamefont {F.~M.}\ \bibnamefont
  {Marchetti}}, \bibinfo {author} {\bibfnamefont {J.}~\bibnamefont {Keeling}},
  \ and\ \bibinfo {author} {\bibfnamefont {M.~M.}\ \bibnamefont {Parish}},\
  }\bibfield  {title} {\bibinfo {title} {\emph {Spectroscopic Signatures of
  Quantum Many-Body Correlations in Polariton Microcavities}},\ }\href
  {\doibase 10.1103/PhysRevLett.123.266401} {\bibfield  {journal} {\bibinfo
  {journal} {Phys. Rev. Lett.}\ }\textbf {\bibinfo {volume} {123}},\ \bibinfo
  {pages} {266401} (\bibinfo {year} {2019}{\natexlab{b}})}\BibitemShut
  {NoStop}%
\bibitem [{\citenamefont {Bastarrachea-Magnani}\ \emph
  {et~al.}(2019)\citenamefont {Bastarrachea-Magnani}, \citenamefont
  {Camacho-Guardian}, \citenamefont {Wouters},\ and\ \citenamefont
  {Bruun}}]{Bastarrachea-Magnani2020}%
  \BibitemOpen
  \bibfield  {author} {\bibinfo {author} {\bibfnamefont {M.~A.}\ \bibnamefont
  {Bastarrachea-Magnani}}, \bibinfo {author} {\bibfnamefont {A.}~\bibnamefont
  {Camacho-Guardian}}, \bibinfo {author} {\bibfnamefont {M.}~\bibnamefont
  {Wouters}}, \ and\ \bibinfo {author} {\bibfnamefont {G.~M.}\ \bibnamefont
  {Bruun}},\ }\bibfield  {title} {\bibinfo {title} {\emph {Strong interactions
  and biexcitons in a polariton mixture}},\ }\href {\doibase
  10.1103/PhysRevB.100.195301} {\bibfield  {journal} {\bibinfo  {journal}
  {Phys. Rev. B}\ }\textbf {\bibinfo {volume} {100}},\ \bibinfo {pages}
  {195301} (\bibinfo {year} {2019})}\BibitemShut {NoStop}%
\bibitem [{\citenamefont {Khurgin}(2001)}]{Khurgin2001}%
  \BibitemOpen
  \bibfield  {author} {\bibinfo {author} {\bibfnamefont {J.}~\bibnamefont
  {Khurgin}},\ }\bibfield  {title} {\bibinfo {title} {\emph {Excitonic radius
  in the cavity polariton in the regime of very strong coupling}},\ }\href
  {\doibase https://doi.org/10.1016/S0038-1098(00)00469-5} {\bibfield
  {journal} {\bibinfo  {journal} {Solid State Communications}\ }\textbf
  {\bibinfo {volume} {117}},\ \bibinfo {pages} {307 } (\bibinfo {year}
  {2001})}\BibitemShut {NoStop}%
\end{thebibliography}%

\end{document}